\documentclass[a4paper,fleqn,usenatbib]{mnras}

\usepackage{newtxtext,newtxmath}
\usepackage[flushleft]{threeparttable}
\usepackage{times,epsfig,natbib,amssymb,amsmath,graphics,longtable,lscape,threeparttable}

\usepackage[T1]{fontenc}
\usepackage{ae,aecompl}

\usepackage{graphicx}	
\usepackage{amsmath}	
\usepackage{nccmath}
\usepackage{amssymb}	
\usepackage{pdflscape}	
\usepackage{multicol}
\usepackage{soul}

\usepackage{eso-pic}

\AddToShipoutPictureBG*{%
  \AtPageUpperLeft{%
    \hspace{0.75\paperwidth}%
    \raisebox{-3.5\baselineskip}{%
      \makebox[0pt][l]{\textnormal{DES 2018-0399}}
}}}%

\AddToShipoutPictureBG*{%
  \AtPageUpperLeft{%
    \hspace{0.75\paperwidth}%
    \raisebox{-4.5\baselineskip}{%
      \makebox[0pt][l]{\textnormal{Fermilab PUB-20-016-AE}}
}}}%

\title[High-redshift SN~II Hubble diagram using DES-SN]{Studying Type II supernovae as cosmological standard candles using the Dark Energy Survey}

\author[de Jaeger et al.]{
\parbox{\textwidth}{
\Large
T.~de Jaeger,$^{1,2}$\thanks{E-mail: tdejaeger@berkeley.edu},
L.~Galbany,$^{3}$
S. Gonz\'alez-Gait\'an,$^{4}$
R.~Kessler,$^{5,6}$
A.~V.~Filippenko,$^{1,7}$
F.~F\"orster,$^{8,9}$
M.~Hamuy,$^{10}$
P.~J.~Brown,$^{11}$
T.~M.~Davis,$^{12}$
C.~P.~Guti\'errez,$^{13}$
C.~Inserra,$^{14}$
G.~F.~Lewis,$^{15}$
A.~M\"oller,$^{16}$
D.~Scolnic,$^{17}$
M.~Smith,$^{13}$
D.~Brout,$^{18}$
D.~Carollo,$^{19}$
R.~J.~Foley,$^{20}$
K.~Glazebrook,$^{21}$
S.~R.~Hinton,$^{12}$
E.~Macaulay,$^{22}$
B.~Nichol,$^{22}$
M.~Sako,$^{18}$
N.~E.~Sommer,$^{23}$
B.~E.~Tucker,$^{23}$
T.~M.~C.~Abbott,$^{24}$
M.~Aguena,$^{25,26}$
S.~Allam,$^{27}$
J.~Annis,$^{27}$
S.~Avila,$^{28}$
E.~Bertin,$^{29,30}$
S.~Bhargava,$^{31}$
D.~Brooks,$^{32}$
D.~L.~Burke,$^{33,34}$
A.~Carnero~Rosell,$^{35}$
M.~Carrasco~Kind,$^{36,37}$
J.~Carretero,$^{38}$
M.~Costanzi,$^{39,40}$
M.~Crocce,$^{41,42}$
L.~N.~da~Costa,$^{26,43}$
J.~De~Vicente,$^{35}$
S.~Desai,$^{44}$
H.~T.~Diehl,$^{27}$
P.~Doel,$^{32}$
A.~Drlica-Wagner,$^{5,27,6}$
T.~F.~Eifler,$^{45,46}$
J.~Estrada,$^{27}$
S.~Everett,$^{20}$
B.~Flaugher,$^{27}$
P.~Fosalba,$^{41,42}$
J.~Frieman,$^{27,6}$
J.~Garc\'ia-Bellido,$^{28}$
E.~Gaztanaga,$^{42}$
D.~Gruen,$^{47,33,34}$
R.~A.~Gruendl,$^{36,37}$
J.~Gschwend,$^{26,43}$
G.~Gutierrez,$^{27}$
W.~G.~Hartley,$^{32,48}$
D.~L.~Hollowood,$^{20}$
K.~Honscheid,$^{49,50}$
D.~J.~James,$^{51}$
K.~Kuehn,$^{52,53}$
N.~Kuropatkin,$^{27}$
T.~S.~Li,$^{54,55}$
M.~Lima,$^{25,26}$
M.~A.~G.~Maia,$^{26,43}$
F.~Menanteau,$^{36,37}$
R.~Miquel,$^{56,38}$
A.~Palmese,$^{27,6}$
F.~Paz-Chinch\'{o}n,$^{36,37}$
A.~A.~Plazas,$^{54}$
A.~K.~Romer,$^{31}$
A.~Roodman,$^{33,34}$
E.~Sanchez,$^{35}$
V.~Scarpine,$^{27}$
M.~Schubnell,$^{57}$
S.~Serrano,$^{41,42}$
I.~Sevilla-Noarbe,$^{35}$
M.~Soares-Santos,$^{58}$
E.~Suchyta,$^{59}$
M.~E.~C.~Swanson,$^{37}$
G.~Tarle,$^{57}$
D.~Thomas,$^{22}$
D.~L.~Tucker,$^{27}$
T.~N.~Varga,$^{60,61}$
A.~R.~Walker,$^{24}$
J.~Weller,$^{60,61}$
and R.~Wilkinson$^{31}$
\begin{center} (DES Collaboration) \end{center}
}
\\
\\
Affiliations are listed at the end of the paper.
}

\date{Accepted XXX. Received YYY; in original form ZZZ}

\pubyear{2020}

\begin{document}
\label{firstpage}
\pagerange{\pageref{firstpage}--\pageref{lastpage}}

\maketitle

\begin{abstract}
\noindent
Despite vast improvements in the measurement of the cosmological parameters, the nature of dark energy and an accurate value of the Hubble constant (H$_0$) in the Hubble-Lema\^itre law remain unknown. To break the current impasse, it is necessary to develop as many independent techniques as possible, such as the use of Type II supernovae (SNe~II). The goal of this paper is to demonstrate the utility of SNe~II for deriving accurate extragalactic distances, which will be an asset for the next generation of telescopes where more-distant SNe~II will be discovered. More specifically, we present a sample from the Dark Energy Survey Supernova Program (DES-SN) consisting of 15 SNe~II with photometric and spectroscopic information spanning a redshift range up to 0.35. Combining our DES SNe with publicly available samples, and using the standard candle method (SCM), we construct the largest available Hubble diagram with SNe~II in the Hubble flow (70 SNe~II) and find an observed dispersion of 0.27\,mag. We demonstrate that adding a colour term to the SN~II standardisation does not reduce the scatter in the Hubble diagram. Although SNe~II are viable as distance indicators, this work points out important issues for improving their utility as independent extragalactic beacons: find new correlations, define a more standard subclass of SNe~II, construct new SN~II templates, and dedicate more observing time to high-redshift SNe~II. Finally, for the first time, we perform simulations to estimate the redshift-dependent distance-modulus bias due to selection effects.
\end{abstract}



\section{Introduction}

Measuring accurate extragalactic distances is one of the most challenging tasks in astronomy but remains one of the best observational probes to understand the Universe's content. Traditionally, cosmic distances are derived applying the inverse-square law to astrophysical sources with known and fixed absolute magnitudes (i.e., standard candles) or with absolute magnitudes which can be calibrated (i.e., standardisable candles). For more than two decades, Type Ia supernovae (hereafter SNe~Ia; \citealt{min41,elias85,filippenko97,howell11,maguire17}, and references therein) have been used as standardisable candles (e.g., \citealt{phillips93,hamuy95,hamuy96,riess96,perlmutter97}) to measure extragalactic distances with a precision of $\sim 5$\%--6\%\footnote{Using only SNe~Ia but not combined with measurements of the cosmic microwave background radiation.} (e.g., \citealt{betoule14,rubin16,scolnic18,des_cosmoIa19}). In 1998, observations of SNe~Ia led to the measurement of the Universe's expansion history and revealed the surprising accelerated growth rate of the Universe driven by an unknown effect generally attributed to dark energy \citep{riess98,schmidt98,perlmutter99}.

However, although SN~Ia cosmology is one of the most interesting and prolific fields in astronomy, the nature of dark energy remains unknown. Furthermore, recently a new debate \citep[e.g.,][and references therein]{davis19,riess19b} on the precise value of the Universe's expansion rate (the Hubble constant H$_0$ in the Hubble-Lema\^itre law) appeared in the literature, with the disagreement between the local measurement from SNe~Ia calibrated using Cepheid variable stars \citep{riess16,riess18a,riess18b,burns18,riess19}, from strong-lensing SN studies \citep{shajib19} or strong-lensing quasar studies (HoLICOW; \citealt{bonvin17}) and with the high-redshift estimate from baryon acoustic oscillations (BAO; \citealt{blake03,seo03}) calibrated using the cosmic microwave background radiation (CMB; \citealt{fixsen96,jaffe01,spergel07,bennett03,planck18}). The significance of this discrepancy has now increased to $>4.4\sigma$ \citep{riess19}, and surprisingly, this disagreement does not appear to be due to known systematic errors. Thus, further improvement to constrain H$_0$ and the cosmological parameters requires developing as many independent methods as possible, including gravitational wave sources (``standard sirens''; \citealt{abbott17}) or superluminous supernovae \citep{inserra14}. With different systematic errors, those independent values will favour the local measurement or the high-redshift estimate (or perhaps some intermediate value) and will be critical to understanding the current discrepancy.

Another interesting, independent method for deriving accurate distances and measuring cosmological parameters is the use of SNe~II.\footnote{SNe~II refer to the two subgroups, SNe~IIP and SNe~IIL (see \citealt{anderson14a,sanders15,valenti16,galbany16a,dejaeger19}). SNe~IIb, SNe~IIn, and SN~1987A-like are excluded.} SNe~II are characterised by the presence of strong hydrogen (H) features in their spectra (see \citealt{filippenko97,filippenko00} and \citealt{galyam17} for overviews), and a plateau of varying steepness and length in their light curves \citep{barbon79}.

Despite SNe~II being less luminous than SNe~Ia \citep{richardson14}, their use as cosmic distance indicators is motivated by the facts that (1) they are more abundant than SNe~Ia \citep{li2011,graur17b}, and (2) the physics and the nature of their progenitors are better understood. It has been proven that their progenitors are red supergiants in late-type galaxies which have retained a significant fraction of their H envelopes \citep[e.g.,][]{grassberg71,chevalier76,falk77,vandyk03,smartt09a}. Unlike SNe~Ia, for which no direct progenitors have been detected, SN~II~progenitors have been constrained, and the understanding of the explosion mechanisms of SN~II has made remarkable progress in the past few decades \citep[e.g.,][]{woosley95,janka01,janka07}.

In the last 20 years, after being overshadowed by the well-studied SNe~Ia owing to the difficulty in getting a large sample of sufficiently high-quality data, different distance measurement methods using SNe~II have been proposed and tested \citep[e.g.,][and references therein]{nugent17}: the expanding photosphere method (EPM; \citealt{kirshner74,gall17}), the standard (actually standardisable) candle method (SCM; \citealt{hamuy02}), the photospheric magnitude method (PMM; \citealt{rodriguez14,rodriguez19a}), and most recently the photometric colour method (PCM; \citealt{dejaeger15b,dejaeger17a}). In this paper, we focus our effort on two methods: the SCM, which is the most common and most accurate technique used to derive SN~II distances, and the PCM, being the unique purely photometric method in the literature and a potential asset for the next generation of surveys such as those with the Large Synoptic Survey Telescope \citep[LSST;][]{ivezic09} and the Subaru/Hyper Suprime-Cam \citep[HSC;][]{miyazaki12,aihara18a}.

The SCM is an empirical method based on the observed correlation between SN~II luminosity and photospheric expansion velocity during the plateau phase: more luminous SNe~II have higher velocities \citep{hamuy02}. This relation is physically well-understood: more luminous SNe have their hydrogen recombination front at a larger radius and thus the velocity of the photosphere is greater \citep{kasen09}. Currently, many other studies have refined the SCM by (1) using a colour correction to perform an extinction correction \citep{nugent06,poznanski09,maguire10,olivares10,andrea10,dejaeger15b,gall17}, (2) measuring the velocity through the absorption minimum of P-Cygni features of different lines (e.g., H$\beta$ $\lambda$4861, \ion{Fe}{II} $\lambda 5169$), (3) measuring the velocity using cross-correlation techniques \citep{poznanski10,dejaeger17a}, and (4) using hierarchical Gaussian processes to interpolate the magnitudes and colours at different epochs \citep{dejaeger17b}. All of these works have confirmed the utility of SNe~II as distance indicators, constructing a Hubble diagram with a dispersion of $\sim 10$--14\% in distance up to a redshift $z \approx 0.35$.

Unlike the SCM for which a spectrum is required to measure the velocity, the PCM is a purely photometric method with no input of spectral information. However, we supplement the photometric distance measurement with redshifts of the host galaxy as they are more accurate than the photometric redshifts. PCM is based on the empirical correlation between the slope of the light-curve plateau (hydrogen recombination phase) and the intrinsic brightness: more-luminous SNe~II have a steeper decline (\citealt{anderson14a} and see \citealt{Pejcha15a} for a theoretical explanation). First applied at low redshift ($z = 0.01$--0.04) by \citet{dejaeger15b}, PCM was successfully extended to higher redshifts ($z < 0.5$) by \citet{dejaeger17a}.

In this paper, we use a new sample from the Dark Energy Survey (DES) Supernova Program (DES-SN) to construct the largest SN~II Hubble diagram in the Hubble flow ($z > 0.01$) and to assess and develop the possibility of using SNe~II as distance indicators. We motivate the necessity for the SN community to dedicate specific programs for SN~II cosmology -- the current surveys are mostly designed for SN~Ia cosmology -- to improve methods and compare with SN~Ia results. Future deep surveys (e.g., with LSST) and ground-based telescopes for spectroscopy such as the Keck telescopes or the next generation of 25--39\,m telescopes (European Extremely Large Telescope, E-ELT, \citealt{gilmozzi07}; Giant Magellan Telescope, GMT, \citealt{johns12}; Thirty Meter Telescope, TMT, \citealt{sanders13}) will be extremely useful for high-redshift SN~II observations.\footnote{In this paper, ``high redshift'' refers to $z \gtrsim 0.3$, which is considered to be medium redshift by the wider community.}

This paper is organised as follows. Section 2 contains a description of the data sample, and in Section 3 we briefly discuss the methods used to derive the Hubble diagram. We discuss our results using the SCM in Section 4, while in Section 5 those using the PCM are presented. Section 6 summarises our conclusions.

\section{Data Sample}

In this work, we update the Hubble diagram published by \citet{dejaeger17b} with SNe~II from DES-SN\footnote{\url{https://portal.nersc.gov/des-sn/}} \citep{bernstein12,brout19a,brout19b}. For completeness, readers are reminded that the sample from \citet{dejaeger17b} consists of SNe~II from four different surveys: the Carnegie Supernova Project-I [CSP-I\footnote{\url{http://csp.obs.carnegiescience.edu/}};]\citep{ham06}, the Sloan Digital Sky Survey-II SN Survey [SDSS-II\footnote{\url{http://classic.sdss.org/supernova/aboutsupernova.html}};]\citep{frieman08}, the Supernova Legacy Survey [SNLS\footnote{\url{http://cfht.hawaii.edu/SNLS/}};]\citep{astier06,perrett10}, and the Subaru HSC Survey \citep{miyazaki12,aihara18a}.

\subsection{Previous sample}\label{txt:AKS}

The previous sample used by \citet{dejaeger17b} consists of a total of 93 SNe~II. This includes 61 from CSP-1 (58 of which have spectra\footnote{Three (SN~2005es, SN~2005gk, and SN~2008F) have no spectrum older than 15\,d after the explosion, needed to measure the expansion velocity.}) (Contreras et al., in prep.), 16 from SDSS-II \citep{andrea10}, 15 unpublished SNe~II from SNLS (5 with spectroscopic information), and one from HSC \citep{dejaeger17b}. For more information about the different surveys and data-reduction procedures, the reader is referred to \citet{andrea10}, \citet{dejaeger17b,dejaeger17a}, \citet{stritzinger18}, and references therein. Note that in this work, we update the Hubble diagram of 93 SNe~II published by \citet{dejaeger17b}, with 15 new SNe~II from DES-SN (see Section \ref{txt:SCM_sample}).

All of the magnitudes were simultaneously corrected for Milky Way extinction ($A_{V,G}$; \citealt{schlafly11}), redshifts due to the expansion of the Universe (K-correction; \citealt{oke68,hamuy93,kim96,nugent02}), and differences between the photometric systems (S-correction; \citealt{stritzinger02}) using the cross-filter K-corrections defined by \citet{kim96}. For more details about these corrections, the reader is referred to \citet{nugent02}, \citet{hsiao07}, \citet{dejaeger17a}, and references therein.

Finally, in this work, we use the recalibrated CSP-I photometry that will be published in a definitive CSP-I data paper by Contreras et al. (in prep.), and the explosion dates for the CSP-I sample were updated using the new values published by \citet{gutierrez17a}.

\subsection{DES-SN 5-year survey}

The DES-SN was dedicated to search for astrophysical transients using the $\sim 3$ square degree Dark Energy Camera (DECam; \citealt{flaugher15}) mounted on the 4\,m Blanco telescope at the Cerro Tololo Inter-American Observatory in Chile. During 5 years (2013--2018), from August to January and with a typical cadence of 4--7 nights \citep{diehl16,diehl18}, 10 fields (see Table \ref{tab:DES_field}) were observed in the $g$, $r$, $i$, and $z$ passbands with a median limiting magnitude (respectively) of 23.7, 23.6, 23.5, and 23.3\,mag for the shallow fields (C1, C2, E1, E2, S1, S2, X1, and X2) and 24.6, 24.8, 24.7, and 24.4\,mag for deep fields (C3, X3). A survey overview can be found in \citet{kessler15}, and an overview of spectroscopic targeting of the first 3 years is given by \citet{dandrea18}.

\begin{table}
\center
\caption{Locations of the 10 DES-SN fields.} 
\begin{tabular}{lcc}
\hline
Field	& $\alpha$ (J2000) & $\delta$ (J2000) \\
Name &$\fh$ $\fm$ $\fs$ &$\circ$ $\farcm$ $\farcs$\\
\hline
E1 &00:31:29.9	&$-$43:00:34.6 \\
E2 &00:38:00.0	&$-$43:59:52.8 \\
S1 &02:51:16.8	&00:00:00.0 \\
S2 &02:44:46.7	&$-$00:59:18.2 \\
C1 &03:37:05.8	&$-$27:06:41.8 \\
C2 &03:37:05.8	&$-$29:05:18.2 \\
C3 &03:30:35.6	&$-$28:06:00.0 \\
X1 &02:17:54.2	&$-$04:55:46.2 \\
X2 &02:22:39.5	&$-$06:24:43.6 \\
X3 &02:25:48.0	&$-$04:36:00.0 \\
\hline
\label{tab:DES_field}
\end{tabular}
\end{table}
 
The 5-year photometric data were reduced using the Difference Imaging (DIFFIMG) pipeline following the \citet{kessler15} prescriptions. Final photometric points were obtained via point-spread-function (PSF) photometry after host-galaxy subtraction using deep template images from each individual SN image. 

Although the main science driver was to obtain high-quality light curves of thousands of SNe~Ia with the goal of measuring cosmological parameters, some SN~II spectroscopic follow-up observations were achieved. Spectra were obtained using the Magellan 6.5\,m Clay telescope at the Las Campanas Observatory in Chile, the Anglo-Australian 3.9\,m telescope situated at the Siding Spring Observatory in Australia, and the 10\,m Keck-II telescope on Maunakea in Hawaii. The Anglo-Australian 3.9\,m telescope spectra were obtained under the OzDES program \citep{yuan15} and reduced with 2dFDR \citep{AAT2015}, while the other spectra were reduced following standard procedures (bias subtraction, flat-field correction, one-dimensional extraction, wavelength calibration, and flux calibration) using \textit{IRAF}\footnote{\textit{IRAF} is distributed by the National Optical Astronomy Observatory, which is operated by the Association of Universities for Research in Astronomy (AURA), Inc., under a cooperative agreement with the U.S. National Science Foundation (NSF).} routines. Over 5 years, a total of 56 spectroscopically confirmed SNe~II were discovered by DES-SN.

\subsection{Standard Candle Method sample}\label{txt:SCM_sample}
 
Following \citet{andrea10}, the final DES-5yr SN~II sample adopted for the SCM was selected using five selection requirements (cuts): (1) a well-defined explosion date and a nondetection in the same observing season before the first detection of the SN, (2) photometric data up to 45\,d in the rest frame after the explosion (no light-curve extrapolation), (3) at least one spectrum taken between 13 and 90\,d (rest frame) to measure the H$\beta$ line velocity (see Section \ref{txt:methodology}), (4) their spectra must display clear hydrogen P-Cygni profiles, and (5) the light curves should not exhibit unusual features (such as SNe~IIb). In Appendix \ref{AppendixA}, Table \ref{tab:sample_tot} provides a list of all the spectroscopically confirmed SNe~II, and for each SN we indicate whether it passed the cuts. ``SCM'' is noted if the SN is useful for the SCM, while the SNe that failed are labeled with {\sc PHOT} (no photometric data up to 45\,d after the explosion), {\sc EXP} (no explosion date), {\sc SPEC} (no spectrum), {\sc P-Cygni} (no clear P-Cygni profile), or {\sc LC} (unusual light curves). 

From the 56 spectroscopically confirmed SNe~II, 15 passed the five cuts and are useful for our SCM analysis. One SN was rejected owing to the absence of a spectrum after 13\,d, 24 lack a precise explosion date\footnote{All 24 of these SNe~II do not have a nondetection in the same observing season before the first detection of the SN -- they were detected/observed at the beginning of the run in August.}, three lack sufficient photometry (last photometric point $< 45$\,d), one has a slowly rising light curve typical of SNe~IIb, and 12 do not exhibit clear P-Cygni profiles (generally affected by host-galaxy light). The low success rate (15/56 SNe~II) is not surprising, because out of the 56 spectroscopically confirmed SNe~II only 27 SNe~II are potentially useful for the SCM. As the main goal of the spectroscopy was to classify the object, the majority of spectra have low signal-to-noise ratios (S/N). In the future, with a survey dedicated to SNe~II and spectra of sufficient quality to measure the expansion velocities (see Section \ref{txt:methodology}), the rate of useful SNe~II for SCM will increase.

\begin{figure}
\includegraphics[width=1.0\columnwidth]{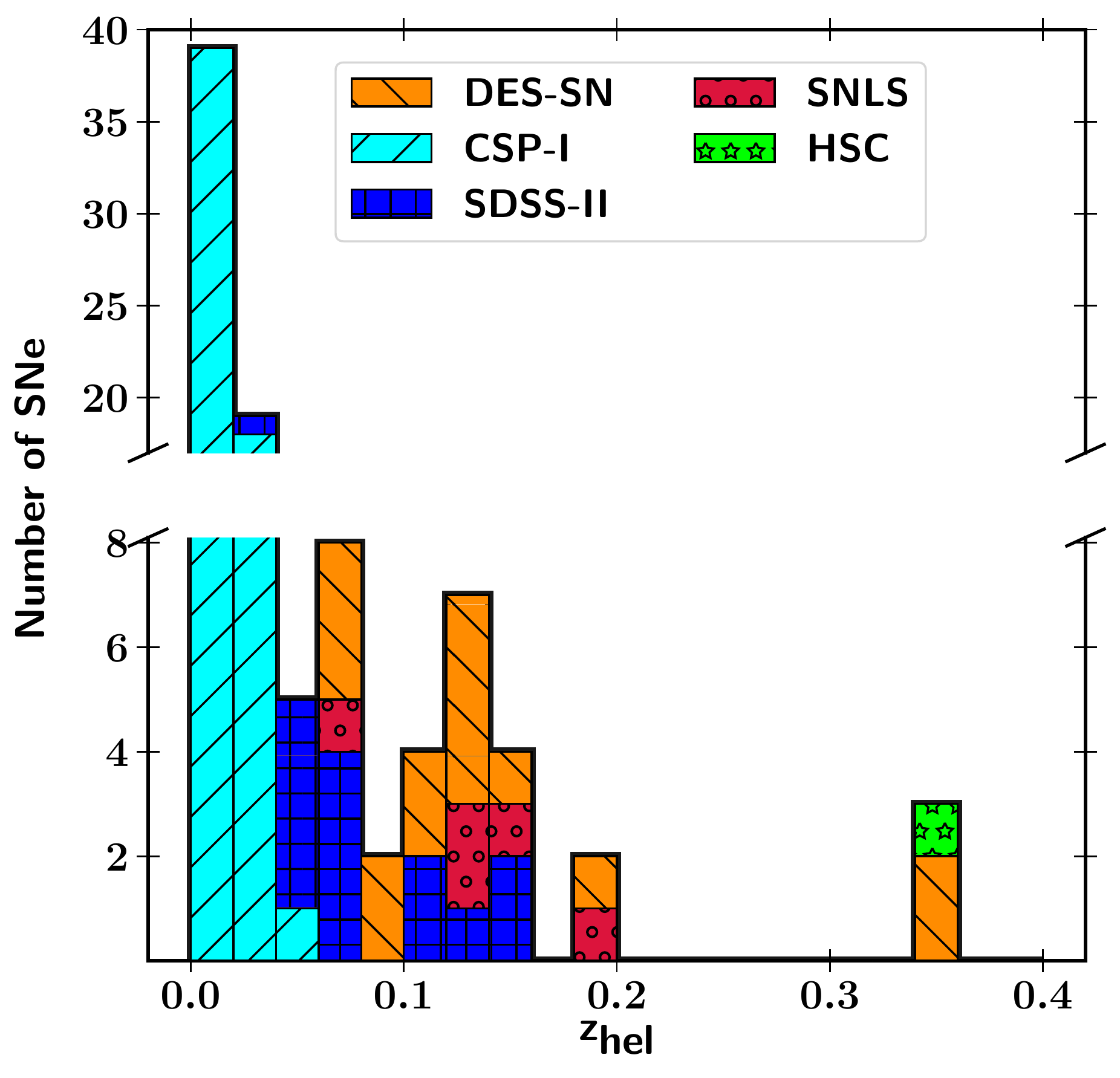}
\caption{Histogram of the SN~II sample redshift distribution. SN from CSP-I, SDSS-II, SNLS, HSC, and DES-SN are respectively displayed in cyan ($/$), blue ($+$), red ($\circ$), lime ($\star$), and orange ($\backslash$). The redshift bin size is 0.02.}
\label{fig:z_distribution}
\end{figure}

The final redshift distribution is presented in Figure \ref{fig:z_distribution}. The SCM DES-SN sample has a concentration of objects with $z = 0.1$--0.2 and only two SNe at high redshift ($\sim 0.35$). The gap in the range $0.2<z<0.35$ is due to our different selection cuts. If we include the 56 spectroscopically confirmed SNe~II, the DES-SN distribution looks different, with nine SNe~II in the range $0.2<z<0.25$ and five SNe~II with $z>0.25$ (only two useful for the SCM). Eight SNe~II with $z>0.2$ have been removed owing to the lack of an explosion date, one for the absence of photometric data after 40\,d, and three owing to the P-Cygni profile cut.

\begin{figure*}
\includegraphics{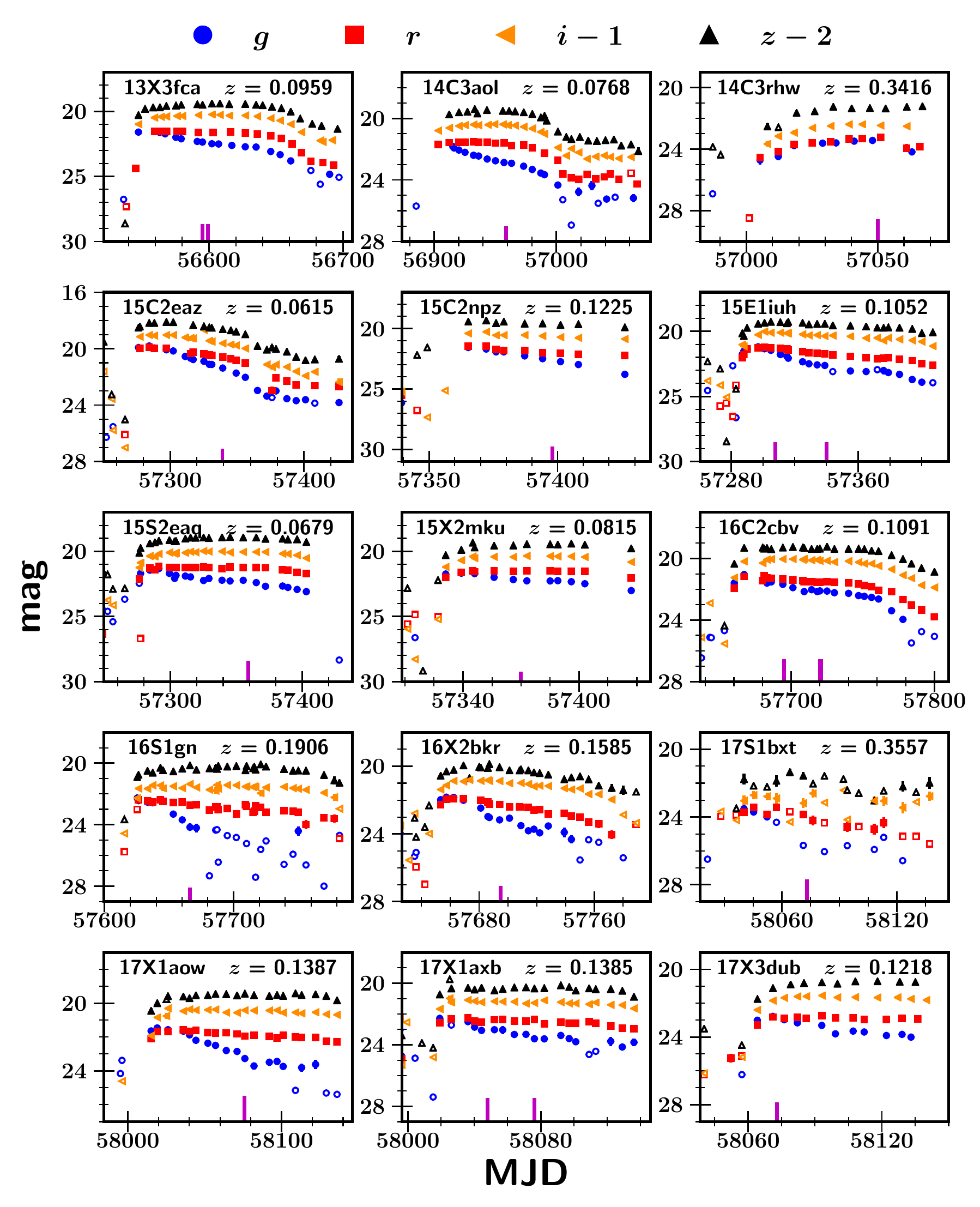}

\caption{Observed light curves of the SNe~II in our SCM sample which were discovered by DES-SN. Blue circles are magnitudes in the $g$ band, red squares are $r$, orange left triangles are $i-1$, and black top triangles are $z-2$. Empty symbols represent real points with flux/err $< 3$, where ``flux/err'' is simply the flux divided by its uncertainty. The abscissa is the Modified Julian Date (MJD). In each panel, the IAU name and the redshift are given in the upper right. Vertical magenta lines indicate the epochs of optical spectroscopy.}
\label{fig:LC_DES}
\end{figure*}

In Figure \ref{fig:LC_DES} we present the DES-SN measured light curves for the 15 SNe~II discovered by DES-SN and chosen for our SCM sample. Figure \ref{fig:spectra_DES} shows all of the spectra used to measure the expansion velocities. The full set of light curves and spectra of SNe~II discovered by DES-SN will be available to the community (see Appendix \ref{AppendixB} and Appendix \ref{AppendixC}) and be can be requested from the authors or for download\footnote{\url{https://github.com/tdejaeger}}. 

The final SCM sample thus consists of 93 SNe~II: 58 (CSP-I) $+$14 (SDSS-II) $+$5 (SNLS) $+$1 (HSC) $+$15 (DES-SN). Note that in contrast to \citep{dejaeger17b}, we use SN~2006iw and SN~2007ld from the CSP-I sample and not from the SDSS-II sample. Both SNe have better-sampled light curves in the new recalibrated CSP-I photometry (Contreras et al., in prep.). In Table \ref{tab:sum_cut}, we define the different samples employed and the different cuts used in this work.

\begin{figure*}
	\includegraphics[scale=0.8]{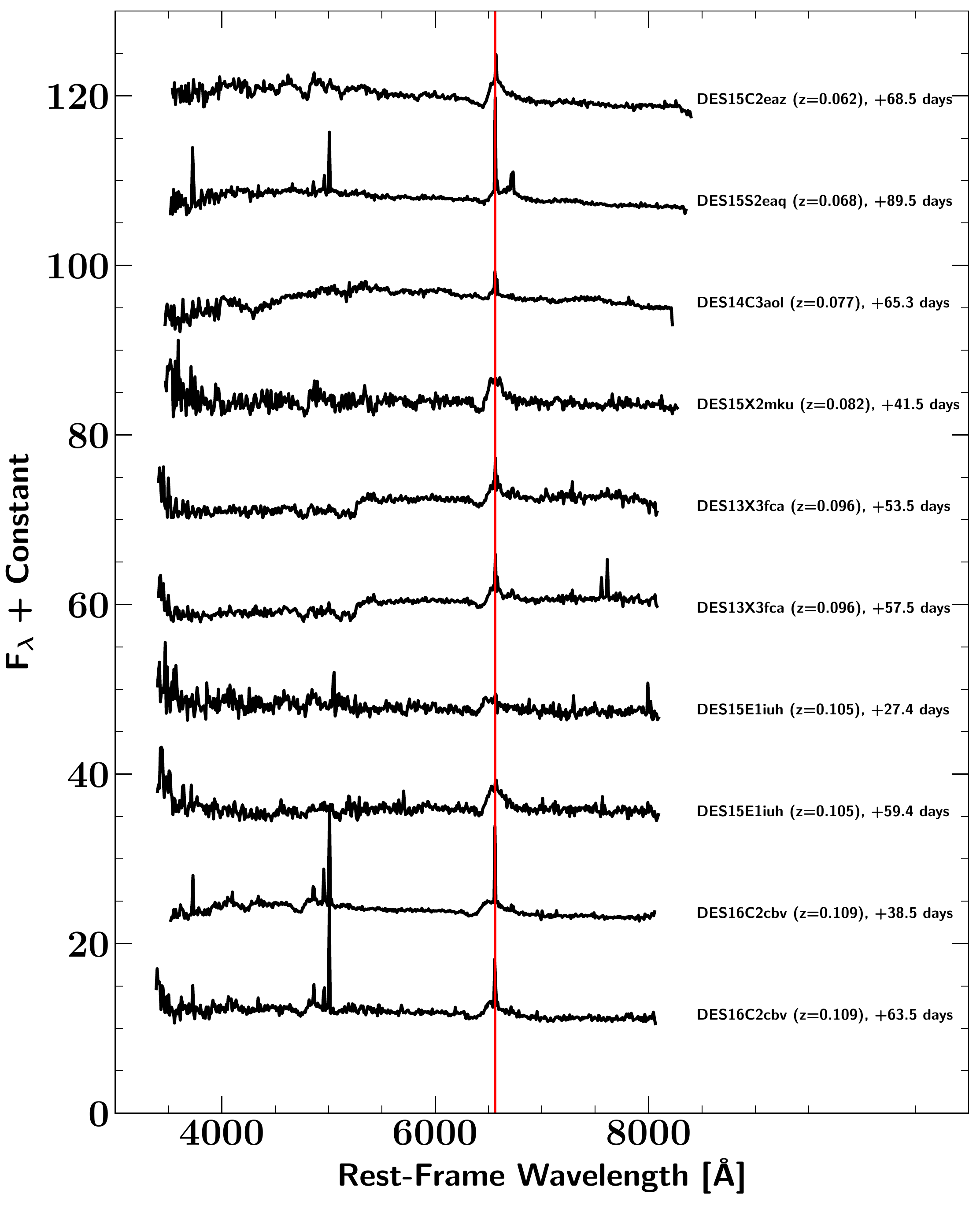}
\caption{Spectra of the 15 SNe~II from the DES-SN sample used for the SCM. The spectra are shown in the rest frame, and the date listed for each SN is the number of days since explosion (rest frame). The redshift of each SN is labelled. The spectra were binned (10\,\AA). The red vertical line corresponds to H$\alpha$ ($\lambda$6563) in the rest frame.}
\label{fig:spectra_DES}
\end{figure*}

\begin{figure*}
	\addtocounter{figure}{-1}
	\includegraphics[scale=0.8]{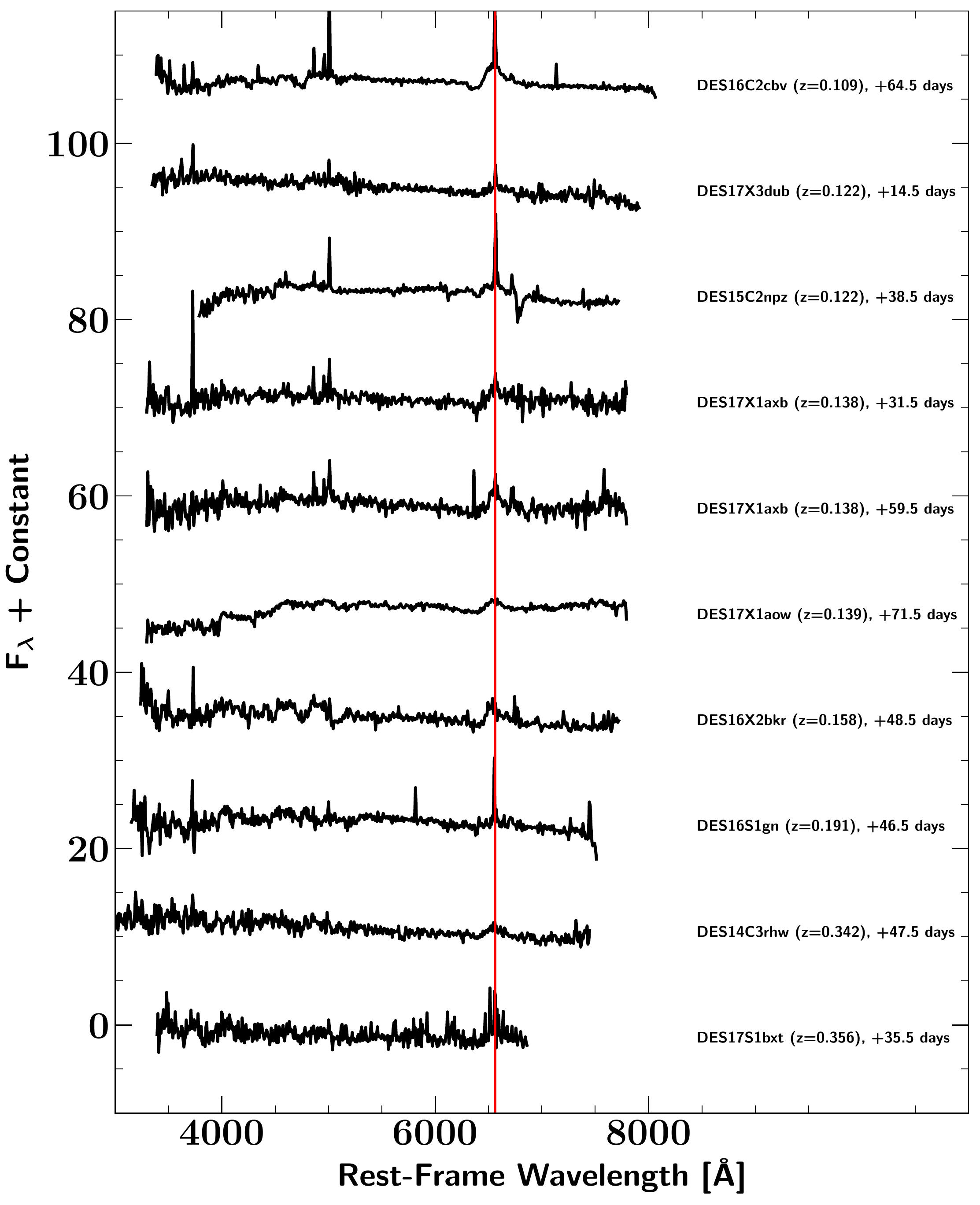}
\caption{\textit{(Cont.)} Spectra of the 15 SNe~II from the DES-SN sample used for the SCM. The spectra are shown in the rest frame, and the date listed for each SN is the number of days since explosion (rest frame). The redshift of each SN is labelled. The spectra were binned (10\,\AA). The red vertical line corresponds to H$\alpha$ ($\lambda$6563) in the rest frame.}
\end{figure*}

\subsection{Photometric Colour Method sample}

The sample used for the PCM includes the SCM sample plus 12 SNe~II for which no clear P-Cygni profile is seen in their spectra. After light-curve inspection, we removed three SNe~II: DES15X3nad, whose light curve is short and looks like that of a SN~IIb, and DES17C3aye and DES17C3bei, whose $g$-band light curves exhibit a second bump perhaps caused by ejecta interacting with circumstellar matter (relatively narrow lines are present in their spectra). All of the light curves and spectra are shown in Appendix \ref{AppendixB}. The final PCM sample is thus composed of 115 SNe~II (61 + 14 + 15 + 1 + 24; CSP-I + SDSS-II + SNLS + HSC + DES-SN, respectively). A summary of all the SNe~II available and the different cuts can be found in Table \ref{tab:sum_cut}.

\begin{table*}
\caption{Summary of all the SNe~II available and used per survey.}
\begin{threeparttable}
\begin{tabular}{cccccccccc}
\hline
Survey & All & Unique & Spectrum &Outliers &$z>0.01$ &$Texp$ &Photo &3$\sigma$ clipping & Used\\
\hline
\hline
CSP-I & 61 (61) & 61 (61) & 58 (61) & 58 (61) & 44 (47) & 39 (42) &37 (40) &37 (40) & 37 (40)\\
SDSS-II & 16 (16) & 14 (14) & 14 (14) & 14 (14) & 14 (14) & 13 (13) & 13 (13) & 13 (13) &13  (13)\\
SNLS & 15 (15) & 15 (15) & 5 (15) & 5 (15) & 5 (15) & 4 (14) & 4 (14)  & 4 (14) & 4 (14) \\
HSC & 1 (1) & 1 (1) & 1 (1) & 1 (1) & 1 (1) & 1 (1) & 1 (1)  & 1 (1) & 1 (1) \\
DES-SN & 27 (27) & 27 (27) & 15 (27)  & 15 (24) & 15 (24) & 15 (23) & 15 (23) & 15 (22) & 15 (22) \\
\hline
Total & 120 (120) & 118 (118) & 93 (118) & 93 (115) & 79 (101) & 72 (93) & 70 (91) & 70 (90) & 70 (90) \\

\hline
\hline
\end{tabular}
\textit{Notes:} For each survey, the number of SNe~II used for the SCM and the PCM (written in parentheses) is shown for different selection cuts. Unique: we removed two SNe~II from SDSS-II in common with the CSP sample, spectrum: for the SCM we need at least one spectrum, outliers: from the PCM sample after light-curve inspection we removed three
SNe~II from the DES-SN sample, $z>0.01$: only select SNe~II in the Hubble flow, $Texp$: SNe~II with explosion date with an uncertainty $leq 10$\,d, photo: photometry data at 43\,d after the explosion, and finaly, $3\sigma$ clipping: one SN~II from DES-SN is identified as an outlier.
\label{tab:sum_cut}
\end{threeparttable}
\end{table*}

\section{Methodology}\label{txt:methodology}

In this section, we describe how the quantities (expansion velocities, magnitudes, and colours) required to derive the Hubble diagram are obtained. As the methodology is exactly the same as that used by \citet{dejaeger17b}, only a brief description is presented here.

\subsection{Photospheric velocities}

The vast majority of DES-SN follow-up spectroscopy was performed to provide host-galaxy redshifts and classifications, so the average S/N of the spectra is low. A direct measurement of the H$\beta$ velocity from the minimum flux of the absorption component of the P-Cygni profile is difficult. However, \citet{poznanski10} and \citet{dejaeger17b} (at low-$z$ and at high-$z$, respectively) have demonstrated that for noisy spectra the H$\beta$ velocity can be determined by computing the cross-correlation between the observed spectra and a library of high S/N SN~II spectra (templates) using the Supernova Identification code (SNID; \citealt{blondin07}). Velocities from direct measurement or using SNID have shown a dispersion of only 400\,km\,s$^{-1}$, the same order of magnitude as the uncertainties (see Figure 3, \citealt{dejaeger17b}).

We cross-correlated each observed spectrum with the SN~II template library spectra (for which the H$\beta$ $\lambda$4861 velocities have been measured precisely from the minimum flux of the absorption component), constraining the wavelength range to 4400--6000\,\AA\ (rest frame). For each spectrum, the resulting velocities are the sum of the template velocities (measured from the minimum flux) and the relative Doppler shift between the observed spectrum and the template. Finally, the velocities of the top 10\% best-fitting templates are selected; the final velocity and its uncertainty correspond to the weighted mean and standard deviation of those selected templates. We add to the velocity error derived from the cross-correlation technique a value of 150\,km\,s$^{-1}$, in quadrature, to account for the rotational velocity of the galaxy at the SN position \citep{sofue01}. For example, in Figure 4 of \citet{galbany14}, we can see that the rotational velocity of the host galaxy reaches $\sim 150$\,km\,s$^{-1}$ with respect to the centre, measured from integral field spectroscopy of a large sample of SN Ia host galaxies. Additionally, for a SN located farther from the centre, larger differences are seen between the redshift at the SN position and the redshift of the host-galaxy nucleus. Note that all of the CMB redshifts were corrected to account for peculiar flows induced by visible structures using the model of \citet{carrick15}.

\subsection{Light-curve parameters}

\indent To derive the magnitude and the colour at different epochs, we model the light curves using hierarchical Gaussian processes (GP). This method has been successfully applied in different SN studies \citep{mandel09,mandel11,burns14,lochner16,dejaeger17b,inserra18}. To apply the GP method we use the fast and flexible Python library \textsc{George} developed by \citet{ambikasaran14}. For a more quantitative comparison between the GP and linear interpolation methods, the reader is referred to \citet{dejaeger17b}.

\indent To measure the slope of the plateau during the recombination phase ($s_{2}$), we use a Python program which performs a least-squares fitting of the light curves corrected for Milky Way extinction and K/S-corrections. The choice between one or two slopes is achieved using the statistical method F-test\footnote{Fast-declining SN light curves generally exhibit one slope, while the slow-declining SN light curves also show the cooling phase called $s_{1}$ by \citet{anderson14a}}. A full analysis of these slopes for our sample together with SNe from the literature will be published in a forthcoming paper.

\subsection{Hubble diagram}

The SCM is based on the correlation between the SN absolute magnitude and the photospheric expansion velocity and the colour. The observed magnitude can be modelled as

\begin{ceqn}
\begin{align}
\begin{split}
m^\mathrm{model}_{i}=&\mathcal{M}_{i}-\alpha\, \mathrm{log_{10}}\left(\frac{v_\mathrm{H\beta}}{<v_\mathrm{H\beta}>}\right) \\ +& \beta [(r-i) - <(r-i)>] + 5\, \mathrm{log_{10}} (\mathcal{D}_{L}(z_\mathrm{CMB}|\Omega_\mathrm{m},\Omega_\mathrm{\Lambda})),
\end{split}
\label{m_model}
\end{align}
\end{ceqn}

where $i$ is the $i$-band filter, $(r-i)$ is the colour ($<(r-i)> \approx -0.04$\,mag; the average colour), $v_\mathrm{H\beta}$ is the velocity measured using H$\beta$ absorption ($<v_\mathrm{H\beta}>$ $\sim 6000$\,km\,s$^{-1}$; the average value), $\mathcal{D}_{L}(z_\mathrm{CMB}|\Omega_\mathrm{m},\Omega_\mathrm{\Lambda}$) is the luminosity distance ($\mathcal{D}_{L}$ = H$_{0}d_{L}$) for a cosmological model depending on the cosmological parameters $\Omega_{m}$, $\Omega_{\Lambda}$, the CMB redshift $z_{\rm CMB}$, and the Hubble constant. Finally, $\alpha$, $\beta$, and $\mathcal{M}_{i}$ are free parameters, with $\mathcal{M}_{i}$ corresponding to the ``Hubble-constant-free'' absolute magnitude ($\mathcal{M}_{i} = M_{i} - 5\,{\rm log}_{10}({\rm H}_{0}) + 25$).

To determine the best-fitting parameters and to derive the Hubble diagram, a Monte Carlo Markov Chain (MCMC) simulation is performed using the Python package \textsc{EMCEE} developed by \citet{foreman13}. As discussed by \citet{poznanski09}, \citet{andrea10}, and \citet{dejaeger17b}, the minimised likelihood function is defined as

\begin{ceqn}
\begin{align}
-2\,\mathrm{ln}(\mathcal{L})=\sum_\mathrm{SN} \left \{ \frac{\left [m^\mathrm{obs}_{i}- m^\mathrm{model}_{i} \right ]^{2}}{\sigma^{2}_\mathrm{tot}} +\mathrm{ln}(\sigma^{2}_\mathrm{tot}) \right \},
\label{likelihood}
\end{align}
\end{ceqn}
where we sum over all SNe~II available, $m_{i}^\mathrm{obs}$ is the observed $i$-band magnitude corrected for Milky Way extinction and K/S-corrections, and $m_{i}^\mathrm{model}$ is the model defined in Equation \ref{m_model}. The total uncertainty $\sigma_\mathrm{tot}$ is defined as
\begin{ceqn}
\begin{align}
\begin{split}
\sigma^{2}_\mathrm{tot}=&\sigma^{2}_{m_{i}} + \left(\frac{\alpha}{\mathrm{ln} 10}\frac{\sigma_{v_\mathrm{H\beta}}}{v_\mathrm{H\beta}}\right)^2 + (\beta \sigma_{(r-i)})^2\\&+\left (\sigma_{z} \frac{5(1+z)}{z(1+z/2)\,\mathrm{ln}(10)}\right)^{2}+\sigma^{2}_\mathrm{obs}+\sigma^2_\mathrm{lensing}+\sigma^2_\mathrm{lc},
\end{split}
\label{eq:likelihood}
\end{align}
\end{ceqn}

\noindent where $\sigma_{m_{i}}$, $\sigma_{v_{H\beta}}$, $\sigma_{(r-i)}$, and $\sigma_{z}$ are the apparent $i$-band magnitude, velocity, colour, and redshift uncertainties. The quantity $\sigma_\mathrm{obs}$ includes the true scatter in the Hubble diagram and any misestimates of observational uncertainties.

Unlike the case of \citet{dejaeger17a}, the total uncertainty $\sigma_\mathrm{tot}$ includes two new terms: a statistical uncertainty caused by the gravitational lensing ($\sigma_\mathrm{lensing} = 0.055z$; \citealt{jonsson10}) and a covariance term ($\sigma_\mathrm{lc}$) to account for correlations between magnitude, colour, and velocity. The covariance term is a function of $\alpha$ and $\beta$; following \citet{amanullah10},

\begin{ceqn}
\begin{align}
\sigma^2_\mathrm{lc}=2\alpha C_{m,{\rm vel}}-2\beta C_{m,{\rm col}}-2\alpha\beta C_{{\rm vel,col}}.
\label{eq:coVar}
\end{align}
\end{ceqn}

\noindent To derive $C_{m,{\rm vel}}$, $C_{m,{\rm col}}$, and $C_{{\rm vel,col}}$, we run 3000 simulations where for each simulated SN, the magnitude, colour, and velocity are taken at an epoch of 43\,d (see Section \ref{txt:fixed_cosmo_SCM}) plus a random error (Gaussian distribution) from their uncertainties. The covariance for each SN using the 3000 magnitudes, colours, and velocities is then calculated.

For the PCM, the methodology is identical to that used for the SCM, except that instead of using a velocity correction, we use the $s_{2}$ slope correction. The observed magnitudes can be modeled as

\begin{ceqn}
\begin{align}
m^\mathrm{model}_{i}=&\mathcal{M}_{i}-\alpha s_{2} + \beta (r-i) + 5\, \mathrm{log_{10}} (\mathcal{D}_{L}(z_\mathrm{CMB}|\Omega_\mathrm{m},\Omega_\mathrm{\Lambda})),
\label{m_model_PCM}
\end{align}
\end{ceqn}
\noindent 
where all of the quantities are described above (see Eq. \ref{m_model}). As for the SCM, the best-fitting PCM parameters are derived using a MCMC simulation by minimising a similar likelihood function as defined in Eq. \ref{eq:likelihood}, except that in the total uncertainty, $\sigma_\mathrm{tot}$, $\left(\frac{\alpha}{\mathrm{ln} 10}\frac{\sigma_{v_\mathrm{H\beta}}}{v_\mathrm{H\beta}}\right)^2$ is replaced by $(\alpha \sigma_{s_{2}})^2$.

\section{SCM RESULTS}

First, in Section \ref{txt:fixed_cosmo_SCM}, we assume a $\Lambda$CDM cosmological model ($\Omega_{m}=0.3$, $\Omega_{\Lambda} = 0.7$) and present an updated SN~II Hubble diagram using the SCM. Then, in Section \ref{txt:cosmo_SCM}, assuming a flat universe ($\Omega_{m} + \Omega_{\Lambda} = 1$), we constrain the matter density ($\Omega_{m}$). Finally, we discuss differences between the samples and the effect of systematic errors on the distance modulus in Sections \ref{txt:bias}, \ref{txt:err_SCM}, and \ref{txt:sim_bias}.

\subsection{Fixed cosmology: $\Omega_{m}=0.3$, $\Omega_{\Lambda} = 0.7$}\label{txt:fixed_cosmo_SCM}

\begin{figure}
\centering
\includegraphics[width=1.0\columnwidth]{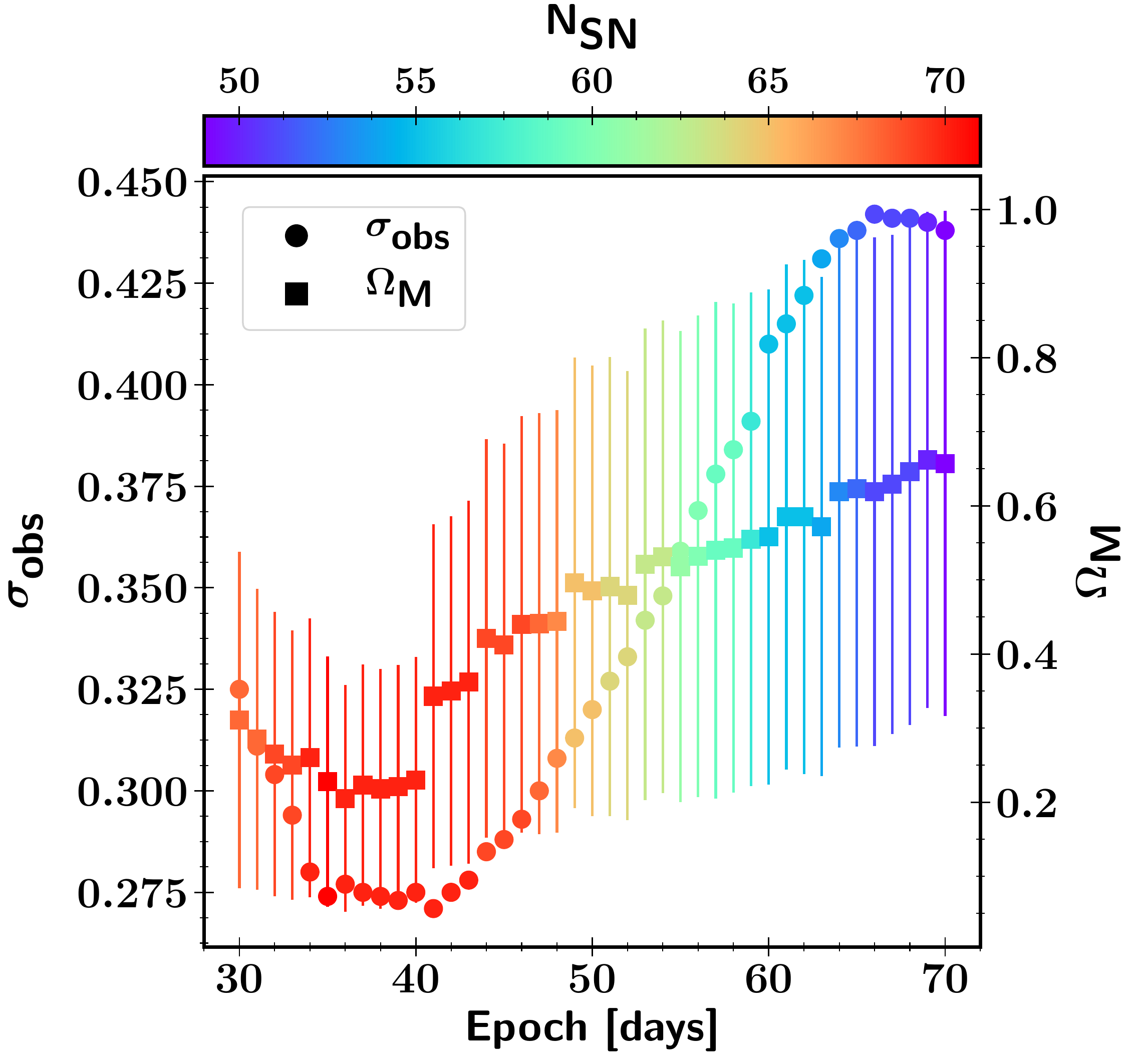}
\caption{Variation by epoch of the intrinsic dispersion in the Hubble
diagram (circles and left ordinate axis) and $\Omega_{m}$ (squares and right ordinate axis) using the SCM. The colour bar at top represents the different sample sizes. For clarity, only the $\Omega_{m}$ uncertainties are plotted.}
\label{fig:sig_ep_SCM}
\end{figure}

\begin{figure*}
	\includegraphics[width=17cm]{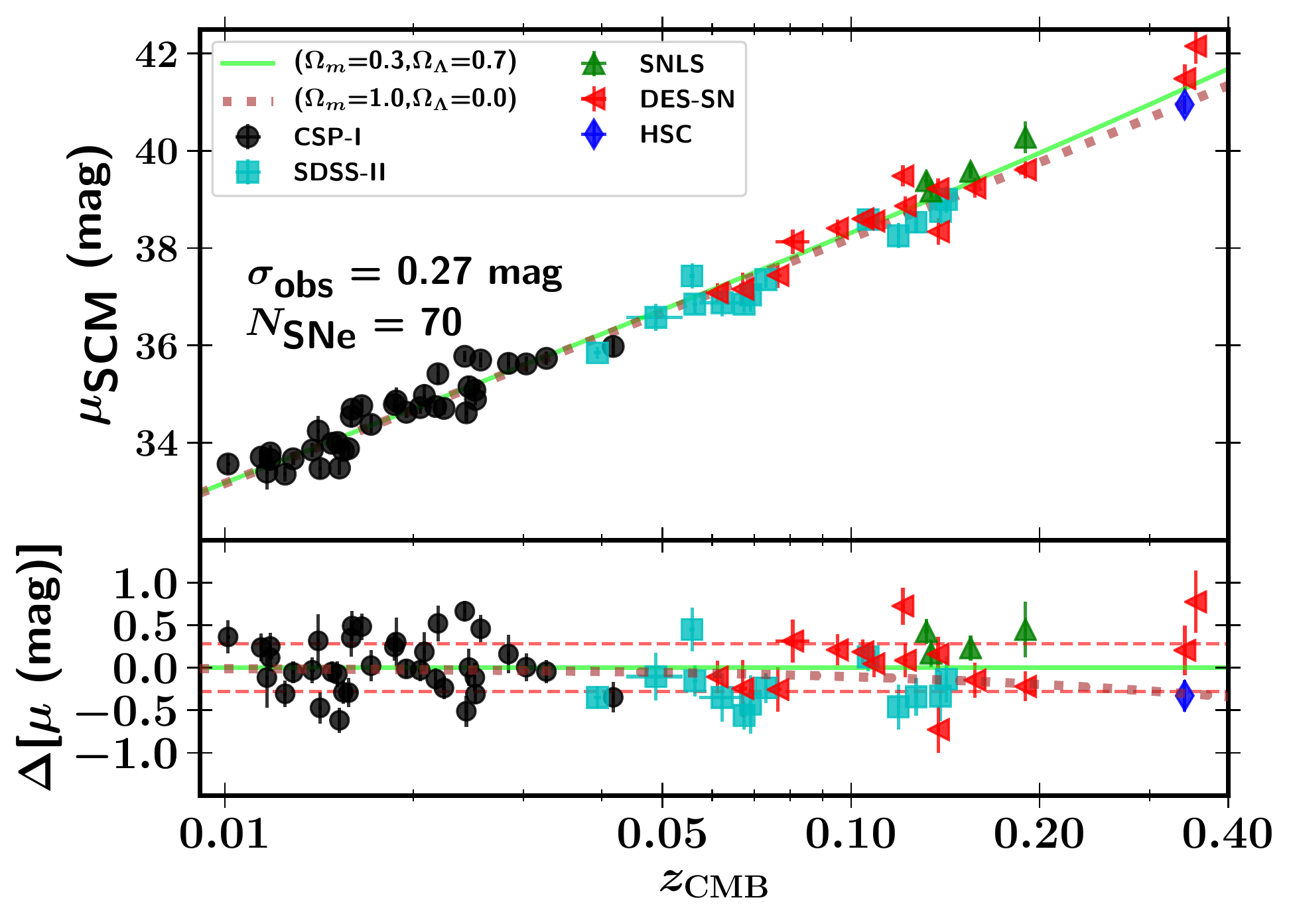}
\caption{Hubble diagram (top) and residuals from the $\Lambda$CDM model (bottom) using the SCM as applied to the data taken from CSP-I (black circles; \citealt{dejaeger17a}), SDSS-II (cyan squares; \citealt{andrea10}), SNLS (green triangles; \citealt{dejaeger17a}), HSC (blue diamond; \citealt{dejaeger17b}), and DES-SN (red left triangles; this work). The lime solid line is the Hubble diagram for the $\Lambda$CDM model ($\Omega_{m}=0.3$, $\Omega_{\Lambda} = 0.7$), while the brown dot line is for an Einstein-de Sitter cosmological model ($\Omega_{m}=1.0$, $\Omega_{\Lambda} = 0.0$). In both models, we use H$_0= 70$\,km\,s$^{-1}$ Mpc$^{-1}$ to standardise the SN~II brightness. We present the number of SNe~II available at this epoch ($N_{\rm SNe}$), the epoch after the explosion, and the observed dispersion ($\sigma_{\rm obs}$).}
\label{fig:HD_SCM}
\end{figure*}

\begin{figure}
	\includegraphics[width=1.0\columnwidth]{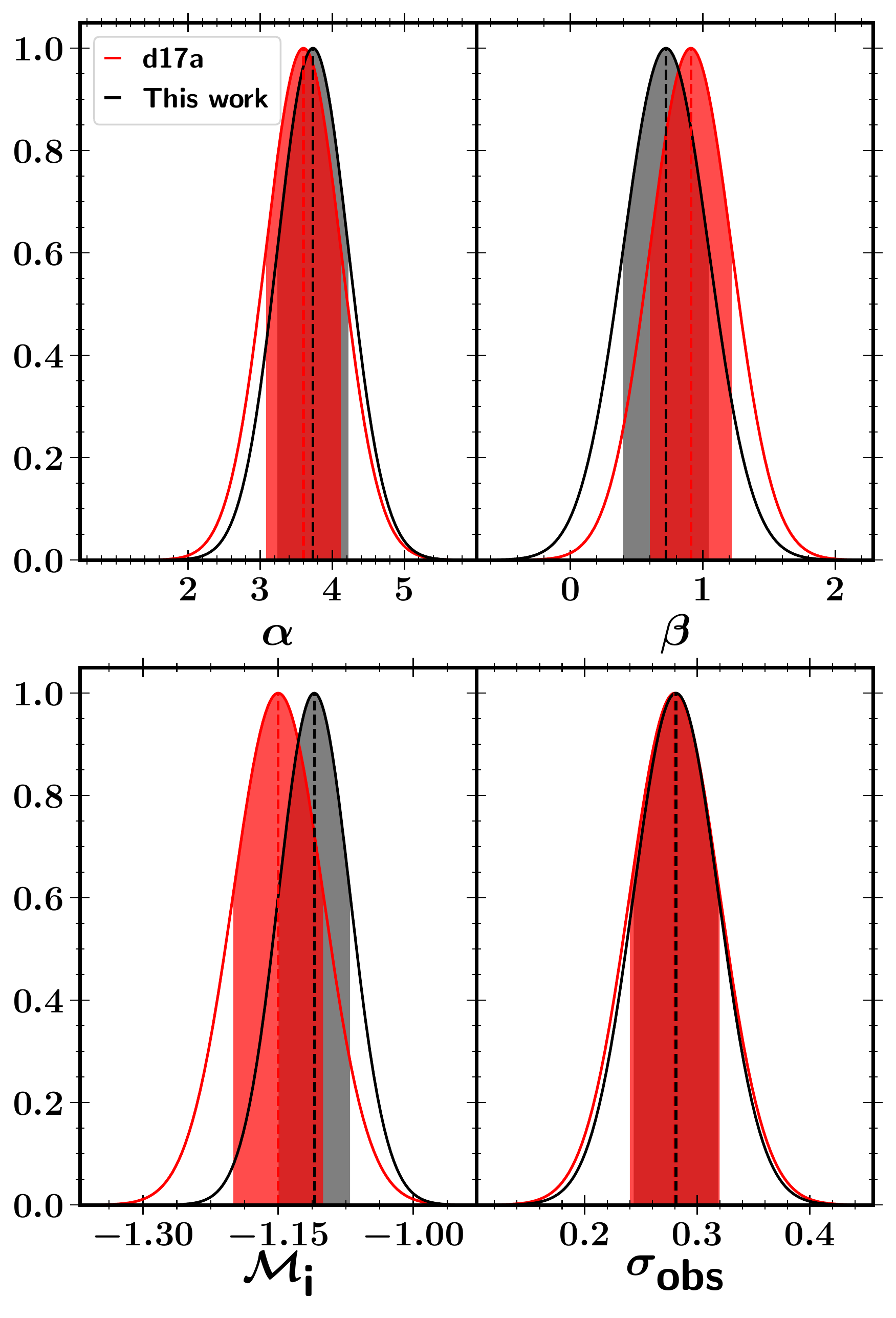}
	\caption{Comparison of the best-fitting parameters using the SCM derived by \citet{dejaeger17b} in red and those obtained in this work (in black) with the DES-SN sample. \textit{Top left:} Distributions of $\alpha$. \textit{Top right:} Distributions of $\beta$. \textit{Bottom left:} Distributions of ``Hubble-constant-free'' absolute magnitude ($\mathcal{M}_{i}$). \textit{Bottom right:} Distributions of observed dispersion ($\sigma_{\rm obs}$). In each panel, the vertical dashed line represents the average value, while the filled region represents the $1\sigma$ uncertainty.}
\label{fig:compa_parameters_SCM}
\end{figure}

To minimise the effect of peculiar-galaxy motions, we select SNe~II located in the Hubble flow, with $z_{\rm CMB} > 0.01$. After this cut, our available sample consists of 79 SNe~II (see Table \ref{tab:sum_cut}). We use SNe~II regardless of their plateau slope since \citet{dejaeger15b} and \citet{gall17} have demonstrated that slowly and rapidly declining SNe~II can be used as distance indicators. We select the SNe~II with an explosion date uncertainty smaller than 10\,d as the explosion date has an influence on the distance modulus (see Section \ref{txt:sys_Texp}). Among the 79 SNe~II, seven SNe~II have an explosion date with an uncertainty $\geq 10$\,d: five from CSP-I (SN~2005lw, SN~2005me, SN~2006bl, SN~2007ab, SN~2008aw), one from SDSS-II (SN~2006jl), and one from SNLS (06D2bt).

The best epoch to apply the SCM is chosen as the one which minimises the intrinsic dispersion in the Hubble diagram as well as maximises the number of objects. In Figure \ref{fig:sig_ep_SCM}, the minimal dispersion is found around 40\,d after the explosion. All these epochs correspond to the recombination phase and are consistent with the epoch (50\,d) used in previous SN~II cosmology studies \citep{hamuy02,nugent06,poznanski09,andrea10}. In this work, we applied the method at 43\,d after the explosion (even if the minimum is at 42\,d) to facilitate the comparison with \citet{dejaeger17b}. At this specific epoch 70 of 72 SNe~II have photometric/spectroscopic information and can be used to build the SN~II Hubble diagram.\footnote{Two SNe~II (SN~2006it and SN~2008il) have no photometric data at 43\,d.} The SCM total sample thus consists of 37 SNe~II from CSP-I, 13 SNe~II from SDSS-II, 4 SNe~II from SNLS, 1 SN~II from HSC, and 15 SNe~II from DES-SN (see Table \ref{tab:sum_cut}). Note that with respect to the sample used by \citet{dejaeger17b}, three CSP-I SNe~II are added: SN~2004fb (explosion date has been updated by \citealt{gutierrez17a}), SN~2006iw, and SN~2007ld (recalibrated CSP-I photometry). The relevant information for our SN~II SCM sample is given in Appendix \ref{AppendixD}, Table \ref{SN_sample}. 

Figure \ref{fig:HD_SCM} shows the updated SCM SN~II Hubble diagram with the Hubble residuals of the combined data. This Hubble diagram was built by finding the best-fitting values ($\alpha$, $\beta$, $\mathcal{M}_{i}$, and $\sigma_{\rm obs}$) assuming a $\Lambda$CDM cosmological model, with $\Omega_{m}=0.3$. We find $\alpha = 3.71 \pm 0.49$, $\beta = 0.72 \pm 0.32$, and $\mathcal{M}_{i} = -1.10 \pm 0.04$, with an observed dispersion $\sigma_{\rm obs} = 0.27^{+0.04}_{-0.03}$\,mag. As seen in Figure \ref{fig:compa_parameters_SCM}, these values are consistent with those from \citet{dejaeger17b} ($\alpha = 3.60^{+0.52}_{-0.51}$, $\beta = 0.91^{+0.31}_{-0.30}$, and $\mathcal{M}_{i} = -1.15 \pm 0.05$, and $\sigma_{\rm obs} = 0.28$\,mag). Despite the large uncertainties, the fact that the best-fitting parameters do not change significantly with the additional DES-SN sample suggests that our study does not seem biased toward brighter or fainter objects (see Sections \ref{txt:bias} and \ref{txt:sim_bias} for a discussion). 

Despite the small differences in the best-fitting parameters and the use of the recalibrated CSP-I photometry, the majority of distance moduli derived in this work are consistent with those derived by \citet{dejaeger17b}. An average difference of $-0.05$\,mag is seen, which is much smaller than the uncertainty of each distance modulus (0.19\,mag average). This small discrepancy could arise from the fitting parameter shifts and changes in the CSP-I photometry. As a test, if instead of using the observed parameters from the new photometry (magnitude, colour, velocity) we used those from \citet{dejaeger17b} with the fitting parameters derived in this work, the average distance modulus difference drops from $-0.05$\,mag to $-0.007$\,mag.

The observed dispersion found in this work using the SCM (0.27\,mag) is consistent to those from previous studies (0.26\,mag, \citealt{nugent06}; 0.25\,mag, \citealt{poznanski09,poznanski10}; 0.29\,mag, \citealt{andrea10}; and 0.27\,mag, \citealt{dejaeger17b}) and corresponds to a 14\% distance uncertainty. It is interesting to note that the majority of studies in the literature (applying the SCM), despite using different samples and techniques, all found a similar intrinsic dispersion of 0.25--0.30\,mag. This consistency suggests that using current techniques, we are reaching the limit of SCM. To break this current impasse, new correlations (e.g., host-galaxy properties, metallicity) or templates (for the K-correction) are needed.

To attempt to reduce the scatter, we investigate the possible effect of the host-galaxy extinction even though recent work \citep{dejaeger18a} suggests that the majority of SN~II colour diversity is intrinsic and not due to host-galaxy extinction. We divide our SN~II sample into two subsamples based on their observed colour 43\,d after the explosion: 35 SNe~II have $r-i < -0.036$\,mag (blue subgroup) and 35 SNe~II have $r-i > -0.036$\,mag (red subgroup). For both subsamples, a similar dispersion of 0.25--0.26\,mag is found. If we apply only the velocity correction (i.e., $\beta = 0$), the scatter of the reddest subsample slightly increases (0.29\,mag), while the bluest subsample dispersion does not change. This test shows that the colour-term correction is not useful for standardising the SN~II brightness; hence, one band is sufficient to derive accurate distances, an asset in terms of observation time. If only the colour correction is applied (i.e., $\alpha = 0$), the dispersion is similar to those obtained by including Milky Way extinction, K-correction, and S-correction: 0.45\,mag for both subsamples. \citet{poznanski09} found that the dust correction has little impact, suggesting that his sample was biased toward dust-free objects. It could also be due to the existence of an intrinsic colour--velocity relation or because differences in colour are mostly intrinsic \citep{dejaeger18a}. If we remove the 20\% reddest SNe~II (i.e., thus potentially highly affected by dust), the total scatter does not significantly improve (0.26\,mag), suggesting that the differences in colour are already taken into account with the velocity correction. 

\begin{figure}
\centering
\includegraphics[width=1.0\columnwidth]{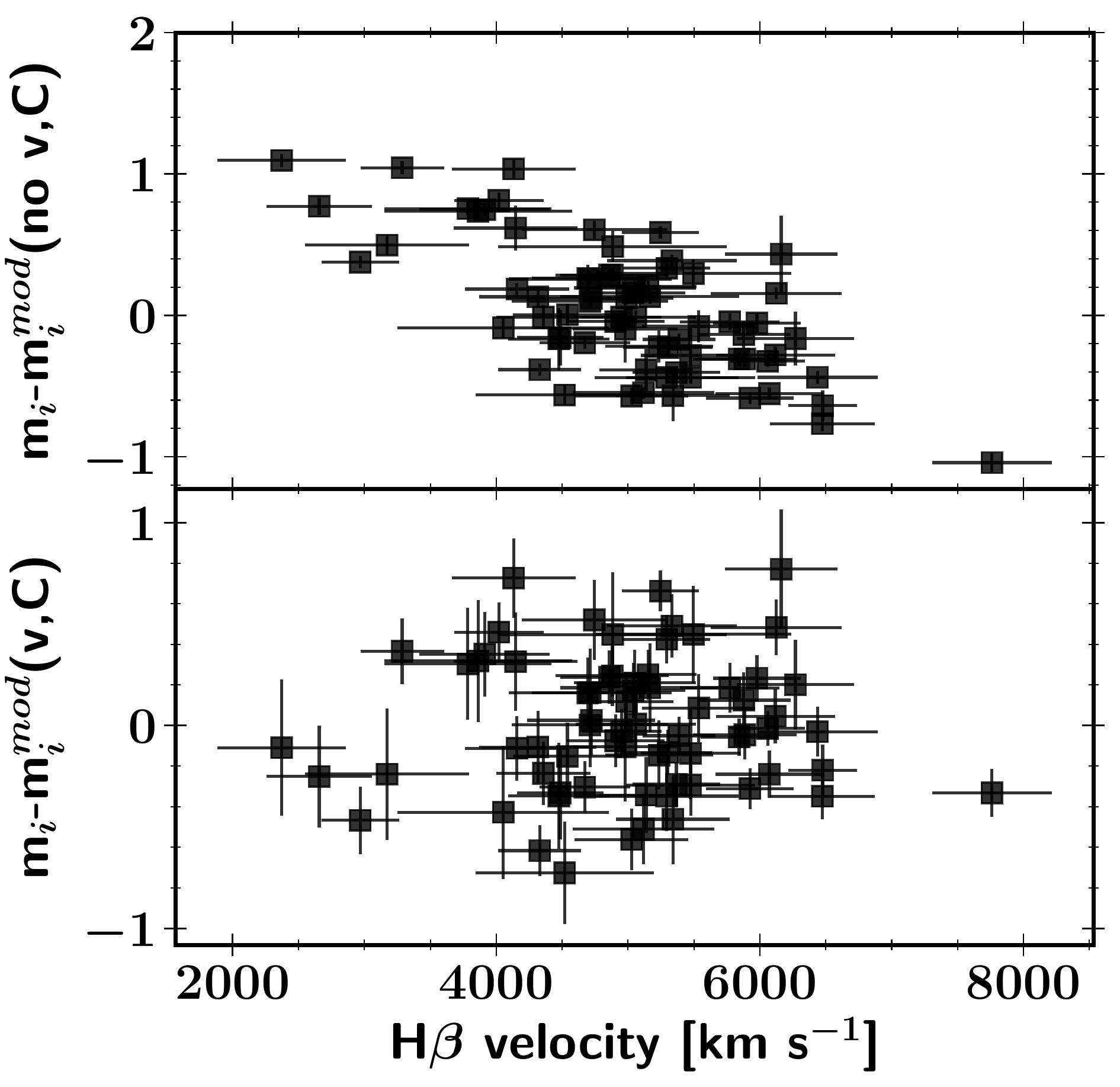}
\caption{The relationship between SN~II luminosity and the ejecta expansion velocity 43\,d after the explosion. The \textit{upper panel} shows the relationship (SN~II magnitudes are corrected for distances and colours), while the \textit{lower panel} shows the trend between luminosity and velocity after correcting the magnitudes for velocities ($\alpha\, \mathrm{log_{10}} v_{\mathrm{H\beta}}$).}
\label{fig:vel_relationship_SCM}
\end{figure}

\begin{figure}
\centering
\includegraphics[width=1.0\columnwidth]{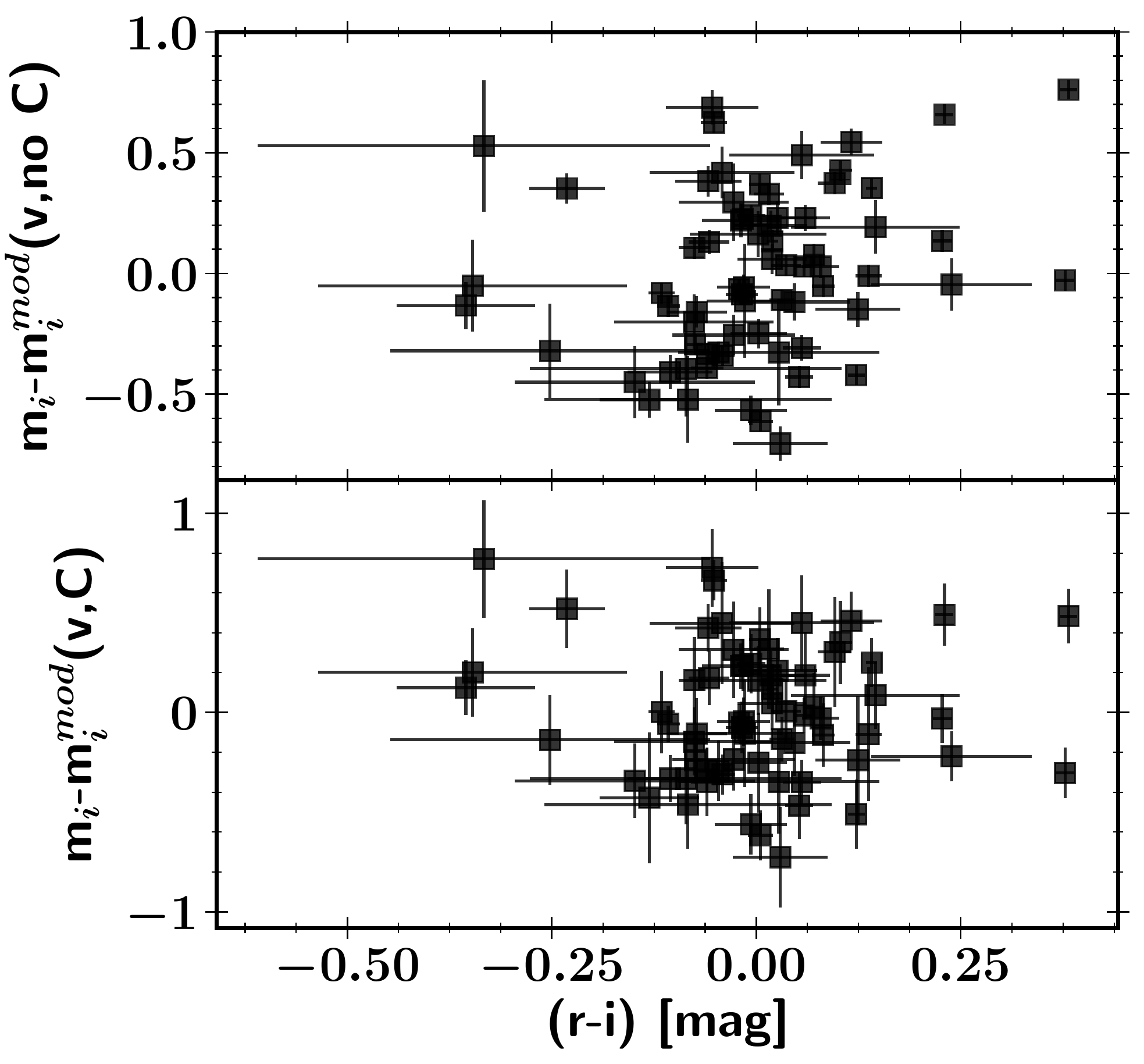}
\caption{The relationship between SN~II luminosity and the colour 43\,d after the explosion. The \textit{upper panel} shows the relationship (SN~II magnitudes are corrected for distances and velocities), while the \textit{lower panel} shows the trend between luminosity and colour after correcting the magnitudes for colours ($\beta(r-i)$).}
\label{fig:color_relationship_SCM}
\end{figure}

The upper panel of Figure \ref{fig:vel_relationship_SCM} shows the relation between the SN~II luminosity corrected for distance+colour and the ejecta velocity (at 43\,d). In the lower panel, the same relation is presented but with the luminosity corrected for distance, colour, and velocity (see Eq. \ref{m_model}). Figure \ref{fig:color_relationship_SCM} is similar to Figure \ref{fig:vel_relationship_SCM} but includes the relation between the luminosity and the colour. Figure \ref{fig:vel_relationship_SCM} clearly shows a correlation between the distance+colour corrected magnitudes and the ejecta velocity (Pearson factor $\sim 0.70$) which disappears when the magnitude is corrected for velocity (Pearson factor $\sim 0.03$). This demonstrates that the velocity correction is useful for standardising SNe~II. On the other hand, in Figure \ref{fig:color_relationship_SCM} there is no statistically significant correlation between the distance+velocity corrected magnitudes and colour (Pearson factors of $\sim 0.24$ and $\sim 0.05$ before and after colour correction, respectively). This confirms the result found above: dust correction is not significant for the SCM.

For the purpose of reducing the scatter in the Hubble diagram, and as suggested by \citet{poznanski09}, we investigate a possible relation between the Hubble residuals and the slope of the plateau. \citet{poznanski09} found that SNe~II with positive decline rates in the $I$ band have the largest Hubble residuals. However, we do not find a correlation between these quantities. Therefore, the slope of the plateau cannot be used to identify a more standard SN~II subsample \citep{andrea10}, confirming the results of \citet{dejaeger15b} and \citet{gall17} that both slowly and rapidly declining SNe~II can be used as distance indicators. Therefore, more work should be done to identify a SN~II subsample and reduce the scatter in the Hubble diagram (e.g., host-galaxy properties).

\subsection{Sample comparisons}\label{txt:bias}

We note that in the Hubble diagram plotted in Figure \ref{fig:HD_SCM}, there is an average systematic offset of $\sim 0.28$\,mag between SDSS-II and DES-SN: for SDSS-II, the average residual from the $\Lambda$CDM model is $-0.22$\,mag while for DES-SN it is 0.06\,mag. First, as suggested by \citet{andrea10} and \citet{poznanski10}, this offset could be due to a selection effect where brighter objects were favoured by SDSS-II. SDSS-II was built for SN~Ia cosmology; thus, the spectroscopic follow-up program was designed for SNe~Ia, which are more luminous than SNe~II, so only the brightest SNe~II would have been spectroscopically followed. In Section 4.5 we calculate a $\mu$-bias correction to account for effects such as Malmquist bias, based on simulations of each survey. In an ideal world, that correction would remove selection effects like those caused by the SDSS-II follow-up strategy. However, our $\mu$-bias simulation is only an approximation, and in the future it should be calculated more accurately using better SN~II templates, and with the infrastructure to model SN~II spectral features and their correlations with brightness. Second, this discrepancy could arise from photometric calibration errors (e.g., zero-points). We investigated possible calibration errors by checking the photometric system zero-point using different spectrophotometric standard stars. We also checked our methodology (Milky Way extinction, K/S correction; see Section \ref{txt:AKS}) by comparing two SN~II magnitudes observed by CSP-I and SDSS-II. In their natural photometric systems, a clear offset is seen between the CSP-I and SDSS-II photometry (e.g., $i$ band: $\sim -0.12$\,mag), while after applying our correction (Milky Way extinction, K/S correction; see Section \ref{txt:AKS}), the offset disappears and the photometry is consistent (e.g., $i$ band: $\sim -0.02$\,mag). Third, we compare the SDSS distance moduli derived in this work and those by \citet{poznanski10}. We find good agreement and an average difference of 0.05\,mag which is much lower than the uncertainties. All of our tests confirm our methodology; hence, as with \citet{andrea10} and \citet{poznanski10}, we believe that this offset is due to a selection effect where only bright SNe~II have been spectroscopically observed and our current $\mu$-bias simulation cannot correct it. We have estimated the potential cosmological impact of this offset and found it to be significantly smaller than the current uncertainties, but it will become important for future analyses.

We show that SDSS is biased toward bright objects; thus, to investigate if the DES-SN sample comes from a progenitor population similar to that of the other SN~II samples, we compare their velocity and absolute magnitude (without applying velocity or dust correction and assuming a $\Lambda$CDM model) distributions to those of the other samples.

\begin{figure}
\includegraphics[width=\columnwidth]{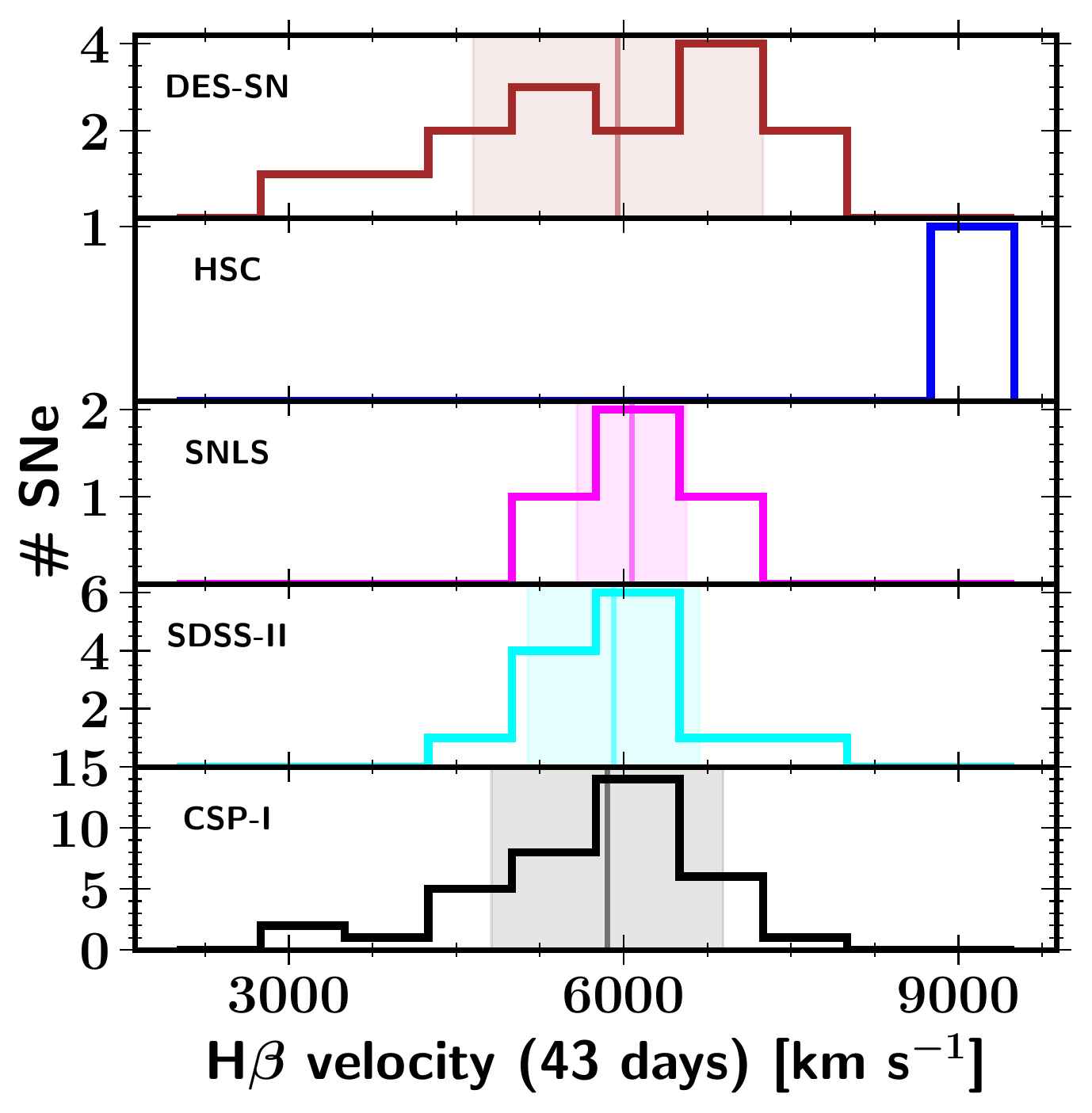}
\includegraphics[width=\columnwidth]{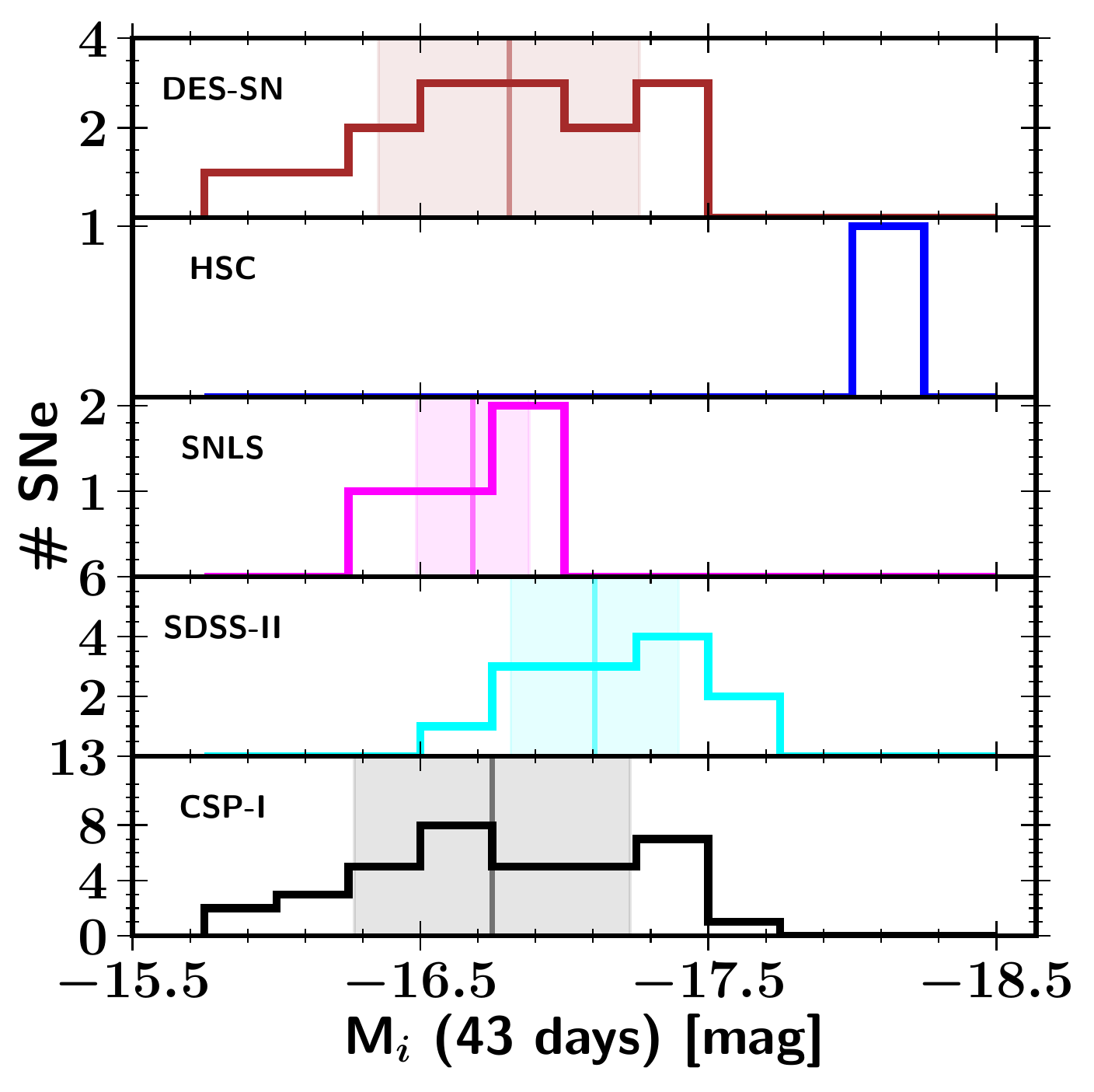}
\caption{{\it Upper:} Histograms of the H$\beta$ velocities in km\,s$^{-1}$ measured using the cross-correlation technique for the CSP-I (black), SDSS-II (cyan), SNLS (magenta), HSC (blue), and DES-SN (brown) surveys. {\it Lower:} Distribution of the absolute $i$-band magnitude at 43\,d of the CSP-I (black), SDSS-II (cyan), SNLS (magenta), HSC (blue), and DES-SN (brown) surveys. The absolute magnitudes were calculated assuming a $\Lambda$CDM model ($\Omega_{m}=0.30$, $\Omega_{\Lambda} = 0.70$) and a Hubble constant of 70\,km\,s$^{-1}$ Mpc$^{-1}$. Readers are reminded that the HSC sample had only one SN. In both figures, the vertical line and the filled region represent the median and their 1$\sigma$ uncertainties, respectively.}
\label{fig:compa_vel_mag}
\end{figure}

Figure \ref{fig:compa_vel_mag} (upper) shows the H$\beta$ velocity distribution. Although the DES-SN sample distribution looks slightly different from the CSP-I (no peak around 6000\,km\,s$^{-1}$), a Kolomogorov-Smirnov test does not reject the null hypothesis that both groups are sampled from populations with identical distributions ($p = 0.66$). Therefore, all of the velocity distributions are consistent with coming from the same distribution. In addition, all of the surveys have similar average velocities. Figure \ref{fig:compa_vel_mag} (lower) shows the absolute magnitude distribution. There we see that the DES-SN sample distribution is similar to the CSP-I sample, while the SDSS-II sample distribution statistically differs ($p = 0.012$) with an average absolute magnitude brighter than for CSP-I. As discussed above, \citet{andrea10} and \citet{poznanski10} suggested that the SDSS-II sample is biased toward brighter objects. 

This is also seen in the $\Omega_{m}$ values obtained using CSP-I + SDSS-II and CSP-I + DES-SN (see Table \ref{tab:params_SCM_samples}). With CSP-I + SDSS-II, because SDSS-II is biased toward brighter objects, $\Omega_{m}$ is larger than using CSP-I + DES-SN ($0.66^{+0.25}_{-0.37}$ versus $0.30^{+0.35}_{-0.21}$). These distributions show that the DES-SN sample has a different or less extreme bias than SDSS~II (see Section \ref{txt:sim_bias}), explaining why the best-fitting parameters are consistent with or without the DES-SN sample. 

To determine whether we see any evolution effects on the fitting parameters, we fit our data using different samples (see Section \ref{txt:sim_bias} for bias simulation). All of the best-fitting values and their associated uncertainties are displayed in Table \ref{tab:params_SCM_samples}. The easiest way to look for potential redshift effects is to compare the parameters derived using only the local CSP-I sample and the most distant SNe~II from a combination of the SDSS-II, SNLS, DES-SN, and HSC samples. Both subsamples have roughly the same size (37 versus 33 SNe~II). Even if the best-fitting parameters are consistent at 1$\sigma$ owing to their large uncertainties, we see variations between the two subsamples, suggesting possible redshift effects. However, we think that the differences could be explained by a bias selection (Malmquist) rather than by redshift evolution. This trend was also found in previous studies \citep{nugent06,andrea10,poznanski10} when they compared their low-$z$ and high-$z$ samples. For example, \citet{andrea10} and \citet{poznanski10} found that the SDSS-II sample was overluminous and favoured a smaller value of $\alpha$. 

Regarding the effect of host-galaxy extinction, we do not find a statistically significant correlation between the $(r-i)$ colour at 43\,d and the redshift. However, even if the colour scatter is large and the order of magnitude of the K-correction is $\sim 0.02$\,mag for CSP-I or $\sim 0.12$\,mag for SDSS-II/DES-SN (depending on the SN redshift and filters), a possible trend is seen. Most distant SNe~II seem to have smaller $(r-i)$ values (bluer objects). We find an average colour of $0.003 \pm 0.119$\,mag ($N=39$), $-0.072 \pm 0.106$\,mag ($N=24$), and $-0.134 \pm 0.185$\,mag ($N=7$) for $z<0.05$, $0.05<z<0.15$, and $z>0.15$, respectively. This could be an effect of the Malmquist bias; at high-$z$ we observe the brightest events, those less affected by host-galaxy extinction. However, as demonstrated in the previous paragraph, the colour has a tiny effect on the SN~II standardisation, suggesting that the trend is more due to noise than a correlation between the redshift and the colour. Nonetheless, it could also be caused by intrinsic properties \citep{dejaeger18a}.

\begin{table*}
\caption{SCM-fit parameters: samples}
\begin{threeparttable}
\begin{tabular}{ccccccc}
\hline
Dataset & $\alpha$ & $\beta$ & $M_{i}$ &$\sigma_{\rm int}$ &$\Omega_{m}$ & $N$(SNe) \\
\hline
\hline
CSP-I & 3.82 $^{+0.68}_{-0.64}$ & 0.97 $\pm$ 0.45 & $-$16.79 $\pm$ 0.06 & 0.29 $^{+0.05}_{-0.04}$ & 0.50 $^{+0.33}_{-0.34}$ &37\\
CSP-I$+$SDSS-II & 3.78 $^{+0.61}_{-0.59}$ & 0.93 $^{+0.35}_{-0.34}$ & $-$16.87 $\pm$ 0.05 & 0.28 $\pm$ 0.05 & 0.66 $^{+0.25}_{-0.37}$ &50\\
CSP-I$+$SNLS & 3.68 $^{+0.64}_{-0.62}$ & 0.82 $^{+0.35}_{-0.34}$ & $-$16.79 $^{+0.05}_{-0.05}$ & 0.29 $^{+0.05}_{-0.04}$ & 0.28 $^{+0.37}_{-0.21}$ &41\\
CSP-I$+$DES-SN & 3.64 $^{+0.54}_{-0.53}$ & 0.56 $^{+0.35}_{-0.34}$ & $-$16.80 $\pm$ 0.05 & 0.29 $\pm$ 0.05 & 0.30 $^{+0.35}_{-0.21}$ &52\\
CSP-I$+$SDSS-II$+$SNLS & 3.64 $^{+0.61}_{-0.58}$ & 0.90 $^{+0.35}_{-0.34}$ & $-$16.86 $\pm$ 0.05 & 0.29 $\pm$ 0.04 & 0.44 $^{+0.34}_{-0.30}$ &54\\
CSP-I$+$SDSS-II$+$SNLS+HSC & 3.79 $\pm$ 0.55 & 0.89 $^{+0.35}_{-0.33}$ & $-$16.88 $\pm$ 0.05 & 0.29 $\pm$ 0.04 & 0.51 $^{+0.33}_{-0.30}$ &55\\
\citet{dejaeger17b} & 3.60 $^{+0.52}_{-0.51}$ & 0.91 $^{+0.31}_{-0.30}$ & $-$16.92 $\pm$ 0.05 & 0.29 $^{+0.04}_{-0.03}$ & 0.38 $^{+0.31}_{-0.25}$ &61\\
CSP-I$+$SDSS-II$+$DES-SN & 3.63 $^{+0.51}_{-0.49}$ & 0.74 $^{+0.33}_{-0.32}$ & $-$16.87 $\pm$ 0.05 & 0.29 $\pm$ 0.04 & 0.39 $^{+0.35}_{-0.27}$ &65\\
CSP-I$+$SNLS$+$DES-SN & 3.59 $^{+0.53}_{-0.52}$ & 0.47 $^{+0.39}_{-0.37}$ & $-$16.80 $\pm$ 0.05 & 0.29 $\pm$ 0.04 & 0.23 $^{+0.30}_{-0.17}$ &56\\
CSP-I$+$SDSS$+$SNLS$+$DES-SN$+$HSC & 3.71 $^{+0.51}_{-0.49}$ & 0.71 $^{+0.32}_{-0.33}$ & $-$16.88 $\pm$ 0.05 & 0.29 $^{+0.04}_{-0.03}$ & 0.35 $^{+0.33}_{-0.23}$ &70\\
SDSS-II & 3.84 $^{+1.39}_{-1.91}$ & 0.02 $^{+0.64}_{-0.58}$ & $-$17.15 $\pm$ 0.09 & 0.20 $\pm$ 0.12 & 0.57 $^{+0.30}_{-0.36}$ &13\\
SDSS-II$+$SNLS & 2.22 $^{+1.88}_{-1.95}$ & 0.48 $^{+0.75}_{-0.77}$ & $-$17.04 $\pm$ 0.10 & 0.34 $^{+0.10}_{-0.07}$ & 0.38 $^{+0.38}_{-0.27}$ &17\\
SDSS-II$+$DES-SN & 3.36 $^{+0.87}_{-0.85}$ & 0.26 $^{+0.52}_{-0.53}$ & $-$16.97 $^{+0.09}_{-0.08}$ & 0.3 $^{+0.07}_{-0.06}$ & 0.31 $^{+0.37}_{-0.23}$ &28\\
SDSS-II$+$SNLS$+$DES-SN & 3.18 $\pm$ 0.83 & 0.25 $^{+0.50}_{-0.52}$ & $-$16.94 $\pm$ 0.08 & 0.31 $\pm$ 0.06 & 0.27 $^{+0.36}_{-0.20}$ &32\\
SDSS-II$+$SNLS$+$HSC$+$DES-SN & 3.56 $^{+0.80}_{-0.78}$ & 0.26 $^{+0.52}_{-0.54}$ & $-$16.97 $^{+0.09}_{-0.08}$ & 0.30 $\pm$ 0.06 & 0.33 $^{+0.37}_{-0.24}$ &33\\
SNLS$+$DES-SN & 3.40 $\pm$ 0.89 & $-$0.46 $^{+0.58}_{-0.62}$ & $-$16.73 $\pm$ 0.10 & 0.25 $^{+0.09}_{-0.08}$ & 0.5 $^{+0.34}_{-0.33}$ &19\\
DES-SN & 3.35 $\pm$ 1.01 & $-$0.38 $^{+0.68}_{-0.65}$ & $-$16.76 $\pm$ 0.11 & 0.30 $^{+0.11}_{-0.09}$ & 0.50 $^{+0.34}_{-0.33}$ &15\\

\hline
\hline
\end{tabular}
Best-fitting values and the associated uncertainties for each parameter of the SCM fit at 43\,d after the explosion and using different samples.
\label{tab:params_SCM_samples}
\end{threeparttable}
\end{table*}

\subsection{Fit for $\Omega_{m}$ in $\Lambda$CDM cosmological model}\label{txt:cosmo_SCM}

\begin{figure}
\centering
\includegraphics[width=1.0\columnwidth]{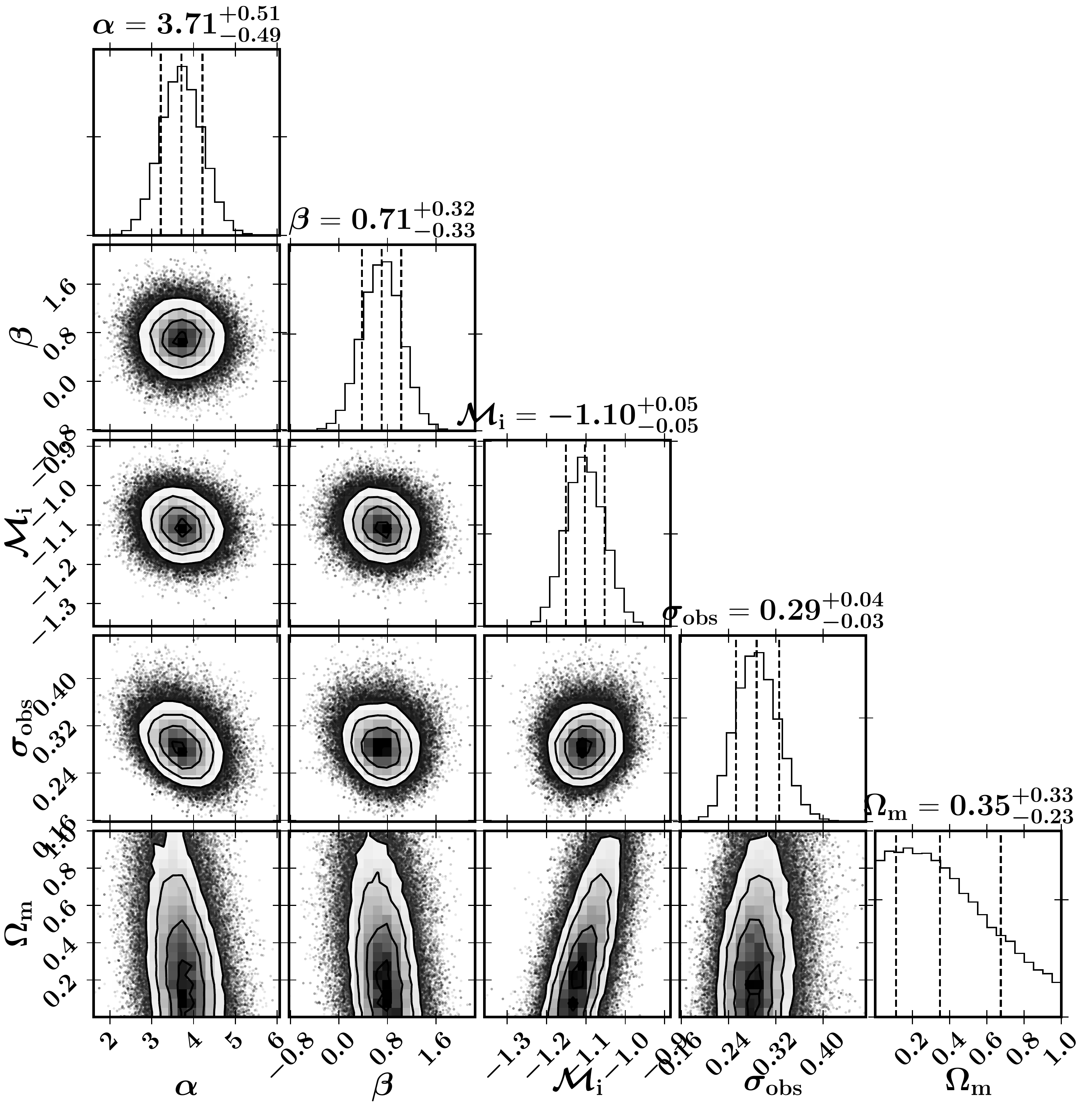}
\caption{Corner plot showing all of the one- and two-dimensional projections. Contours are shown at 0.5$\sigma$, 1$\sigma$, 1.5$\sigma$, and 2$\sigma$ (which, in two dimensions, correspond to the 12\%, 39\%, 68\%, and 86\% of the volume). The five free parameters are plotted: $\alpha$, $\beta$, $\mathcal{M}_{i}$, $\sigma_{\rm obs}$, and $\Omega_{m}$. To make this figure we used the corner plot package (triangle.py v0.1.1. Zenodo. 10.5281/zenodo.11020); we assume a flat universe and use the SCM.}
\label{fig:corner_plot_SCM}
\end{figure}

After constructing a high-$z$ Hubble diagram assuming a fixed cosmology, here we constrain cosmological parameters. We follow the procedure presented in Section \ref{txt:fixed_cosmo_SCM} with the exception of leaving $\Omega_{m}$ as a free parameter together with $\alpha$, $\beta$, $\mathcal{M}_{i}$, and $\sigma_{\rm obs}$.\footnote{As priors we choose $0.0 < \Omega_{m} < 1.00$, $0.0 < \sigma_{\rm obs} < 0.9$, and $\alpha$, $\beta$, $\mathcal{M}_{i} \neq 0$.} The best-fitting parameters ($\alpha$, $\beta$, $\mathcal{M}_{i}$, $\sigma_{\rm obs}$, and $\Omega_{m}$) are shown in Figure \ref{fig:corner_plot_SCM} in a corner plot with all of the one- and two-dimensional posterior distributions. 

The fitted value for the matter density is $\Omega_{m} = 0.35^{+0.33}_{-0.23}$, which corresponds to a dark energy density of $\Omega_{\Lambda}=0.65^{+0.24}_{-0.33}$. The value derived in this work is consistent with that obtained by \citet{dejaeger17b} ($\Omega_{m} = 0.38^{+0.31}_{-0.25}$) and demonstrates evidence of dark energy using SNe~II. Despite this independent measurement, the precision reached with SNe~II is far from that obtained with SNe~Ia \citep{betoule14,scolnic18}. A more precise estimate of the cosmological parameters requires a significant improvement of the SCM (see Section \ref{txt:fixed_cosmo_SCM}) and an increase in the number of high-$z$ SN~II observations. In this sample, only three SNe~II have been observed at $z>0.3$ while many hundreds of high-$z$ SNe~Ia have been used for cosmology \citep{betoule14,scolnic18}.

\subsection{Error budget}\label{txt:err_SCM}

In this section, we analyse the effect of each systematic error on the distance modulus. We run a Monte Carlo (MC) simulation ($N=2000$ realisations), where for each simulation, only one systematic (explosion date, magnitude, velocity, etc.) is offset by a random error (Gaussian distribution) due to its uncertainty. Then, for each iteration, the data are fitted using Eq. \ref{eq:likelihood} (without Bayes' inference as it was performed in Section \ref{txt:fixed_cosmo_SCM} and \ref{txt:cosmo_SCM} -- i.e., only a likelihood minimisation without priors). New values of $\alpha$, $\beta$, $\mathcal{M}_{i}$, $\sigma_{\rm obs}$, and $\Omega_{m}$ are derived, and therefore new distance moduli as well. Finally, we compare the average distance moduli obtained with those derived without MC simulation. The effect on the fitting parameters of each systematic uncertainty is summarised in Table \ref{tab:params_sys_SCM}. Note that the fitting parameters at 43\,d shown in Table \ref{tab:params_sys_SCM} slightly differ from those displayed in Figure \ref{fig:corner_plot_SCM}, as the former are derived only by minimising Eq. \ref{eq:likelihood} without running an MCMC simulation.

\subsubsection{Zero-point uncertainties}

Ground-based photometric zero-point calibration is generally limited to an accuracy of 0.01--0.02\,mag (see Table 10 of \citealt{conley11}). To compute the zero-point uncertainty effects on the distance modulus, for each survey we shift in turn the photometry from each band by 0.015\,mag \citep{amanullah10} and refit. All of the fitting parameters and the distance moduli remain essentially similar. If we use different offset for each survey \citep{conley11}, only $\Omega_{m}$ changes slightly (see Table \ref{tab:params_sys_SCM}).

\subsubsection{Magnitude/colour uncertainties}
The changes of the distance moduli and the fitting parameters due to the uncertainties in the photometry are evaluated by applying a magnitude/colour offset within the errors and refitting the data. The average fitting parameters and their standard deviation are shown in Table \ref{tab:params_sys_SCM}. As expected, $\beta$ is the fitting parameter with the largest difference, as it is the one which multiplies the colour term. The distance modulus residual between the values obtained with and without MC simulation has an average difference of 0.02\,mag. A strong correlation is seen between the distance modulus residuals and the colours in the sense that bluer SNe~II have larger positive residuals.

\subsubsection{Photospheric velocity uncertainties}

Here we investigate the influence of the photospheric velocity uncertainties on the distance moduli. We offset all of the velocities by a random error and refit all the data. We perform a MC analysis with 2000 realisations. The fitting parameter and distance modulus values and uncertainties correspond to the average value and the standard deviation over these 2000 realisations and are displayed in Table \ref{tab:params_sys_SCM}. The most affected fitting parameters are $\alpha$ and $\Omega_{m}$. This is easily explained by the fact that $\alpha$ is the parameter which multiplies the velocity. Regarding the distance modulus residual, the average of the absolute value is 0.038\,mag with a maximum of 0.12\,mag for SN~2008br. A strong correlation is seen between the distance modulus residuals and the velocities, in the sense that SNe with higher velocities have positive and larger residuals, while SNe with smaller velocities have negative and smaller residuals.

\subsubsection{Explosion date}\label{txt:sys_Texp}

Explosion date uncertainties are among the most important systematic errors, as they affect all of the observables: magnitudes, colours, and expansion velocities. In order to quantify the effect on the distance modulus, we compare the distance moduli derived at 43\,d (see Section \ref{txt:fixed_cosmo_SCM}) with those derived at 43\,d plus a random value within a normal distribution due to the uncertainty (MC simulation, $N=2000$). 

In Table \ref{tab:params_sys_SCM}, the average fitting parameters and their standard deviations are displayed. A comparison of the distance moduli obtained at 43\,d and those derived here gives a maximum difference of $\sim 0.1$\,mag, while the average absolute difference is $\sim 0.035$\,mag -- that is, $\sim 18$\% of the average distance modulus uncertainties ($\sim 0.20$\,mag; excluding the observed dispersion of 0.30\,mag). This is not surprising, as the distributions are centred on 43\,d, and thus the average distance moduli are also centred on the correct values derived in Section \ref{txt:fixed_cosmo_SCM}). However, we can look at the SNe~II with the largest differences (SN~2008br, SN~2008hg, DES14C3rhw, DES17S1bxt, and SN~2016jhj). One SN~II (SN~2008br) has a large explosion date uncertainty (9\,d), while for the other SNe~II, the uncertainties are all $\leq 5$\,d. However, both DES14C3rhw and DES17S1bxt have large magnitude/colour uncertainties, and SN~2016jhj has a steeply declining plateau. We can also compare the scatter around the mean value and the uncertainty in the distance modulus itself. Six SNe~II have a scatter larger than the uncertainty: SN~2005dt, SN~2007W, SN~2008ag, SN~2008bu, SN~2009bu, and 04D1pj. All of these SNe~II have relatively large explosion date uncertainties: 9, 7, 8, 7, 8, and 8\,d, respectively.

Finally, it is important to note that with our methodology, two effects affect the distance modulus: the explosion date and the explosion-date uncertainty. In Figure \ref{fig:sig_ep_SCM}, we study the explosion-date effect by showing $\Omega_{m}$ for different epochs. We clearly see that $\Omega_{m}$ varies depending on the epoch at which we apply the method. At 30--40\,d after the explosion, $\Omega_{m} \approx 0.2$--0.3, while at later epoch (70\,d), the value increases to $\sim 0.7$. Even if the value changes, almost all of the values are consistent at 1$\sigma$ owing to their large uncertainties. We also look at the evolution of the fitting parameters using different epochs (between 40 and 70\,d after the explosion). All of the fitting parameters evolve with the reference epoch; for example, $\alpha = 2.42 \pm 0.49$ when the reference epoch is 55\,d. Finally, it is interesting to note that as for the velocity uncertainties, the same correlation is seen between the distance modulus residuals and the velocities. This could be explained by the fact that $\alpha$ is the one of the most affected fitting parameters.

\subsubsection{Gravitational lensing}

Gravitational lensing only affects the high-redshift part of the Hubble diagram, leading to potential bias of the cosmological parameters. However, even if for our sample ($z < 0.36$) gravitational lensing should not have a strong effect, we adopt the approach of \citet{conley11,betoule14,scolnic18} by adding a value of 0.055$z$ \citep{jonsson10} in quadrature to the total uncertainty (see Eq. \ref{eq:likelihood}). Other studies (e.g., \citealt{kowalski08,amanullah10}) treat gravitational lensing using a value of 0.093$z$ \citep{holtz05}. Including the gravitational lensing term in the total uncertainty increases the average distance modulus uncertainties by 0.01\,mag. 

\subsubsection{Minimum redshift}

To evaluate for possible effects from using a given minimum redshift cut ($z_{\rm CMB} > 0.01$), we construct a new sample including all of the SNe~II, with no minimum redshift. The new sample size increases to 82 SNe~II (instead of 70 SNe~II). A systematic offset of $\sim 0.02$\,mag is seen between the distance moduli derived using the whole sample and the cut sample. The fitting parameters $\alpha$ and $\beta$ slightly differ because their velocity and colour distribution centres are similar. However, $\mathcal{M}_{i}$ varies when including all the SNe with a difference of almost 1$\sigma$. If we change the redshift cut to $z_{\rm CMB}> 0.0223$ (the cut used by \citealt{riess16}), the sample decreases to 44 SNe~II and the fitting parameters change as seen in Table \ref{tab:params_sys_SCM}. A difference of $\sim 0.6$, $\sim 0.7$, $\sim 1.4$, $\sim 0.2$, and $0.3\sigma$ for (respectively) $\alpha$, $\beta$, $\mathcal{M}_{i}$, $\sigma_{\rm obs}$, and $\Omega_{m}$ is seen.

\subsubsection{Milky Way extinction}

All of the light curves were corrected for Milky Way extinction using the \citet{car89} law, assuming a total-to-selective extinction ratio of $R_{V}= 3.1$ and using the extinction maps of \citealt{schlafly11}. To quantify the Milky Way extinction uncertainty effects on the distance modulus, we follow the approach of \citep{amanullah10}, increasing the Galactic $E(B-V)$ by 0.01\,mag for each SN and repeating the fit. All of the fitting parameters and the distance moduli are almost identical; the distance modulus residual has an average of $4.0 \times 10^{-4}$\,mag.


\begin{table*}
\normalsize
\caption{SCM-fit parameters: systematics errors.}
\begin{threeparttable}
\begin{tabular}{cccccc}
\hline
Systematic errors & $\alpha$ & $\beta$ & $\mathcal{M}_{i}$ &$\sigma_{\rm int}$ &$\Omega_{m}$\\
\hline
Original	&3.77 $\pm$ 0.51 	&0.76 $\pm$ 0.32 	&$-$1.13 $\pm$ 0.06 	&0.27 $\pm$ 0.04 	&0.17 $\pm$ 0.35\\ 
\hline
ZP	&3.76 $\pm$ 0.51 	&0.77 $\pm$ 0.32 	&$-$1.11 $\pm$ 0.06 	&0.27 $\pm$ 0.03 	&0.19 $\pm$ 0.36\\
Mag/colour	&3.75 $\pm$ 0.51 	&0.61 $\pm$ 0.34 	&$-$1.12 $\pm$ 0.06 	&0.28 $\pm$ 0.04 	&0.24 $\pm$ 0.39\\
Velocity &3.29 $\pm$ 0.56 	&0.75 $\pm$ 0.35 	&$-$1.11 $\pm$ 0.06	&0.30 $\pm$ 0.04 	&0.30 $\pm$ 0.42\\
$t_{\rm exp}$	&3.47 $\pm$ 0.62 	&0.60 $\pm$ 0.37 	&$-$1.11 $\pm$ 0.07	&0.30 $\pm$ 0.04 	&0.31 $\pm$ 0.45\\
All $z$ &3.82 $\pm$ 0.47 	&0.78 $\pm$ 0.32 	&$-$1.06 $\pm$ 0.04	&0.29 $\pm$ 0.04 	&0.21 $\pm$ 0.37\\
$z>0.0223$ &4.14 $\pm$ 0.71 	&0.36 $\pm$ 0.43 	&$-$1.25 $\pm$ 0.05	&0.27 $\pm$ 0.05 	&0.05 $\pm$ 0.42\\
$A_{V,G}$	&3.77 $\pm$ 0.51 	&0.77 $\pm$ 0.32 	&$-$1.11 $\pm$ 0.05 	&0.27 $\pm$ 0.04 	&0.17 $\pm$ 0.35\\
\hline
mean systematic &0.17 $\pm$ 0.18 	&0.11 $\pm$ 0.25 	&0.04 $\pm$ 0.04 	&0.01 $\pm$ 0.01 	&0.08 $\pm$ 0.05\\
\hline
\hline
\end{tabular}

Effect of the systematic errors on the best-fitting values using the SCM. Original line corresponds to the values obtained by minimising Eq. \ref{eq:likelihood} without MCMC (no Bayesian inference). Velocity, $t_{\rm exp}$, All $z$, $z > 0.0223$, $A_{V,G}$, ZP (shift separately for each survey), and Mag/colour correspond to the values derived by changing the velocities, explosion time, including all the redshifts, including only the SNe~II with $z > 0.0223$, the Galactic visual extinction, the filter photometric zero-point, and the colour/magnitude as described in Section \ref{txt:err_SCM}. Note that for each parameter, the total errors correspond to the standard deviation of the 2000 MC simulations added in quadrature to the mean of the 2000 errors obtained for each parameter. The mean systematic uncertainty corresponds to the average of the difference between the original and each systematic, while the error corresponds to the standard deviation.
\label{tab:params_sys_SCM}
\end{threeparttable}
\end{table*}
\subsection{Simulated distance modulus bias versus redshift}\label{txt:sim_bias}

In this section, we will use the public ``SuperNova ANAlysis'' (SNANA)\footnote{\url{http://snana.uchicago.edu/}} software package \citep{kessler09b} to estimate the distance modulus bias ($\mu$-bias) due to selection effects (e.g., Malmquist bias) versus redshift. As seen in Figure \ref{fig:z_distribution} where the overall number of events exponentially declines with redshift, the Malmquist bias could be significant and an important source of uncertainty (see Section \ref{txt:err_SCM}).

To simulate events, SNANA needs three ingredients \citep{kessler18,brout19b}: (1) a source model, to generate a variety of spectral energy distributions (SEDs); (2) a noise model, to convert true magnitudes to true fluxes with a certain cadence, and apply Poisson noise to get measured fluxes; and (3) a trigger model, to define the final sample by applying spectroscopic selection functions or candidate logic (e.g., at least two detections). 

As a source model, we use the ``SNII-NMF'' model used for the Photometric LSST Astronomical Time Series Classification Challenge (PLAsTiCC; \citealt{kessler19}). 
It consists of a SED, which is a linear combination of three ``eigenvectors'' built using hundreds of well-observed SNe~II after applying a non-negative matrix factorisation (NMF) as a dimensionality reduction technique. 
For each simulated SN~II, the multiplicative factors of the three ``eigenvectors'' (``eigenvalues'') are obtained from correlated Gaussian distributions measured from the data. 

Unlike the SN~Ia bias simulation in \cite{kessler18}, for SNe~II we do not have the infrastructure to model spectral features and their correlations with brightness (e.g., expansion velocities vs. brightness); thus, we apply a slightly different methodology. First, we assume that the total rest-frame brightness variation is $\sim 0.95$\,mag, and second, that after standardisation the Hubble scatter is 0.27\,mag. Therefore, to model the magnitude variation, we will use two sources: a known random scatter with a dispersion of 0.815\,mag (the SNII-NMF model by itself includes a scatter of 0.4\,mag) which is exactly corrected in the analysis\footnote{This would correspond to the colour and stretch variation for a SN~Ia simulation.}, and an unknown intrinsic scatter with a dispersion of 0.27\,mag. Note that the combined dispersion is 0.95\,mag. Both scatters are added coherently to all bands and phases (COH model).

SNANA can directly use the image properties (PSF, sky noise, zero point) to simulate the noise; however, other than for DES-SN, we do not have access to the meta data to perform these accurate simulations\footnote{We do not perform simulations for HSC as we have only one object. For the DES sample, to simplify the analysis, we decide to apply the same methodology used for CSP-I, SDSS-II, and SNLS, even if we have access to the meta data.}. Therefore, we follow the procedure described by \citet{kessler18} (in their Section 6.1.1) for their low-$z$ sample. Instead of using the image properties, an approximate cadence is generated directly from the observed data (light curves, redshifts, coordinates, observation dates, etc.). 

\begin{figure}
	\includegraphics[width=1.00\columnwidth]{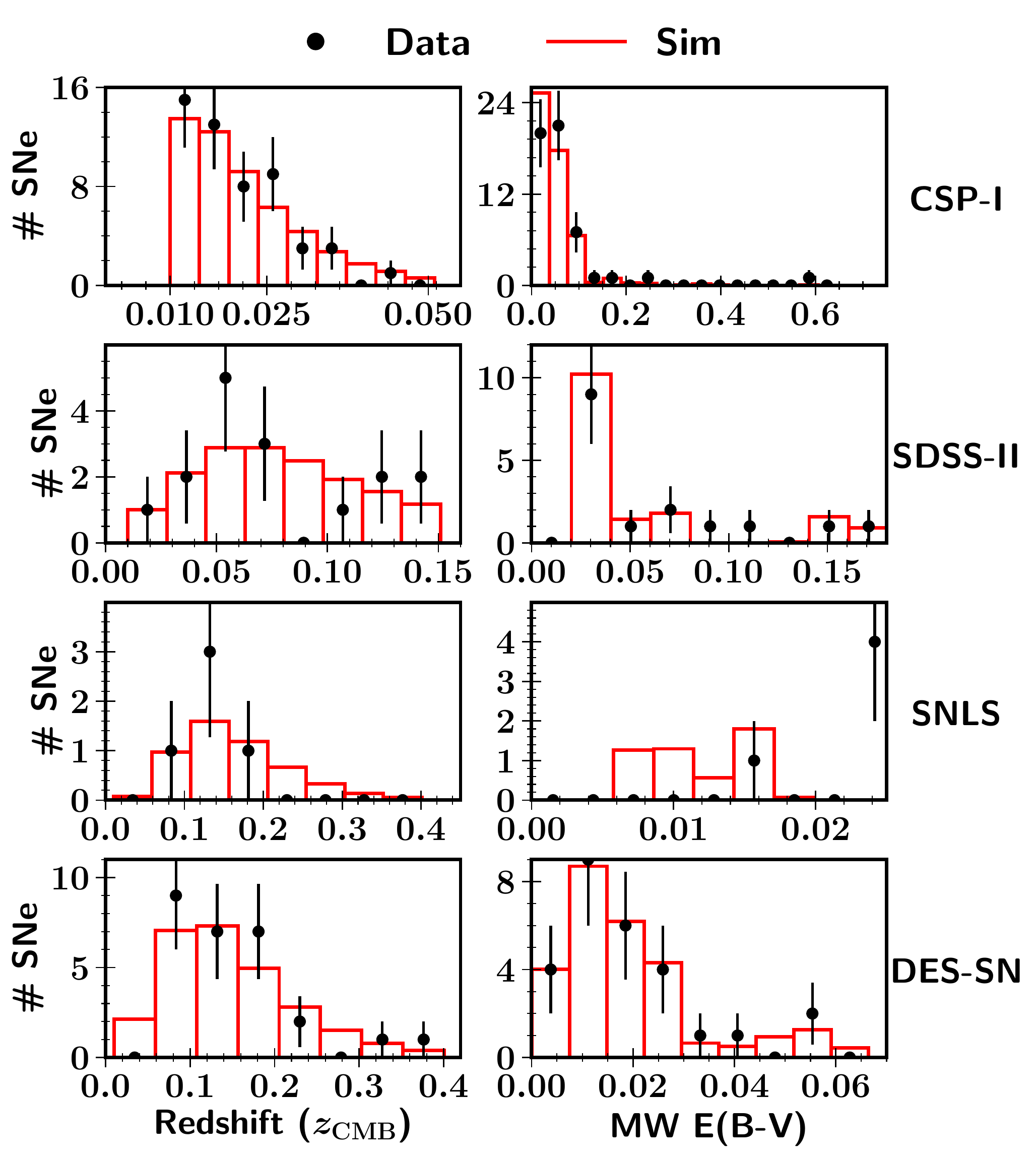}
	\caption{Comparison of data (black dots) and simulation using SNANA (red histogram) for distributions in the CSP-I (\textit{top row}), SDSS-II (\textit{second row}), SNLS (\textit{third row}), and DES-SN (\textit{bottom row}) samples. The simulations have 1,000,000 SNe for each survey, but the histograms were scaled to have the same number of events as the data. The left column shows CMB redshift $z_{\rm CMB}$, while the right column represents Galactic extinction MW $E(B-V)$.}
\label{fig:data_sim}
\end{figure}

The last step (trigger model) is to apply the spectroscopic selection function. For each survey, it consists of a function of peak $i$-band magnitude versus redshift. This function is manually adjusted until good agreement between simulations and data for redshift and Milky Way extinction (MW $E(B-V)$) distributions is obtained. As seen in Figure \ref{fig:data_sim}, we find good agreement between the data and simulations for all surveys and for both redshift and MW $E(B-V)$ parameters. Note that for each survey, we simulated 1,000,000 objects; 2.4\%, 1.3\%, 1.8\%, and 2.6\% (CSP-I, SDSS-II, SNLS, and DES-SN, respectively) of the objects passed the spectroscopic selection.

\begin{figure}
\centering
\includegraphics[width=1.0\columnwidth]{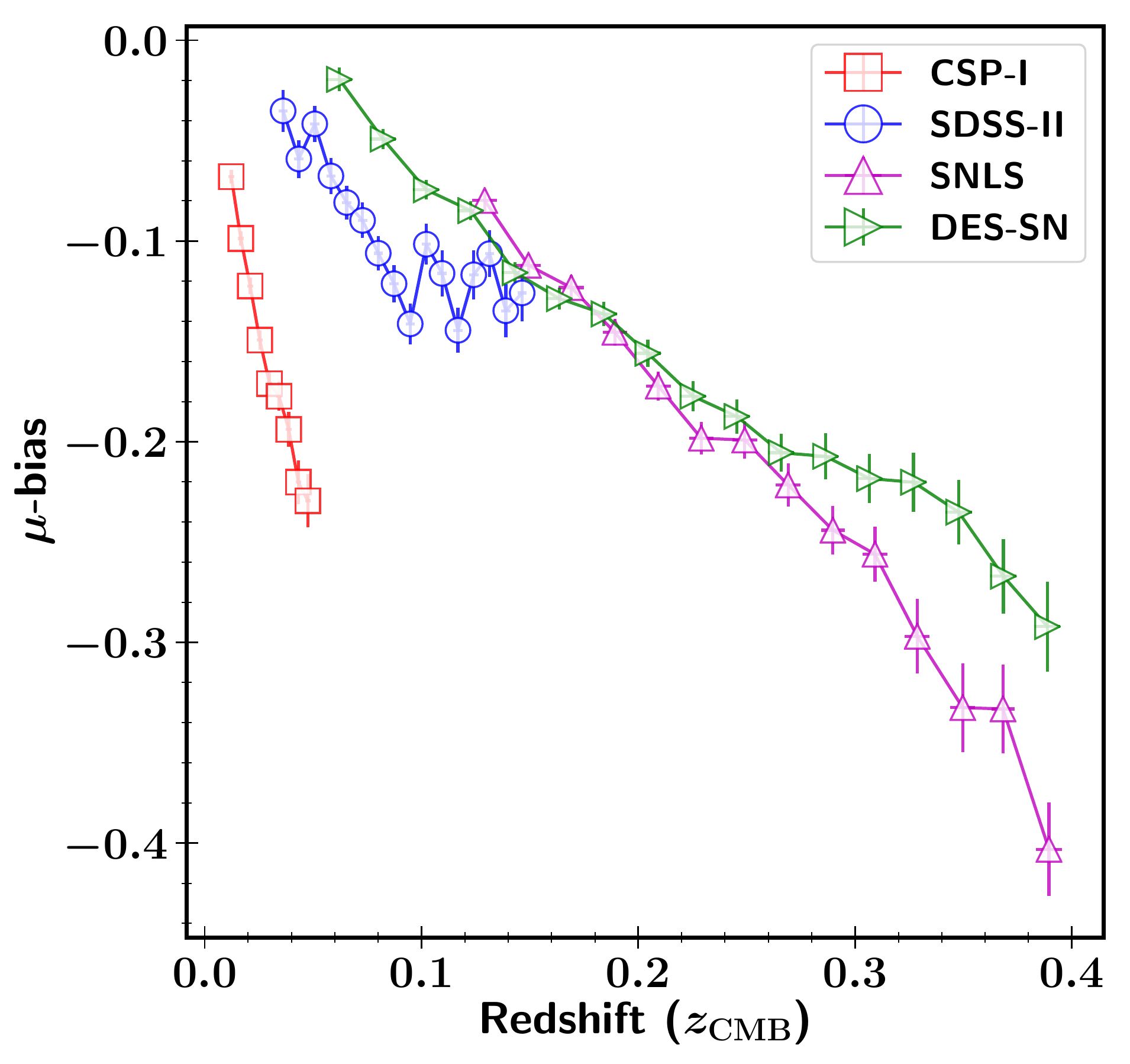}
\caption{Distance modulus bias due to selection effects versus redshift for CSP-I (red squares), SDSS-II (blue circles), SNLS (magenta triangles), and DES-SN (green right-pointed triangles).}
\label{fig:mu_bias}
\end{figure}

Finally, the $\mu$-bias versus redshift is obtained by taking the average value of the random Gaussian smear applied in the simulation corresponding to the unknown scatter (dispersion of 0.27\,mag). In Figure \ref{fig:mu_bias}, $\mu$-bias versus redshift is shown for four surveys: CSP-I, SDSS-II, SNLS, and DES-SN. The average $\mu$-bias for the CSP-I survey is $\sim -0.15$\,mag, while for SDSS, the $\mu$-bias is lower with an average value of $\sim -0.09$\,mag. From these simulations, we see that the SN~II $\mu$-bias increase can be large at high redshifts, with a value of $\sim -0.25$\,mag at $z = 0.3$.

It is important to note that the SN~II bias is much larger than the one obtained for SNe~Ia. With their low-$z$ sample, \citet{kessler18} obtained an average value of $\sim -0.02$\,mag. Even if one expects to obtain a larger bias for SNe~II than for SNe~Ia because SN~II are less luminous (by $\sim 2$\,mag), the large difference is also due to a difference in the methodology. If the same technique used in this work is applied to the low-$z$ SN~Ia sample from \citet{kessler18}, the average SN~Ia bias increases to $-0.10$\,mag. To obtain a more accurate $\mu$-bias simulation, the spectral features and their correlations with brightness need to be modeled, as well as the use of a better SN~II template; this is matter for future work.

Even though our method is an approximation, we apply the $\mu$-bias to each SN~II and refit the cosmology. Note that for the HSC sample, we use the SNLS bias. The best-fitting parameters obtained with bias correction are consistent with those obtained without. For example, we derive $\Omega_{m} = 0.29^{+0.32}_{-0.20}$ versus $\Omega_{m} = 0.35^{+0.33}_{-0.23}$ (see Section \ref{txt:cosmo_SCM}). Regarding the other parameters, we get $\alpha = 3.52 \pm 0.49$ (versus $\alpha = 3.71^{+0.51}_{-0.49}$), $\beta = 0.66 \pm 0.33$ (versus $\beta = 0.71 ^{+0.32}_{-0.33}$), and $\mathcal{M}_{i} = -1.00 \pm 0.05$ (versus $\mathcal{M}_{i} = -1.10 \pm 0.05$), with an observed dispersion $\sigma_{\rm obs} = 0.29^{+0.04}_{-0.03}$\,mag (versus $\sigma_{\rm obs} = 0.29^{+0.04}_{-0.03}$\,mag). With these new fitting-parameter values, the offset between SDSS and DES seen in Figure \ref{fig:HD_SCM} remains the same ($\sim 0.28$\,mag). If we fixed $\alpha$, $\beta$, and $\mathcal{M}_{i}$, and apply the $\mu$-bias correction, the SDSS average offset reduces to $-0.13$\,mag but the DES average offset increases to 0.15\,mag, and therefore the offset between SDSS and DES remains almost identical.

\section{PCM Results}

In this section, we will first assume a $\Lambda$CDM cosmological model and present an updated SN~II Hubble diagram using the PCM. Second, assuming a flat universe, we will constrain the matter density ($\Omega_{m}$). In both cases, a comparison with photometric Hubble diagrams from the literature is presented. Note that, unlike for the SCM, in this Section we do not perform $\mu$-bias simulation. We leave a detailed modelisation of the photometric features and their correlations with brightness to a future paper as it will require a significant effort to update SNANA.

\begin{figure}
\centering
\includegraphics[width=1.0\columnwidth]{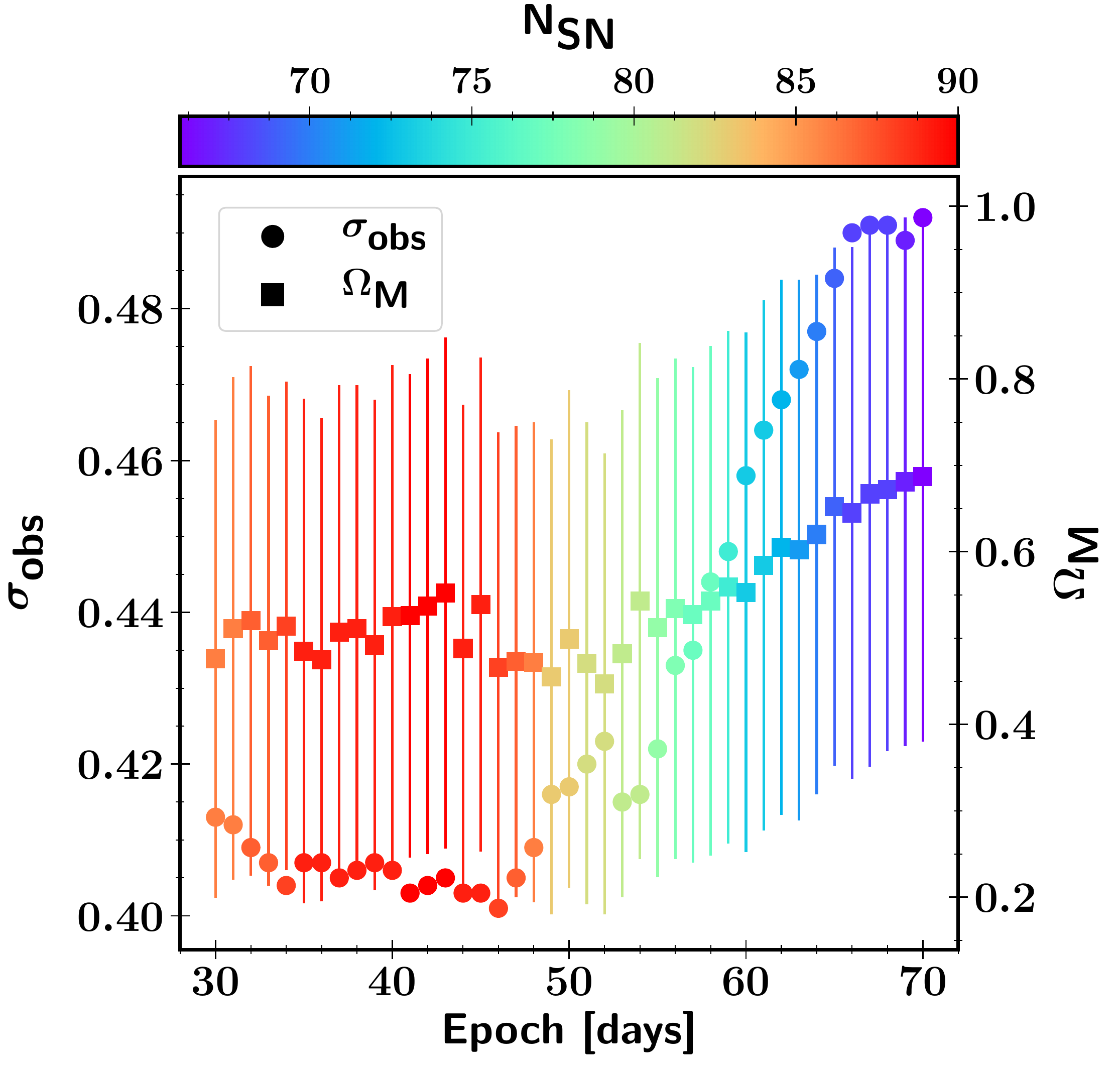}
\caption{Variation by epoch of the intrinsic dispersion in the Hubble
diagram (circles and left ordinate axis) and $\Omega_{m}$ (squares and right ordinate axis) using the PCM. The colour bar at the top represents the different sample sizes. For clarity, only the $\Omega_{m}$ uncertainties are plotted.}
\label{fig:sig_ep_PCM}
\end{figure}

\subsection{Fixed cosmology}\label{txt:fixed_cosmo_PCM}

\begin{figure*}
	\includegraphics[width=17cm]{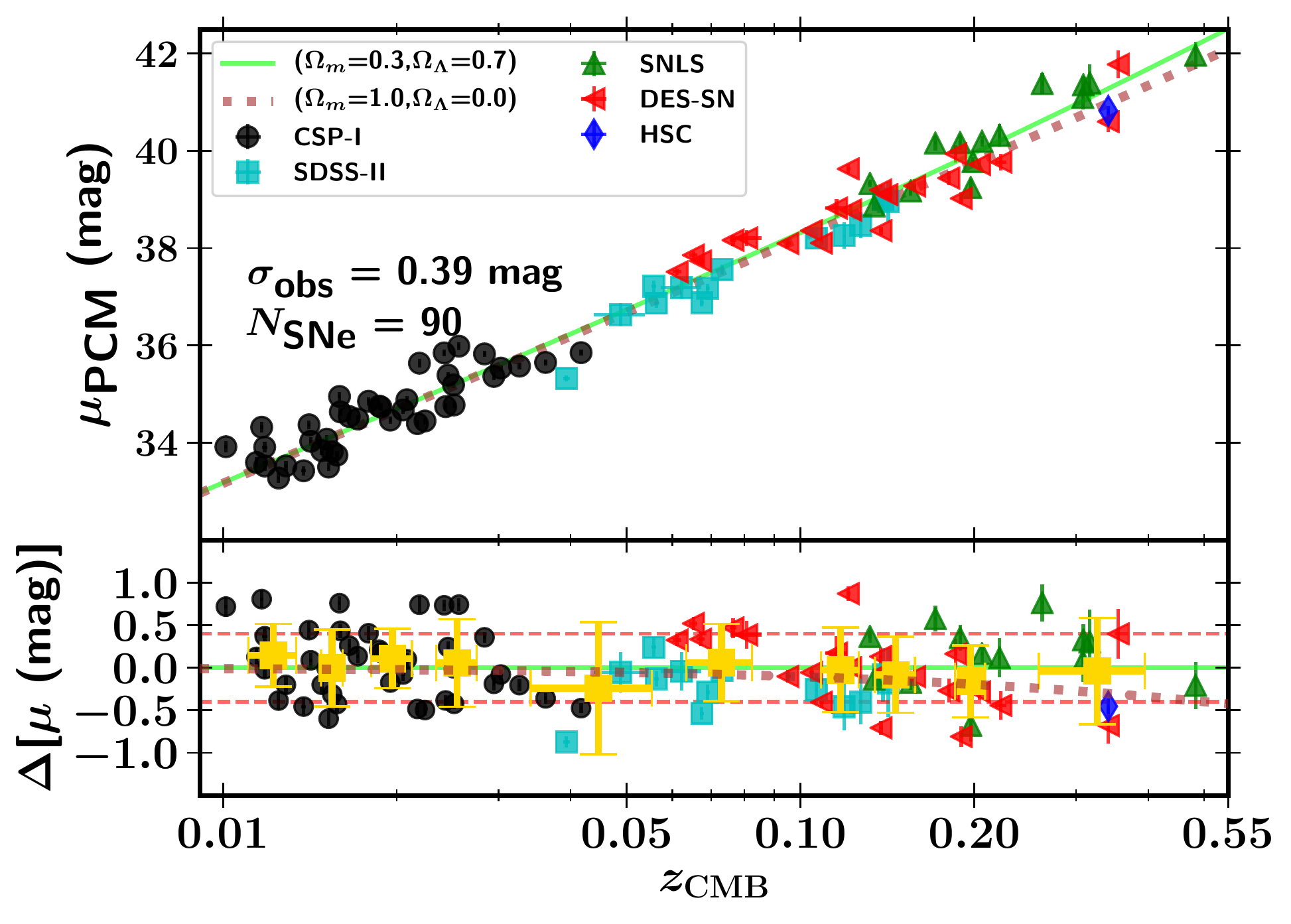}
\caption{Hubble diagram (top) and residuals from the $\Lambda$CDM model (bottom) using the PCM as applied to the data taken from CSP-I (black circles; \citealt{dejaeger17a}), SDSS-II (cyan squares; \citealt{andrea10}), SNLS (green triangles; \citealt{dejaeger17a}), HSC (blue diamond; \citealt{dejaeger17b}), and DES-SN (red left triangles; this work). The lime solid line is the Hubble diagram for the $\Lambda$CDM model ($\Omega_{m}=0.3$, $\Omega_{\Lambda} = 0.7$), while the brown dot line is for an Einstein-de Sitter cosmological model ($\Omega_{m}=1.0$, $\Omega_{\Lambda} = 0.0$). In both models, we use H$_0= 70$\,km\,s$^{-1}$ Mpc$^{-1}$ to standardise the SN~II brightness. We present the number of SNe~II available at this epoch ($N_{\rm SNe}$), the epoch after the explosion, and the observed dispersion ($\sigma_{\rm obs}$). The yellow squares in the Hubble residual plot represent the binned data using 10 SNe~II per bin.}
\label{fig:HD_PCM}
\end{figure*}

As for the SCM, we select SNe~II in the Hubble flow -- a total of 101 SNe~II (47 CSP-I $+$ 14 SDSS-II $+$ 15 SNLS $+$ 1 HSC $+$ 24 DES-SN). We apply the PCM at 43\,d after the explosion even if at 46\,d the scatter is slightly smaller as shown in Figure \ref{fig:sig_ep_PCM}. This choice is motivated by the fact that between the two epochs the intrinsic dispersion differs by 0.003\,mag but at 43\,d the comparison with the SCM will be straightforward. From the total sample we cut 8 SNe~II because their explosion date uncertainties are larger than 10\,d (see Table \ref{tab:sum_cut}), 2 SNe~II for a lack of photometry, and 1 SN~II (DES13C2jtx) identified as an outlier ($3\sigma$ clipping).

Finally, the PCM total sample at 43\,d is composed of 90 SNe~II: 40 SNe~II from CSP-I, 13 SNe~II from SDSS-II, 14 SNe~II from SNLS, 1 SN~II from HSC, and 22 SNe~II from DES-SN.
 
\begin{figure}
	\includegraphics[width=1.0\columnwidth]{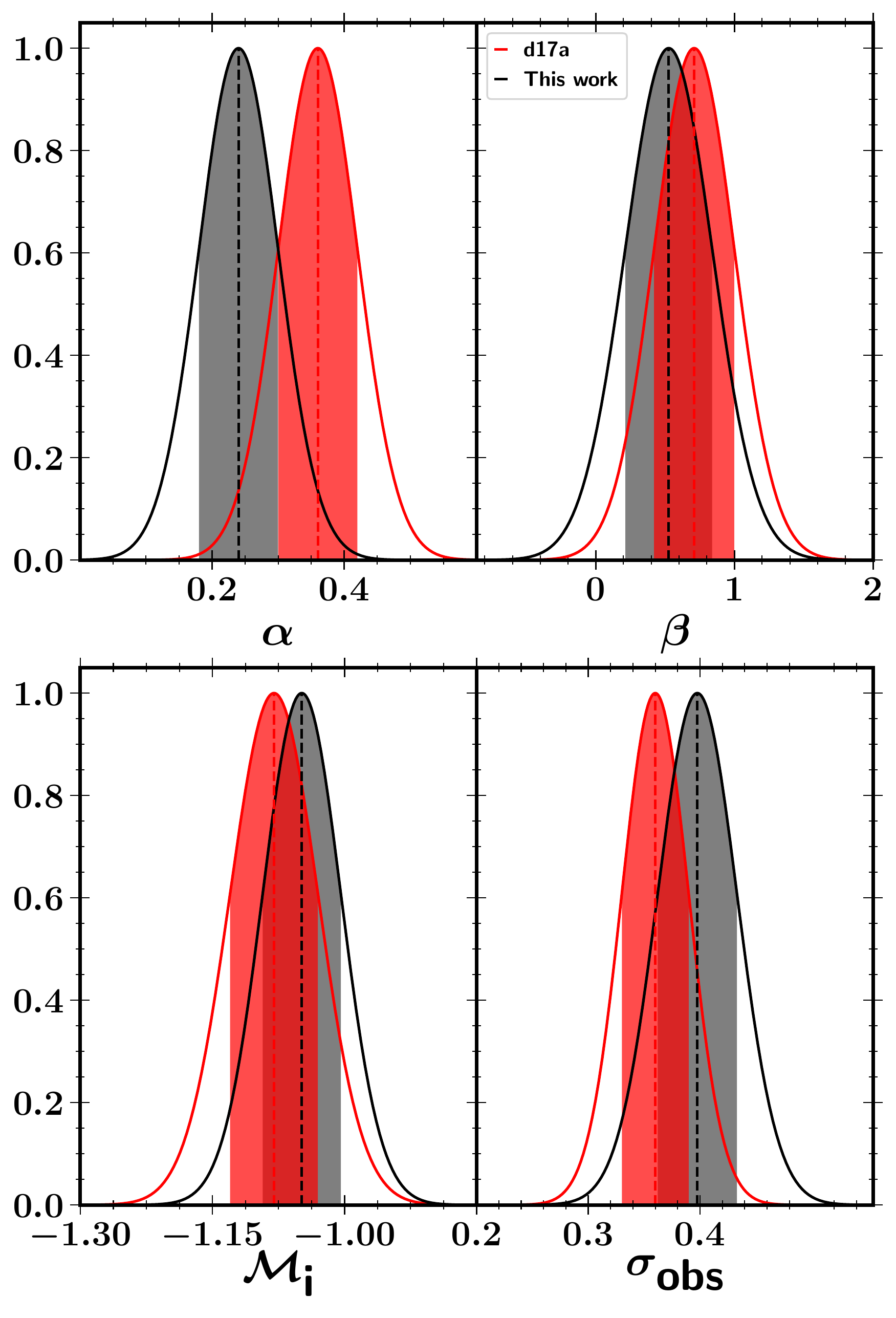}
	\caption{Comparison of the best-fitting parameters using the PCM derived by \citet{dejaeger17a} in red and those obtained in this work (in black) with the DES-SN sample.  \textit{Top left:} Distributions of $\alpha$. \textit{Top right:} Distributions of $\beta$. \textit{Bottom left:} Distributions of ``Hubble-constant-free'' absolute magnitude ($\mathcal{M}_{i}$). \textit{Bottom right:} Distributions of observed dispersion ($\sigma_{\rm obs}$). In each panel, the vertical dashed line represents the average value, while the filled region represents the 1$\sigma$ uncertainty.}
\label{fig:compa_parameters_PCM}
\end{figure}

\begin{figure}
	\includegraphics[width=\columnwidth]{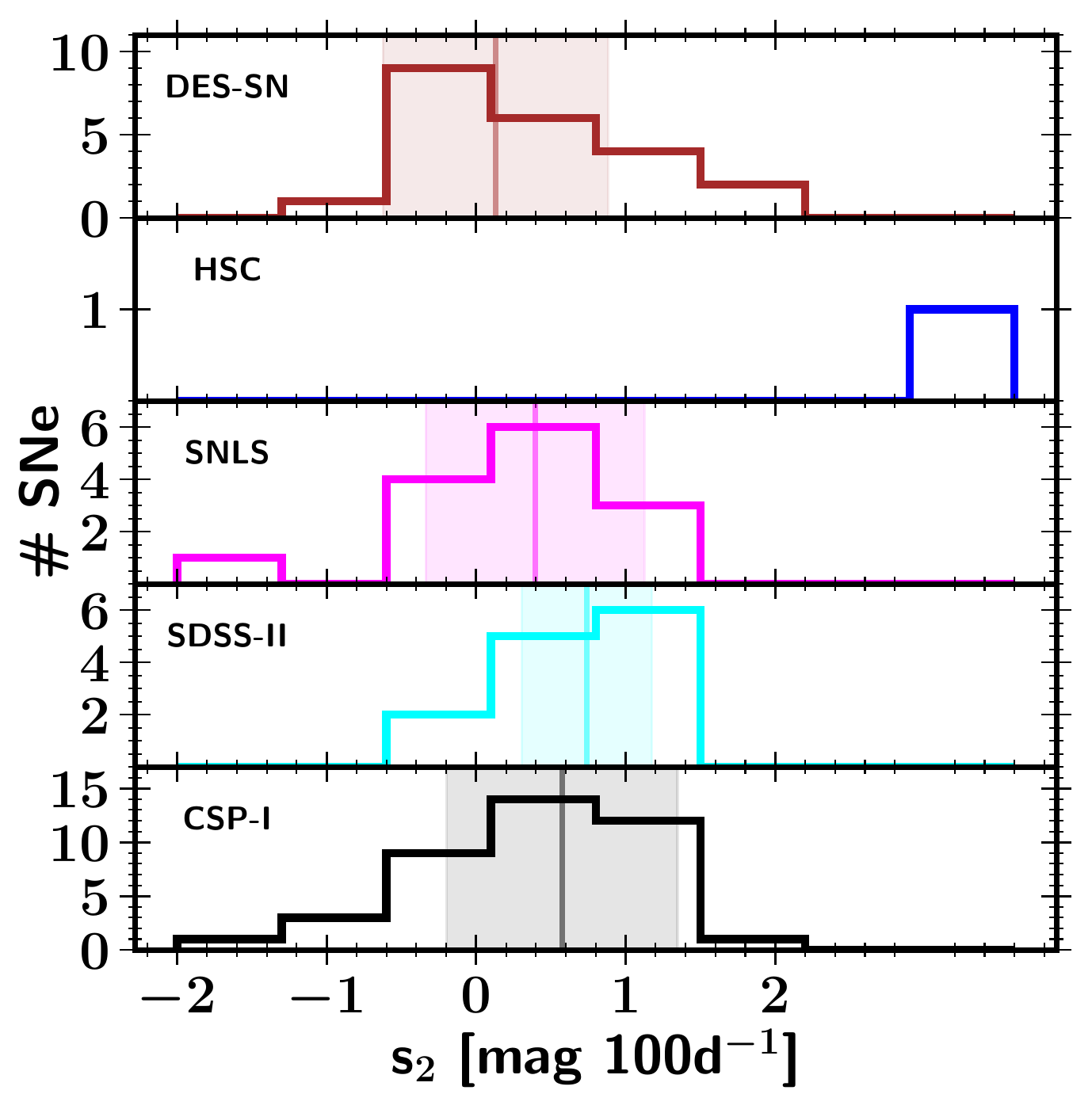}
\caption{Histograms of the $s_{2}$ slope in mag\,(100\,d)$^{-1}$ for the CSP-I (black), SDSS-II (cyan), SNLS (magenta), HSC (blue), and DES-SN (brown) surveys. Readers are reminded that the HSC sample had only one SN. The vertical lines and the filled regions represent the medians and their 1$\sigma$ uncertainties, respectively.}
\label{fig:compa_s2}
\end{figure}

In Figure \ref{fig:HD_PCM} the SN~II Hubble diagram and the Hubble residuals of the combined data are shown. Assuming a $\Lambda$CDM cosmological model, the best-fitting parameters are $\alpha = 0.24 \pm 0.06$, $\beta = 0.53 \pm 0.31$, and $\mathcal{M}_{i} = -1.05 \pm 0.05$, with an observed dispersion $\sigma_{\rm obs} = 0.39 \pm 0.04$\,mag, or 17--18\% in distance. As shown in Figure \ref{fig:compa_parameters_PCM}, almost all the fitting parameters are consistent at 1$\sigma$ with those derived by \citet{dejaeger17a} ($\alpha = 0.36 \pm 0.06$, $\beta = 0.71^{+0.29}_{-0.28}$, $\mathcal{M}_{i} = -1.08 \pm 0.05$, and $\sigma_{\rm obs} = 0.36 \pm 0.03$\,mag). However, difference are seen in $\alpha$ and could be explained by the newly reanalysed $s_{2}$ values for the whole sample (Galbany et al., in prep.). For all the surveys, the $s_{2}$ distributions are displayed in Figure \ref{fig:compa_s2}. The DES-SN sample distribution is statistically (KS test) consistent with the other distributions. Using the PCM, the average systematic offset between SDSS-II and DES-SN ($\sim 0.28$\,mag) seen in Figure \ref{fig:HD_SCM} is smaller. For SDSS, the median residual from the $\Lambda$CDM model is $-0.17$\,mag while for DES-SN it is $-0.01$\,mag.

We also compare the distance moduli derived in this work and those by \citet{dejaeger17a}. A mean difference of $-0.02$\,mag with a standard deviation of 0.24\,mag is found. 14 SNe~II (9 from CSP-I, 3 from SDSS-II, and 2 from SNLS) have distance moduli not consistent at 1$\sigma$, 5 SNe~II (4 from CSP-I and 1 from SNLS) at 2$\sigma$, and 2 SNe~II from CSP-I at 3$\sigma$. These differences could be attributed to a difference of methodology (linear interpolation versus Gaussian Process), to a fine-tuned measurement of $s_{2}$ (mean average difference of $-0.05$\,mag\,(100\,d)$^{-1}$), but mostly by the use of the recalibrated CSP-I photometry (14/19 SNe~II are from CSP). However, it is important to note that if we take into account the minimum uncertainty in distance determination using the PCM ($\sim 0.40$\,mag), all the distances are consistent.

\begin{figure}
\centering
\includegraphics[width=1.0\columnwidth]{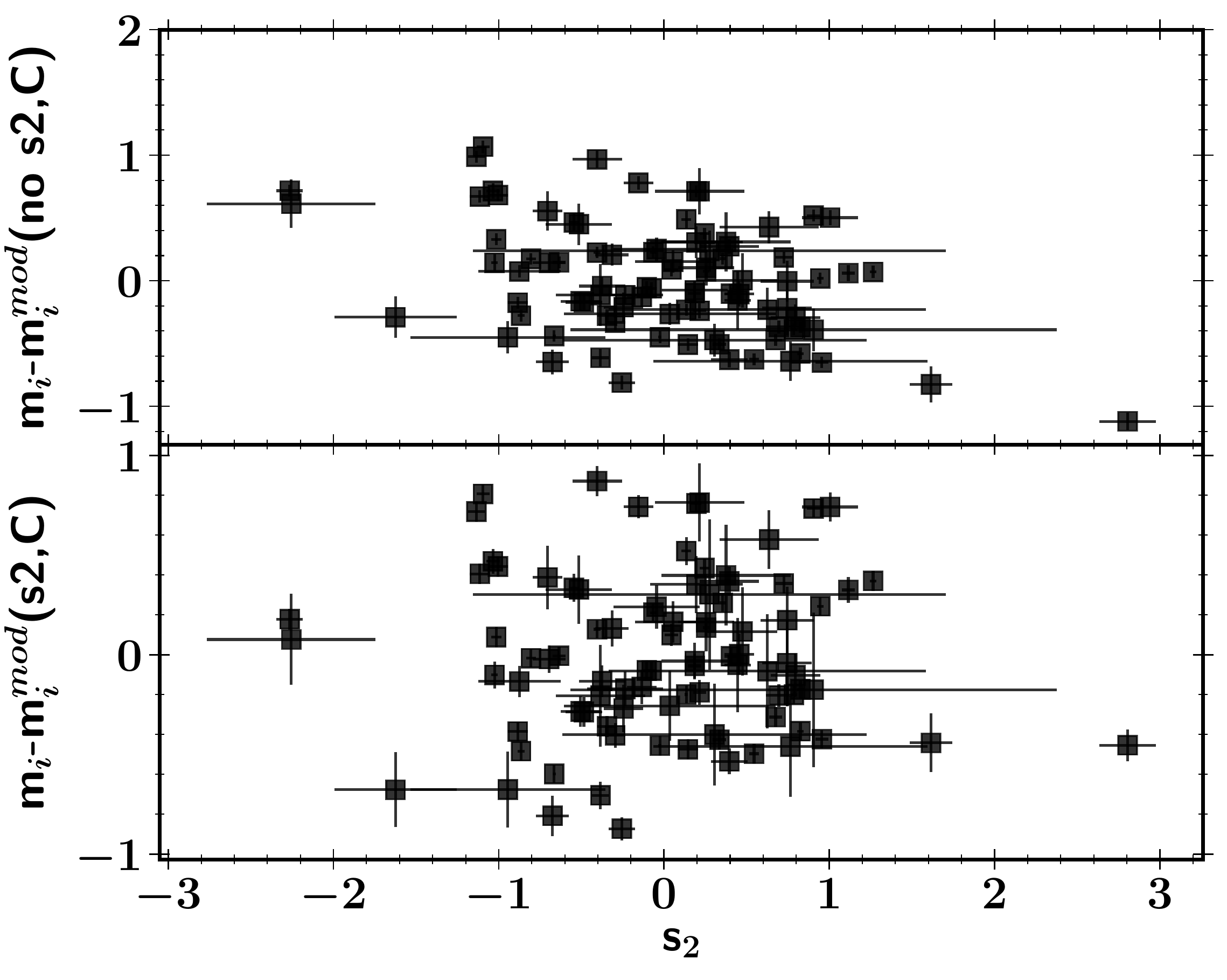}
\caption{The relationship between SN~II luminosity and $s_{2}$. The \textit{upper panel} illustrates the relationship (SN~II magnitudes are corrected for distances and colours), while the \textit{lower panel} shows the trend between luminosity $s_{2}$ after correcting the magnitudes for $s_{2}$ ($\alpha s_{2}$).}
\label{fig:s2_relationship_PCM}
\end{figure}

\begin{figure}
\centering
\includegraphics[width=1.0\columnwidth]{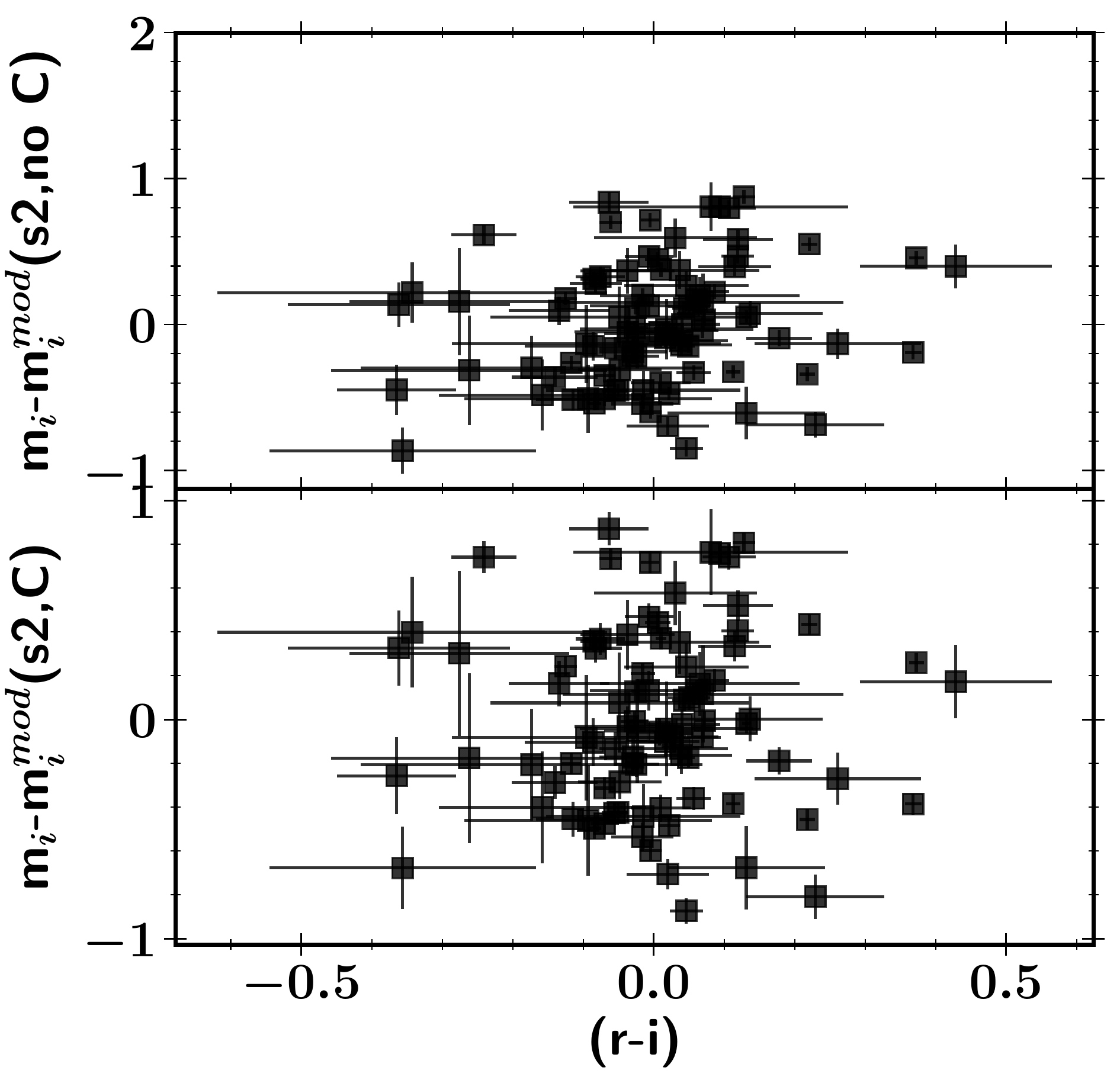}
\caption{The relationship between SN~II luminosity and the colour 43\,d after the explosion. The \textit{upper panel} shows the relationship (SN~II magnitudes are corrected for distances and $s_{2}$), while the \textit{lower panel} shows the trend between luminosity colour after correcting the magnitudes for colours ($\beta$$(r-i)$).}
\label{fig:color_relationship_PCM}
\end{figure}

Finally, in Figure \ref{fig:s2_relationship_PCM} and Figure \ref{fig:color_relationship_PCM}, the relationship between the two parameters ($s_{2}$ and colour) that have been used to standardise SNe~II and the luminosity are shown. From these figures as seen with the SCM, the colour does not improve the standardisation. The Pearson factor between the colour and the luminosity corrected for distance and $s_{2}$ is 0.21 and decreases to 0.03 after correction. On the other hand, a correlation is seen between $s_{2}$ and the magnitude corrected for distance and colour with a Pearson factor of $-0.43$. The $s_{2}$ coefficient is efficient, as the Pearson factor drops to $-0.02$ when a $s_{2}$ correction is applied.

\subsection{$\Omega_{m}$ derivation}

\begin{figure}
\centering
\includegraphics[width=1.0\columnwidth]{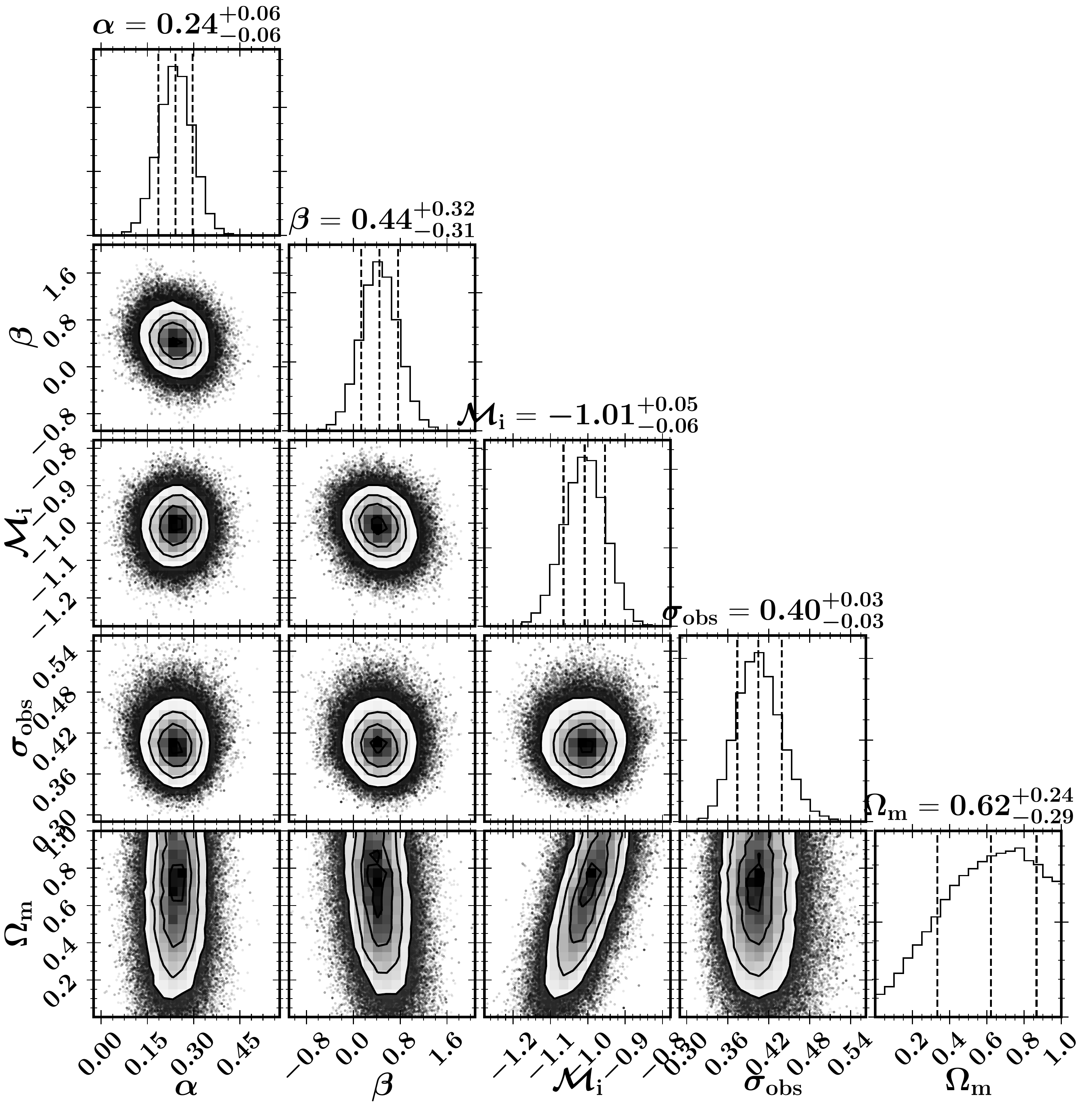}
\caption{Corner plot showing all of the one- and two-dimensional projections. Contours are shown at 0.5$\sigma$, 1$\sigma$, 1.5$\sigma$, and 2$\sigma$ (which, in two dimensions, correspond to the 12\%, 39\%, 68\%, and 86\% of the volume). The five free parameters are plotted: $\alpha$, $\beta$, $\mathcal{M}_{i}$, $\sigma_{\rm obs}$, and $\Omega_{m}$. To make this figure we use the corner-plot package (triangle.py v0.1.1. Zenodo. 10.5281/zenodo.11020). In deriving this figure, we assume a flat universe.}
\label{fig:corner_plot_PCM}
\end{figure}

Following the procedure described in Section \ref{txt:cosmo_SCM}, we also derive an $\Omega_{m}$ value assuming a flat universe. In Figure \ref{fig:corner_plot_PCM}, a corner plot with all the one- and two-dimensional projections is shown. Assuming a flat universe, we derive a value for the matter density of $\Omega_{m} = 0.62^{+0.24}_{-0.29}$ -- that is, a dark energy density value $\Omega_{\Lambda}=0.38^{+0.29}_{-0.24}$. Even if this result is almost consistent at 1$\sigma$ with the latest SN~Ia results (\citealt{scolnic18}; $\Omega_{m} = 0.298 \pm 0.022$), our $\Omega_{m}$ value is much larger. It is also important to note that this result appears to be affected by the priors. If we choose less restrictive priors for $\Omega_{m}$, $0.0 < \Omega_{m} < 2.5$ instead of $0.0 <\Omega_{m} < 1.0$, the value and the uncertainties increase to $\Omega_{m} = 0.77^{+0.46}_{-0.36}$. Both values are consistent owing to their large uncertainties; however, the fact that $\Omega_{M}$ depends on the priors could suggest that currently with our small sample of high-$z$ SNe~II, SNe~II cannot play a key role in the $\Omega_{M}$ determination and should be used only at low-$z$ to derive H$_0$. In the future, though, more SNe~II will be observed at high-$z$, and this larger set of SNe~II will be useful for estimating $\Omega_{M}$.

With respect to \citet{dejaeger17a} -- that is, the same sample except the DES-SN and HSC samples -- in this work we found a higher value (but still consistent) for the matter density ($\Omega_{m} = 0.32^{+0.30}_{-0.21}$ in \citealt{dejaeger17a}). This might be explained by the fact that the DES-SN sample could be biased toward brighter objects, implying smaller distances and thus, by definition, favouring a Universe with more matter. However, as discussed in Section \ref{txt:bias}, it does not seem to be the case. Even if using the PCM our results are larger than the current best-fit values from other probes, we think that this method is still encouraging as it allows us to use more objects (only those with photometric information). However, future work should focus on reducing the intrinsic dispersion by (for example) developing a new SN~II template for the K-correction or to fit the light curves and measure more precisely the $s_{2}$ slopes and the magnitudes. Finally, new improvements could also be possible by adding another parameter which correlates with the intrinsic brightness or by finding a SN~II subgroup which is better standardisable. Note that if we use the velocity and the slope term, the dispersion does not decrease and remains around 0.28--0.30\,mag.

\subsection{Redshift bias}

As done with the SCM, here we determine if there is any bias effect as a function of the redshift. We fit our data using different samples and all the best-fitting values are shown in Table \ref{tab:params_PCM}. From this table, a possible redshift evolution is seen in $\alpha$. A value of $0.30 \pm 0.09$ is found for the low-$z$ sample (CSP-I; 40 SNe~II) while $\alpha = 0.19 \pm 0.07$ using the rest of the sample (SDSS-II $+$ SNLS $+$ DES-SN $+$ HSC; 49 SNe~II) or $\alpha = 0.07^{+0.10}_{-0.09}$ for SNLS $+$ DES-SN. Values at low-$z$ and high-$z$ differ by $\sim 1\sigma$; therefore, this difference could be explained by a redshift evolution or by a Malmquist bias (at high-$z$ the brightest objects are observed). In any case, further investigations with better statistics at high-$z$ should be done to confirm or invalidate this result. Regarding the $\beta$ value, the large uncertainties prevent a definitive conclusion; however, at first sight, the values remain around 0.4 except for SNLS $+$ DES-SN where a smaller but still consistent value is found. Finally, the $\Omega_{m}$ values obtained using CSP-I $+$ SDSS-II and CSP-I $+$ DES-SN are more similar for the PCM than the SCM which confirms the absence of an offset in the Hubble diagram for the PCM (see Figure \ref{fig:HD_PCM}). 

\begin{table*}
\caption{PCM-fit parameters.}
\begin{threeparttable}
\begin{tabular}{ccccccc}
\hline
Dataset & $\alpha$ & $\beta$ & $M_{i}$ &$\sigma_{\rm int}$ &$\Omega_{m}$ &SNe\\
\hline
\hline

CSP-I & 0.29 $\pm$ 0.09 & 0.33 $^{+0.62}_{-0.60}$ & $-$16.75 $\pm$ 0.07 & 0.43 $^{+0.06}_{-0.05}$ & 0.52 $^{+0.33}_{-0.35}$ &40\\
CSP-I$+$SDSS-II & 0.26 $\pm$ 0.08 & 0.65 $\pm$ 0.47 & $-$16.83 $\pm$ 0.06 & 0.41 $^{+0.05}_{-0.04}$ & 0.64 $^{+0.26}_{-0.37}$ &53\\
CSP-I$+$SNLS & 0.24 $\pm$ 0.08 & 0.30 $\pm$ 0.46 & -16.73 $\pm$ 0.07 & 0.41 $^{+0.05}_{-0.04}$ & 0.38 $^{+0.33}_{-0.25}$ &54\\
CSP-I$+$DES-SN & 0.25 $\pm$ 0.07 & 0.39 $\pm$ 0.41 & $-$16.73 $\pm$ 0.06 & 0.42 $\pm$ 0.04 & 0.72 $^{+0.20}_{-0.30}$ &62\\
CSP-I$+$SDSS-II$+$SNLS & 0.23 $\pm$ 0.07 & 0.41 $\pm$ 0.39 & $-$16.79 $\pm$ 0.06 & 0.40 $^{+0.04}_{-0.03}$ & 0.37 $^{+0.33}_{-0.25}$ &67\\
CSP-I$+$SDSS-II$+$SNLS+HSC & 0.26 $^{+0.06}_{-0.07}$ & 0.51 $^{+0.39}_{-0.40}$ & $-$16.82 $\pm$ 0.06 & 0.40 $\pm$ 0.04 & 0.44 $^{+0.32}_{-0.27}$ &68\\
CSP-I$+$SDSS-II$+$DES-SN &0.24 $\pm$ 0.07 & 0.56 $\pm$ 0.35 & $-$16.79 $\pm$ 0.05 & 0.41 $^{+0.04}_{-0.03}$ & 0.74 $^{+0.19}_{-0.31}$ &75\\
CSP-I$+$SNLS$+$DES-SN & 0.22 $^{+0.07}_{-0.06}$ & 0.28 $^{+0.36}_{-0.34}$ & $-$16.71 $\pm 0.06$ & 0.41 $^{+0.04}_{-0.03}$ & 0.59 $^{+0.26}_{-0.29}$ &76\\
CSP-I$+$SDSS$+$SNLS$+$DES-SN$+$HSC & 0.24 $\pm$ 0.06 & 0.44 $^{+0.32}_{-0.31}$ &$-$16.78 $\pm$ 0.05 & 0.40 $\pm$ 0.03 & 0.62 $^{+0.24}_{-0.29}$ &90\\
SDSS-II & 0.07 $^{+0.19}_{-0.18}$ & 0.40 $^{+0.76}_{-0.79}$ & $-$17.14 $\pm$ 0.12 & 0.33 $^{+0.11}_{-0.07}$ & 0.50 $^{+0.33}_{-0.34}$ &13\\
SDSS-II$+$SNLS & 0.11 $\pm$ 0.11 & 0.45 $^{+0.55}_{-0.61}$ & $-$16.94 $^{+0.10}_{-0.09}$ & 0.37 $^{+0.07}_{-0.06}$ & 0.17 $^{+0.29}_{-0.13}$ &27\\
SDSS-II$+$DES-SN & 0.18 $\pm$ 0.10 & 0.57 $^{+0.47}_{-0.45}$ & $-$16.87 $\pm$ 0.09 & 0.41 $^{+0.06}_{-0.05}$ & 0.60 $^{+0.28}_{-0.34}$ &35\\
SDSS-II$+$SNLS$+$DES-SN & 0.15 $\pm$ 0.08 & 0.39 $\pm$ 0.39 & $-$16.83 $^{+0.09}_{-0.08}$ & 0.40 $^{+0.05}_{-0.04}$ & 0.40 $^{+0.35}_{-0.27}$ &49\\
SDSS-II$+$SNLS$+$DES-SN$+$HSC & 0.21 $\pm$ 0.07 & 0.44 $\pm$ 0.39 & $-$16.83 $\pm$ 0.09 & 0.40 $^{+0.05}_{-0.04}$ & 0.50 $^{+0.31}_{-0.30}$ &50\\
SNLS$+$DES-SN & 0.09 $^{+0.10}_{-0.09}$ & 0.04 $\pm$ 0.46 & $-$16.66 $^{+0.09}_{-0.10}$ & 0.39 $^{+0.06}_{-0.05}$ & 0.68 $^{+0.23}_{-0.33}$ &36\\
DES-SN & 0.163 $\pm$ 0.14 & 0.27 $^{+0.64}_{-0.62}$ & $-$16.71 $^{+0.11}_{-0.12}$ & 0.44 $^{+0.09}_{-0.07}$ & 0.70 $^{+0.22}_{-0.35}$ &22\\
\hline
\hline
\end{tabular}
Best-fitting values and the associated uncertainties for each parameter of the PCM fit at 43\,d after the explosion and using different samples.
\label{tab:params_PCM}
\end{threeparttable}
\end{table*}

\subsection{Error budget}\label{txt:err_PCM}

As previously done in Section \ref{txt:err_SCM}, in this section, we analyse the effect of each systematic error on the distance modulus. We follow the same procedure explained above, running an MC simulation where for each simulation each observable is offset by a value according to its uncertainty. The effect on the fitting parameters of each systematic error is summarised in Table \ref{tab:params_sys_PCM}. For reference, we use the fitting parameters obtained at 43\,d derived by minimising \ref{eq:likelihood} (without MCMC).

\subsubsection{Zero-point uncertainties}

Similarly to the method used for the SCM, here we compute the zero-point uncertainty effects on the distance modulus by shifting in turn the photometry from each band by 0.015\,mag and refit \citep{amanullah10}. Almost all the fitting parameters and the distance moduli remain identical, only $\mathcal{M}_{i}$ change to $-$98 $\pm$ 0.07. If instead of adding a constant value for all the photometric systems, we add a different offset for each survey \citep{conley11}, $\Omega_{m}$ evolves slightly but not statistically significantly (increase of 0.02, $< 3\%$) as seen in Table \ref{tab:params_sys_PCM},.

\subsubsection{Magnitude/colour uncertainties}

To estimate the influence of the photometry uncertainty, we apply a magnitude/colour offset within the error uncertainties and refit the data (2000 simulations). The average fitting parameters and their associated standard deviations are shown in Table \ref{tab:params_sys_PCM}. The only fitting parameter statistically affected by the photometric uncertainties is $\beta$. An absolute average difference of 0.013\,mag is seen in the distance moduli, and as for the SCM, the distance modulus residuals and the colours are correlated in the sense that bluer SNe~II have larger positive residuals.

\subsubsection{Slope uncertainties}

In this paragraph, the effect of the plateau slope uncertainties on distance moduli are investigated. We offset the slope by a number withing the slope uncertainty and refit all the data. The average and standard deviation of the 2000 fitting parameters are displayed in \ref{tab:params_sys_PCM}. All of the fitting parameters remain mostly identical. Even $\alpha$ which multiplies the slope almost does not change. Therefore, the absolute average distance modulus difference is very small (0.01\,mag) and the maximum value is 0.05\,mag. A strong correlation is seen between the distance modulus residuals and the slope, in sense that SNe with steeper slope have positive and larger residuals.

\subsubsection{Explosion date}\label{txt:sys_Texp_PCM}

Following the procedure described in Section \ref{txt:sys_Texp}, we investigate the explosion date uncertainty effects on the fitting parameters and the distance moduli. For this purpose, we apply the PCM not at 43\,d but at 43\,d plus a random value within a normal distribution due to the explosion date uncertainty which is different for each SN. In Table \ref{tab:params_sys_PCM}, the averaged fitting parameters and their standard deviation are displayed. The distance moduli derived using the PCM are less affected by the explosion date uncertainty than those obtained with the SCM. The average absolute difference in the distance moduli is $\sim 0.012$\,mag against 0.035\,mag for the SCM. This is easily explained as for the SCM, the expansion velocities are strongly affected by the explosion date while for the PCM, the plateau slope is not. This is seen in Figure \ref{fig:sig_ep_SCM} where the $\Omega_{m}$ values at different epoch is displayed. For the PCM, the $\Omega_{m}$ evolves from $\sim 0.50$ at early epochs to $\sim 0.70$ at late time, while for the SCM the variation was larger ($\sim 0.20$ to $\sim 0.70$). Regarding the fitting parameters, only $\beta$ and $\Omega_{m}$ evolve, but they are still consistent at 1$\sigma$ with the ``original'' values.

\subsubsection{Gravitational lensing}

As for the SCM, the gravitational lensing effects are treated by adding a value of 0.055$z$ \citep{jonsson10} in quadrature to the total uncertainty (see Eq. \ref{eq:likelihood}). If we choose another value (e.g., 0.093$z$ \citealt{kowalski08}), the total uncertainty on the distance modulus increase by 0.01\,mag.

\subsubsection{Minimum redshift}

In this subsection, the effects on the fitting parameters on using a give redshift cut ($z_{\rm CMB} > 0.01$) are analysed. For this purpose, we change the redshift cut to $z_{\rm CMB} > 0.0223$, the same cut used by \citet{riess16}. The sample decreases from 90 SNe~II to 63 SNe~II. As seen in Table \ref{tab:params_sys_PCM}, all of the fitting parameters change: $\sim 12$\% for $\alpha$, $\sim 10$\% for $\mathcal{M}_{i}$, and $\sim 40$\% for $\Omega_{m}$. The distance moduli are different with an absolute average difference of 0.08\,mag.

\subsubsection{Milky Way extinction}

Following \citep{amanullah10}, we evaluate the Milky Way extinction uncertainty effects on the distance modulus by increasing the Galactic $E(B-V)$ by 0.01\,mag for each SN and repeat the fit. As shown in Table \ref{tab:params_sys_PCM}, all the fitting parameters remain almost identical, and therefore, the distance moduli too.


\begin{table*}
\normalsize
\caption{PCM-fit parameters: systematics errors.}
\begin{threeparttable}
\begin{tabular}{cccccc}
\hline
Systematic errors & $\alpha$ & $\beta$ & $\mathcal{M}_{i}$ &$\sigma_{\rm int}$ &$\Omega_{m}$\\
\hline
Original	& 0.24 $\pm$ 0.05	& 0.42 $\pm$ 0.32	& $-$1.00 $\pm$ 0.07	& 0.39 $\pm$ 0.03 & 0.68 $\pm$ 0.39\\ 
\hline
ZP		& 0.24 $\pm$ 0.05	& 0.42 $\pm$ 0.32	& $-$0.98 $\pm$ 0.07	& 0.39 $\pm$ 0.03 & 0.70 $\pm$ 0.39\\
Mag/colour	& 0.24 $\pm$ 0.06 	& 0.32 $\pm$ 0.32	& $-$0.99 $\pm$ 0.07	& 0.40 $\pm$ 0.03 & 0.72 $\pm$ 0.41\\
slope		& 0.22 $\pm$ 0.06	& 0.45 $\pm$ 0.33	& $-$1.00 $\pm$ 0.07	& 0.39 $\pm$ 0.03 & 0.67 $\pm$ 0.39\\
$t_{\rm exp}$  & 0.25 $\pm$ 0.06	& 0.34 $\pm$ 0.35	& $-$1.00 $\pm$ 0.07	& 0.40 $\pm$ 0.04 & 0.61 $\pm$ 0.40\\
All $z$ 	& 0.24 $\pm$ 0.05 	& 0.45 $\pm$ 0.32 	& $-$0.99 $\pm$ 0.07	& 0.38 $\pm$ 0.03 & 0.74 $\pm$ 0.41\\
$z<0.0223$	& 0.21 $\pm$ 0.07	& 0.41 $\pm$ 0.37	& $-$1.10 $\pm$ 0.09	& 0.38 $\pm$ 0.04 & 0.40 $\pm$ 0.39\\
$A_{V,G}$	& 0.24 $\pm$ 0.05	& 0.42 $\pm$ 0.32	& $-$0.98 $\pm$ 0.07	& 0.39 $\pm$ 0.03 & 0.67 $\pm$ 0.38\\
\hline
mean systematic & 0.01 $\pm$ 0.01	& 0.04 $\pm$ 0.04	& 0.02 $\pm$ 0.03	& 0.006 $\pm$ 0.005	& 0.07 $\pm$ 0.09\\
\hline
\hline
\end{tabular}
Effect of the systematic errors on the best-fitting values using the PCM. Original line corresponds to the values obtained by minimising Eq. \ref{eq:likelihood} without MCMC (no Bayesian inference), while slope, $t_{\rm exp}$, All $z$, $z  > 0.0223$, $A_{V,G}$, ZP (shift separately for each survey), and Mag/colour respectively correspond to the values derived by changing the slopes, explosion time, including all the redshifts, including only the SNe~II with $z > 0.0223$, the filter photometric zero-point, and the colour/magnitude as described in Section \ref{txt:err_PCM}. Note that for each parameter, the total errors correspond to the standard deviation of the 2000 MC simulations added in quadrature to the mean of the 2000 errors obtained for each parameter. The mean systematic uncertainty corresponds to the average of the difference between the original and each systematic while the error corresponds to the standard deviation. 
\label{tab:params_sys_PCM}
\end{threeparttable}
\end{table*}

\subsection{SCM versus PCM}

\begin{figure}
\centering
\includegraphics[width=1.0\columnwidth]{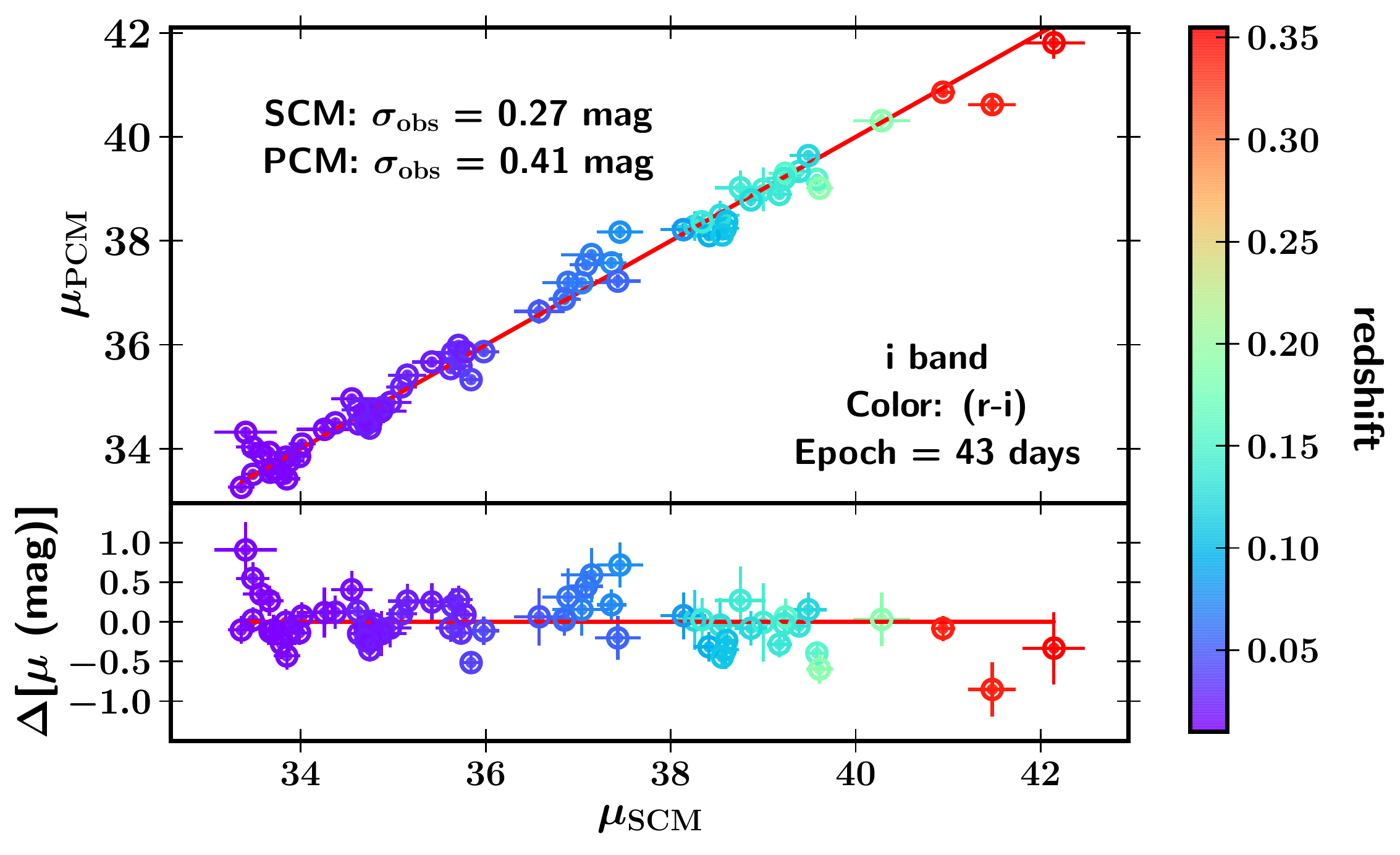}
\caption{Comparison between the distance moduli measured from the SCM and those determined from the PCM. The residuals are plotted in the bottom panel. The red solid line represents a slope of unity while the colour bar on the right side represents
the different redshifts. We present the observed dispersion ($\sigma_{\rm obs}$) of both methods.}
\label{fig:SCM_PCM}
\end{figure}

In this Section, we compare the intrinsic dispersion and the distance moduli obtained applying the SCM and the PCM. For this purpose, we restrict the PCM sample to the SNe~II in common with those used with the SCM: 70 SNe~II. Figure \ref{fig:SCM_PCM} shows a comparison of the Hubble diagrams obtained with both method. As we can see, the distance moduli derived with the SCM and PCM are almost all consistent with a median absolute difference of 0.15\,mag, much lower than the intrinsic dispersion of both methods ($\sim 0.3$ and $\sim 0.4$\,mag). Though the distance moduli are similar, the intrinsic dispersion is different. The SCM is a better method to standardise the SNe~II than the PCM with a difference of $\sim 0.1$\,mag, or $\sim 5$\% in distance. However, spectroscopic follow-up observations for all events discovered by the next generation of surveys will be impossible, and more work should be done to try to improve a photometric method as for example developing a new SN~II template for SN~II light-curve fitting.

\section{Conclusions}

Using the DES-SN combined with four other surveys (CSP-I, SDSS-II, SNLS, and HSC), we perform the most complete SN~II cosmology analysis and construct the two largest Hubble diagrams with SNe~II in the Hubble flow. First, using the SCM at 43\,d after the explosion -- epoch which minimises the intrinsic dispersion and maximises the number of objects -- and 70 SNe~II we find an intrinsic dispersion in the Hubble diagram of 0.27\,mag which is consistent with previous studies. We derive cosmological parameters ($\Omega_{m} = 0.35^{+0.33}_{-0.23}$) consistent with the $\Lambda$CDM model and the accelerated expansion of the Universe. We demonstrate that the colour term does not improve the SN~II standardisation and solely the expansion velocity correction is enough. This would be an asset as only one photometric band and one spectrum are necessary to calibrate the SN~II. This leaves room for the possibility of a new correlation which will help to improve the standardisation.

For the first time in SN~II cosmology, a SN~II distance modulus bias simulation using SNANA is performed and we show that the best-fitting parameters are not affected. Second, to take advantage of the next generation of surveys and their thousands of thousands SN~II discoveries, we apply a purely photometric method (PCM). We construct a Hubble diagram with a redshift range up to $\sim 0.5$ and an observed scatter of 0.39\,mag, or 17--18\% in distances. Both methods demonstrate a promising future for SNe~II as distance indicators and their utility at low-$z$ to derive H$_0$. However, we address the important needs for building a survey mainly dedicated to SN~II cosmology as the majority of the current surveys were concentrated on SN~Ia cosmology (e.g., noisy spectra). Additionally, future work should focus on building a SN~II template to perform K-corrections and to develop a SN~II light-curve fitter. Currently, SNe~II are not competitive with SN~Ia in term of precision, but with these improvements, we will have the real capacity to compare them with the SNe~Ia and see if they can or cannot play a key role in cosmology.

\section*{Acknowledgements}
The anonymous referee is thanked for their thorough reading of the manuscript, which helped clarify and improve the paper. Support for A.V.F.'s supernova research group at U.C. Berkeley has been provided by the NSF through grant AST-1211916, the TABASGO Foundation, Gary and Cynthia Bengier (T.d.J. is a Bengier Postdoctoral Fellow), the Christopher R. Redlich Fund, the Sylvia and Jim Katzman Foundation, and the Miller Institute for Basic Research in Science (U.C. Berkeley). L.G. was funded by the European Union's Horizon 2020 research and innovation programme under the Marie Sk\l{}odowska-Curie grant agreement No. 839090. This work has been partially supported by the Spanish grant PGC2018-095317-B-C21 within the European Funds for Regional Development (FEDER). CPG acknowledges support from EU/FP7-ERCgrant No. [615929]. The work of the CSP-I has been supported by the U.S. NSF under grants AST-0306969, AST-0607438, and AST-1008343.

This paper is based in part on data collected at the Subaru Telescope and retrieved from the HSC data archive system, which is operated by the Subaru Telescope and Astronomy Data Center at the National Astronomical Observatory of Japan (NAOJ). The Hyper Suprime-Cam (HSC) collaboration includes the astronomical communities of Japan and Taiwan, and Princeton University. The HSC instrumentation and software were developed by the NAOJ, the Kavli Institute for the Physics and Mathematics of the Universe (Kavli IPMU), the University of Tokyo, the High Energy Accelerator Research Organization (KEK), the Academia Sinica Institute for Astronomy and Astrophysics in Taiwan (ASIAA), and Princeton University. Funding was contributed by the FIRST program from the Japanese Cabinet Office, the Ministry of Education, Culture, Sports, Science, and Technology (MEXT), the Japan Society for the Promotion of Science (JSPS), the Japan Science and Technology Agency (JST), the Toray Science Foundation, NAOJ, Kavli IPMU, KEK, ASIAA, and Princeton University. 

The Pan-STARRS1 Surveys (PS1) have been made possible through contributions of the Institute for Astronomy (the University of Hawaii), the Pan-STARRS Project Office, the Max-Planck Society and its participating institutes, the Max Planck Institute for Astronomy, Heidelberg and the Max Planck Institute for Extraterrestrial Physics, Garching, The Johns Hopkins University, Durham University, the University of Edinburgh, Queen's University Belfast, the Harvard-Smithsonian Center for Astrophysics, the Las Cumbres Observatory Global Telescope Network Inc., the National Central University of Taiwan, the Space Telescope Science Institute, the National Aeronautics and Space Administration (NASA) under grant NNX08AR22G issued through the Planetary Science Division of the NASA Science Mission Directorate, the NSF under grant AST-1238877, the University of Maryland, and the Eotvos Lorand University (ELTE). This paper makes use of software developed for the Large Synoptic Survey Telescope (LSST); we thank the LSST Project for making their code available as free software at http://dm.lsst.org. 

Some of the data presented herein were obtained at the W. M. Keck Observatory, which is operated as a scientific partnership among the California Institute of Technology, the University of California, and NASA; the observatory was made possible by the generous financial support of the W. M. Keck Foundation. This work is based in part on data produced at the Canadian Astronomy Data Centre as part of the CFHT Legacy Survey, a collaborative project of the National Research Council of Canada and the French Centre National de la Recherche Scientifique. This research is based in part on observations obtained at the Gemini Observatory, which is operated by the Association of Universities for Research in Astronomy, Inc., under a cooperative agreement with the NSF on behalf of the Gemini partnership: the NSF, the STFC (United Kingdom), the National Research Council (Canada), CONICYT (Chile), the Australian Research Council (Australia), CNPq (Brazil), and CONICET (Argentina). This research used observations from Gemini program numbers GN-2005A-Q-11, GN-2005B-Q-7, GN-2006A-Q-7, GS-2005A-Q-11, GS-2005B-Q-6, and GS-2008B-Q-56. This research has made use of the NASA/IPAC Extragalactic Database (NED), which is operated by the Jet Propulsion Laboratory, California Institute of Technology, under contract with NASA, and of data provided by the Central Bureau for Astronomical Telegrams.

Funding for the DES Projects has been provided by the U.S. Department of Energy, the U.S. NSF, the Ministry of Science and Education of Spain, the Science and Technology Facilities Council of the United Kingdom, the Higher Education Funding Council for England, the National Center for Supercomputing Applications at the University of Illinois at Urbana-Champaign, the Kavli Institute of Cosmological Physics at the University of Chicago, the Center for Cosmology and Astro-Particle Physics at the Ohio State University, the Mitchell Institute for Fundamental Physics and Astronomy at Texas A\&M University, Financiadora de Estudos e Projetos, Fundacao Carlos Chagas Filho de Amparo \'a Pesquisa do Estado do Rio de Janeiro, Conselho Nacional de Desenvolvimento Cient\'ifico e Tecnol\'ogico and the Minist\'erio da Ci\^encia, Tecnologia e Inovacao, the Deutsche Forschungsgemeinschaft, and the Collaborating Institutions in the DES: Argonne National Laboratory, the University of California at Santa Cruz, the University of Cambridge, Centro de Investigaciones Energ\'eticas, Medioambientales y Tecnol\'ogicas-Madrid, the University of Chicago, University College London, the DES-Brazil Consortium, the University of Edinburgh, the Eidgenossische Technische Hochschule (ETH) Zurich, Fermi National Accelerator Laboratory, the University of Illinois at Urbana-Champaign, the Institut de Ciencies de l'Espai (IEEC/CSIC), the Institut de F\'isica d'Altes Energies, Lawrence Berkeley National Laboratory, the Ludwig-Maximilians Universitat Munchen and the associated Excellence Cluster Universe, the University of Michigan, the National Optical Astronomy Observatory, the University of Nottingham, the Ohio State University, the University of Pennsylvania, the University of Portsmouth, SLAC National Accelerator Laboratory, Stanford University, the University of Sussex, Texas A\&M University, and the OzDES Membership Consortium. The DES data management system is supported by the U.S. National Science Foundation under Grant Numbers AST-1138766 and AST-1536171. The DES participants from Spanish institutions are partially supported by MINECO under grants AYA2015-71825, ESP2015-66861, FPA2015-68048, SEV-2016-0588, SEV-2016-0597, and MDM-2015-0509, some of which include ERDF funds from the European Union. IFAE is partially funded by the CERCA program of the Generalitat de Catalunya. Research leading to these results has received funding from the European Research Council under the European Union's Seventh Framework Program (FP7/2007-2013) including ERC grant agreements 240672, 291329, and 306478.
We acknowledge support from the Brazilian Instituto Nacional de Ci\^encia e Tecnologia (INCT) e-Universe (CNPq grant 465376/2014-2). This research uses resources of the National Energy Research Scientific Computing Center, a U.S. Department of Energy (DOE) Office of Science User Facility supported by the Office of Science of the DOE under Contract DE-AC02-05CH11231. This research used resources of the National Energy Research Scientific Computing Center (NERSC), a U.S. DOE Office of Science User Facility operated under Contract DE-AC02-05CH11231.


\section{Affiliations}
$^{1}$ Department of Astronomy, University of California, 501 Campbell Hall, Berkeley, CA 94720-3411, USA.\\
$^{2}$ Bengier Postdoctoral Fellow.\\
$^{3}$ Departamento de F\'isica Te\'orica y del Cosmos, Universidad de Granada, E-18071 Granada, Spain.\\
$^{4}$ CENTRA, Instituto Superior T\'ecnico, Universidade de Lisboa, Av. Rovisco Pais 1, 1049-001 Lisboa, Portugal.\\
$^{5}$ Department of Astronomy and Astrophysics, University of Chicago, Chicago, IL 60637, USA.\\
$^{6}$ Kavli Institute for Cosmological Physics, University of Chicago, Chicago, IL 60637, USA.\\
$^{7}$ Miller Senior Fellow, Miller Institute for Basic Research in Science, University of California, Berkeley, CA 94720, USA.\\
$^{8}$ Millennium Institute of Astrophysics (MAS), Nuncio Monse\~{n}or S\'{o}tero Sanz 100, Providencia, Santiago, Chile.\\
$^{9}$ Center for Mathematical Modeling, Universidad de Chile, Santiago, Chile.\\
$^{10}$ Departamento de Astronom\'{i}a -- Universidad de Chile, Camino el Observatorio 1515, Santiago, Chile.\\
$^{11}$ George P. and Cynthia Woods Mitchell Institute for Fundamental Physics \& Astronomy, Mitchell Physics Building, Texas A. \& M. University, USA.\\
$^{12}$ School of Mathematics and Physics, University of Queensland, Brisbane, QLD 4072, Australia.\\
$^{13}$ School of Physics and Astronomy, University of Southampton, Southampton, SO17 1BJ, UK.\\
$^{14}$ School of Physics \& Astronomy, Cardiff University, Queens Buildings, The Parade, Cardiff, CF24 3AA, UK.\\
$^{15}$ Sydney Institute for Astronomy, School of Physics, A28, The University of Sydney, NSW 2006, Australia.\\
$^{16}$ Universit\'{e} Clermont Auvergne, CNRS/IN2P3, LPC, F-63000 Clermont-Ferrand, France.\\
$^{17}$ Department of Physics, Duke University Durham, NC 27708, USA.\\
$^{18}$ Department of Physics and Astronomy, University of Pennsylvania, Philadelphia, PA 19104, USA.\\
$^{19}$ INAF, Astrophysical Observatory of Turin, I-10025 Pino Torinese, Italy.\\
$^{20}$ Santa Cruz Institute for Particle Physics, Santa Cruz, CA 95064, USA.\\
$^{21}$ Centre for Astrophysics \& Supercomputing, Swinburne University of Technology, Victoria 3122, Australia.\\
$^{22}$ Institute of Cosmology and Gravitation, University of Portsmouth, Portsmouth, PO1 3FX, UK.\\
$^{23}$ The Research School of Astronomy and Astrophysics, Australian National University, ACT 2601, Australia.\\
$^{24}$ Cerro Tololo Inter-American Observatory, National Optical Astronomy Observatory, Casilla 603, La Serena, Chile.\\
$^{25}$ Departamento de F\'isica Matem\'atica, Instituto de F\'isica, Universidade de S\~ao Paulo, CP 66318, S\~ao Paulo, SP, 05314-970, Brazil.\\
$^{26}$ Laborat\'orio Interinstitucional de e-Astronomia - LIneA, Rua Gal. Jos\'e Cristino 77, Rio de Janeiro, RJ - 20921-400, Brazil.\\
$^{27}$ Fermi National Accelerator Laboratory, P. O. Box 500, Batavia, IL 60510, USA.\\
$^{28}$ Instituto de Fisica Teorica UAM/CSIC, Universidad Autonoma de Madrid, 28049 Madrid, Spain.\\
$^{29}$ CNRS, UMR 7095, Institut d'Astrophysique de Paris, F-75014, Paris, France.\\
$^{30}$ Sorbonne Universit\'es, UPMC Univ Paris 06, UMR 7095, Institut d'Astrophysique de Paris, F-75014, Paris, France.\\
$^{31}$ Department of Physics and Astronomy, Pevensey Building, University of Sussex, Brighton, BN1 9QH, UK.\\
$^{32}$ Department of Physics \& Astronomy, University College London, Gower Street, London, WC1E 6BT, UK.\\
$^{33}$ Kavli Institute for Particle Astrophysics \& Cosmology, P. O. Box 2450, Stanford University, Stanford, CA 94305, USA.\\
$^{34}$ SLAC National Accelerator Laboratory, Menlo Park, CA 94025, USA.\\
$^{35}$ Centro de Investigaciones Energ\'eticas, Medioambientales y Tecnol\'ogicas (CIEMAT), Madrid, Spain.\\
$^{36}$ Department of Astronomy, University of Illinois at Urbana-Champaign, 1002 W. Green Street, Urbana, IL 61801, USA.\\
$^{37}$ National Center for Supercomputing Applications, 1205 West Clark St., Urbana, IL 61801, USA.\\
$^{38}$ Institut de F\'{\i}sica d'Altes Energies (IFAE), The Barcelona Institute of Science and Technology, Campus UAB, 08193 Bellaterra (Barcelona) Spain.\\
$^{39}$ Institute for Fundamental Physics of the Universe, Via Beirut 2, 34014 Trieste, Italy.\\
$^{40}$ INAF-Osservatorio Astronomico di Trieste, via G. B. Tiepolo 11, Italy.\\
$^{41}$ Institut d'Estudis Espacials de Catalunya (IEEC), 08034 Barcelona, Spain.\\
$^{42}$ Institute of Space Sciences (ICE, CSIC), Campus UAB, Carrer de Can Magrans, s/n, 08193 Barcelona, Spain.\\
$^{43}$ Observat\'orio Nacional, Rua Gal. Jos\'e Cristino 77, Rio de Janeiro, RJ-20921-400, Brazil.\\
$^{44}$ Department of Physics, IIT Hyderabad, Kandi, Telangana 502285, India.\\
$^{45}$ Department of Astronomy/Steward Observatory, University of Arizona, 933 North Cherry Avenue, Tucson, AZ 85721-0065, USA.\\
$^{46}$ Jet Propulsion Laboratory, California Institute of Technology, 4800 Oak Grove Dr., Pasadena, CA 91109, USA.\\
$^{47}$ Department of Physics, Stanford University, 382 Via Pueblo Mall, Stanford, CA 94305, USA.\\
$^{48}$ Department of Physics, ETH Zurich, Wolfgang-Pauli-Strasse 16, CH-8093 Zurich, Switzerland.\\
$^{49}$ Center for Cosmology and Astro-Particle Physics, The Ohio State University, Columbus, OH 43210, USA.\\
$^{50}$ Department of Physics, The Ohio State University, Columbus, OH 43210, USA.\\
$^{51}$ Center for Astrophysics $\vert$ Harvard \& Smithsonian, 60 Garden Street, Cambridge, MA 02138, USA.\\
$^{52}$ Australian Astronomical Optics, Macquarie University, North Ryde, NSW 2113, Australia.\\
$^{53}$ Lowell Observatory, 1400 Mars Hill Rd, Flagstaff, AZ 86001, USA.\\
$^{54}$ Department of Astrophysical Sciences, Princeton University, Peyton Hall, Princeton, NJ 08544, USA.\\
$^{55}$ Observatories of the Carnegie Institution for Science, 813 Santa Barbara St., Pasadena, CA 91101, USA.\\
$^{56}$ Instituci\'o Catalana de Recerca i Estudis Avan\c{c}ats, E-08010 Barcelona, Spain.\\
$^{57}$ Department of Physics, University of Michigan, Ann Arbor, MI 48109, USA.\\
$^{58}$ Brandeis University, Physics Department, 415 South Street, Waltham MA 02453, USA.\\
$^{59}$ Computer Science and Mathematics Division, Oak Ridge National Laboratory, Oak Ridge, TN 37831, USA.\\
$^{60}$ Max Planck Institute for Extraterrestrial Physics, Giessenbachstrasse, 85748 Garching, Germany.\\
$^{61}$ Universit\"ats-Sternwarte, Fakult\"at f\"ur Physik, Ludwig-Maximilians Universit\"at M\"unchen, Scheinerstr. 1, 81679 M\"unchen, Germany.\\

\appendix
\onecolumn
\section{}\label{AppendixA}
List of the 56 spectroscopically classified SNe~II from the DES-SN survey. For each SN, we indicate whether it did (``SCM'') or did not pass the cut. The SNe that failed are marked with ``PHOT'' (not enough data), ``EXP'' (no explosion date), ``SPEC'' (no spectrum), ``P-Cygni'' (no clear P-Cygni profile), and ``LC'' (unusual light curves). 

\begin{table*}
\caption{Spectroscopically classified SNe~II.}
\begin{threeparttable}
\begin{tabular}{cccc}
\hline
SN & $z_{\rm CMB}$ & Cut & Comments\\
\hline
\hline

DES13C2jtx	&0.2234	&P-Cygni\\
DES13C3ui	&0.0663	&EXP\\
DES13X3fca	&0.0951	&SCM\\
DES14C3aol	&0.0764	&SCM\\
DES14C3nm	&0.3096	&EXP\\
DES14C3rhw	&0.3412	&SCM\\
DES14C3tsg	&0.2096	&PHOT\\
DES14E2ar	&0.0761	&EXP\\
DES14X1qt	&0.1380	&EXP\\
DES14X2cy	&0.2316	&EXP\\
DES14X3ili	&0.1412	&P-Cygni\\
DES15C1okz	&0.0696	&PHOT\\
DES15C1pkx	&0.1564	&PHOT\\
DES15C2eaz	&0.0612	&SCM\\
DES15C2lna	&0.0652	&P-Cygni\\
DES15C2lpp	&0.1806	&P-Cygni &Classified as ``SN~II?'': ATel \#8367\\
DES15C2npz	&0.1221	&SCM\\
DES15C3bj	&0.2870	&EXP &Classified as ``SN~II?'': ATel \#8367\\
DES15E1iuh	&0.1045	&SCM\\
DES15E2ni	&0.2253	&EXP\\
DES15S1by	&0.1283	&EXP\\
DES15S1cj	&0.1661	&EXP &Classified as ``SN~II?'': ATel \#8367\\
DES15S1lrp	&0.2223	&P-Cygni &Classified as ``SN~II?'': ATel \#8658\\
DES15S2eaq	&0.0672	&SCM\\
DES15X1lzp	&0.0792	&SPEC\\	
DES15X2mku	&0.0807	&SCM\\
DES15X3mpq	&0.1872	&P-Cygni\\
DES15X3nad	&0.0998	&P-Cygni\\
DES16C2cbv	&0.1087	&SCM\\
DES16C3at	&0.2171	&EXP\\
DES16E1ah	&0.1480 &EXP\\
DES16E1bkh	&0.1155	&P-Cygni\\
DES16S1gn	&0.1899	&SCM\\
DES16X1ey	&0.0752	&EXP\\
DES16X2bkr	&0.1577	&SCM\\
DES16X3cpl	&0.2042	&P-Cygni &Classified as ``SN~II?'': ATel \#9742\\
DES16X3dvb	&0.3292 &LC &Slow rise and over luminous\\
DES16X3jj	&0.2369	&EXP &Classified as ``SN~II?'': ATel \#9504\\
DES16X3km	&0.0538	&EXP\\
DES17C2pf	&0.1358	&EXP\\
DES17C3aye	&0.1577	&P-Cygni\\
DES17C3bei	&0.1030	&P-Cygni\\
DES17C3de	&0.1070	&EXP\\
DES17C3dw	&0.1632	&EXP\\
DES17E2bhj	&0.1857	&P-Cygni &Classified as ``SN~II?'': ATEL\#11146\\
DES17E2cc	&0.1478	&EXP\\
DES17E2ci	&0.1259	&EXP\\
DES17S1bxt	&0.3550	&PHOT\\
DES17S1lu	&0.0832	&EXP\\
DES17S2oo	&0.2243	&EXP\\
DES17X1aow	&0.1379	&SCM\\
DES17X1axb	&0.1377	&SCM\\
DES17X1gd	&0.1881	&EXP &Classified as ``SN~II?'': ATel \#11146\\
DES17X2ls	&0.2509	&EXP &Classified as ``SN~II?'': ATel \#11147\\
DES17X3bd	&0.1406	&EXP &Classified as ``SN~II?'': ATel \#10759\\
DES17X3dub	&0.1210	&SCM\\
\hline
\hline
\end{tabular}
Notes --- Column 1, SN name; Column 2, heliocentric redshift; Column 3, sample cut: ``SCM'' (useful for cosmology), ``PHOT'' (not enough data), ``EXP'' (no explosion date), ``SPEC'' (no spectrum), ``P-Cygni'' (no clear P-Cygni profile), ``LC'' (unusual light curves); Column 4, comments.
\label{tab:sample_tot}
\end{threeparttable}
\end{table*}

\normalsize
\onecolumn

\section{}\label{AppendixB}
All of the observed light curves for all SNe~II with spectroscopic confirmation discovered by DES-SN and not included in the SCM sample are displayed in this Appendix. The spectra of the SNe~II not used in our SCM sample owing to a lack of clear P-Cygni profiles are shown.
\newpage

\begin{figure*}
\includegraphics[width=1.0\textwidth]{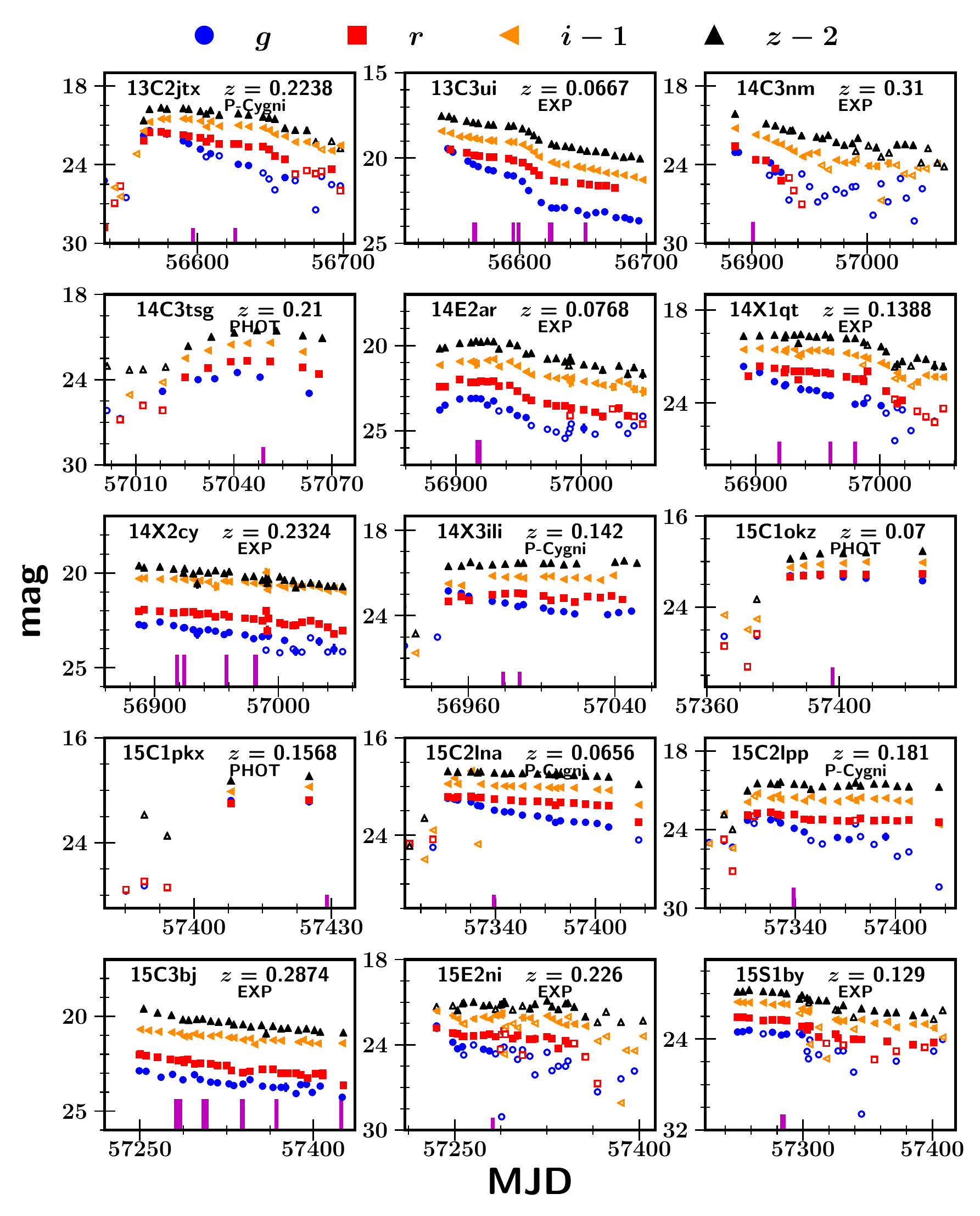}\\

\caption{Observed light curves of all SNe~II discovered by DES-SN with spectroscopic confirmation. Blue circles are $g$-band magnitudes, red squares are $r$, orange left triangles are $i-1$, and black top triangles are $z-2$. Empty symbols represent real points with flux/err $< 3$.} The abscissa is the Modified Julian Date (MJD). In each panel, the IAU name and the redshift are given in the upper right. SNe that failed the cut are marked with ``PHOT'' (not enough data), ``EXP'' (no explosion date), ``SPEC'' (no spectrum), ``P-Cygni'' (no clear P-Cygni profile) , and ``LC'' (unusual light curves). The vertical magenta lines indicate the epochs of optical spectroscopy.
\end{figure*}

\begin{figure*}
\addtocounter{figure}{-1}

\includegraphics[width=1.0\textwidth]{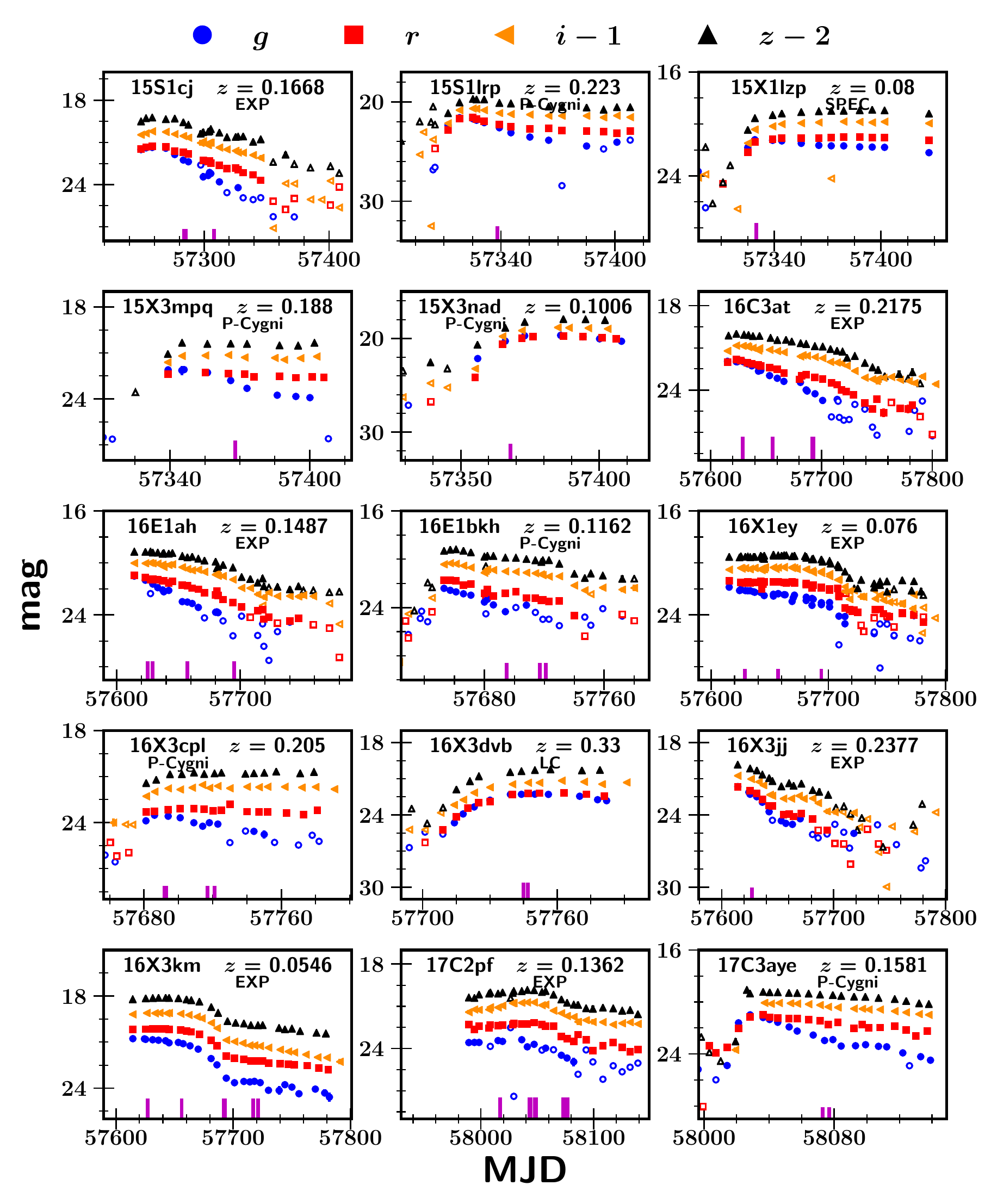}\\

\caption{Observed light curves of all SNe~II discovered by DES-SN with spectroscopic confirmation. Blue circles are $g$-band magnitudes, red squares are $r$, orange left triangles are $i-1$, and black top triangles are $z-2$. Empty symbols represent real points with flux/err $< 3$.} The abscissa is the Modified Julian Date (MJD). In each panel, the IAU name and the redshift are given in the upper right. SNe that failed the cut are marked with ``PHOT'' (not enough data), ``EXP'' (no explosion date), ``SPEC'' (no spectrum), ``P-Cygni'' (no clear P-Cygni profile) , and ``LC'' (unusual light curves). The vertical magenta lines indicate the epochs of optical spectroscopy.
\end{figure*}

\begin{figure*}
\addtocounter{figure}{-1}
\includegraphics[width=1.0\textwidth]{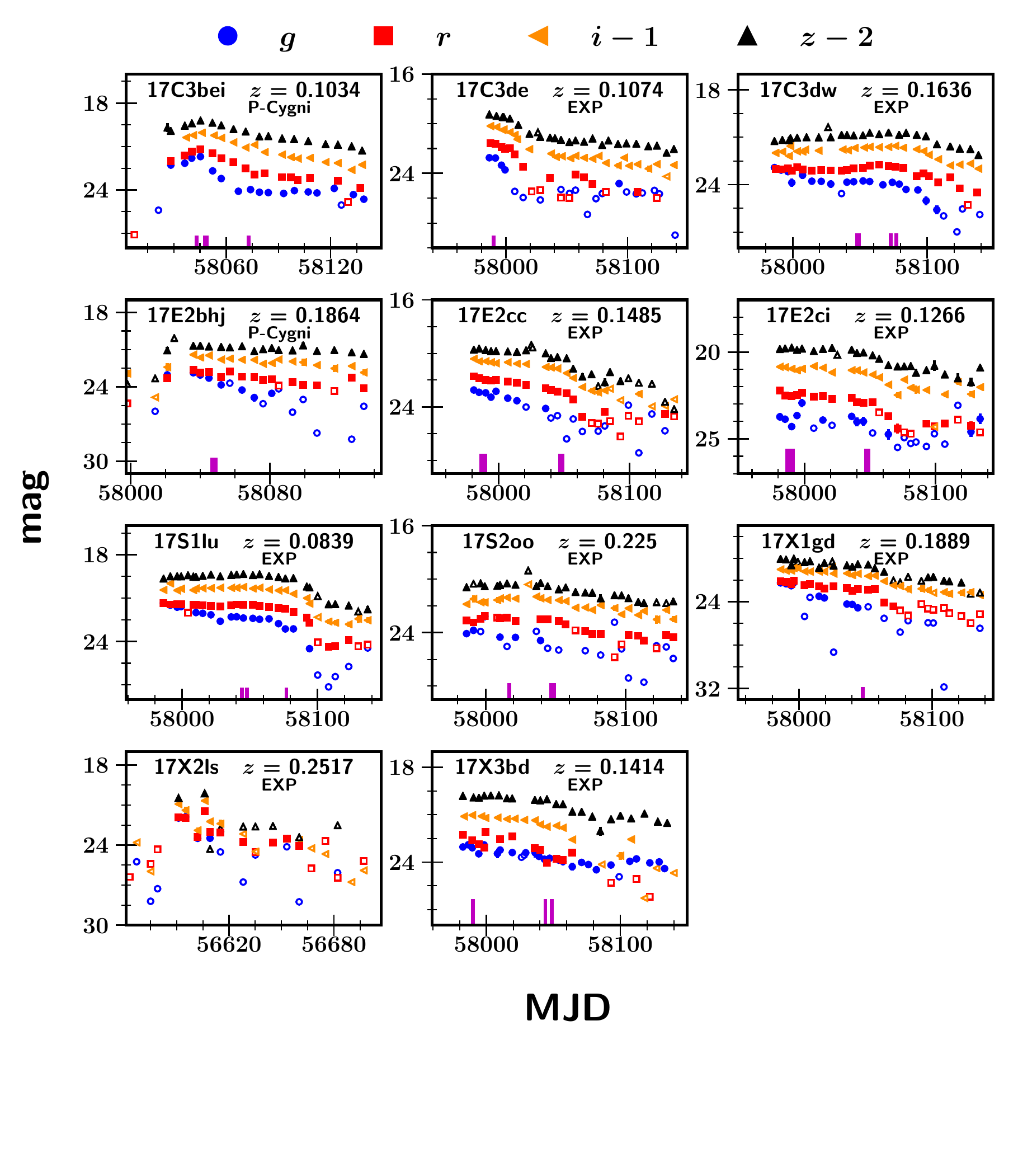}\\
\caption{Observed light curves of all SNe~II discovered by DES-SN with spectroscopic confirmation. Blue circles are $g$-band magnitudes, red squares are $r$, orange left triangles are $i-1$, and black top triangles are $z-2$. Empty symbols represent real points with flux/err $< 3$.} The abscissa is the Modified Julian Date (MJD). In each panel, the IAU name and the redshift are given in the upper right. SNe that failed the cut are marked with ``PHOT'' (not enough data), ``EXP'' (no explosion date), ``SPEC'' (no spectrum), ``P-Cygni'' (no clear P-Cygni profile) , and ``LC'' (unusual light curves). The vertical magenta lines indicate the epochs of optical spectroscopy.
\end{figure*}

\begin{figure*}
	\includegraphics[scale=0.8]{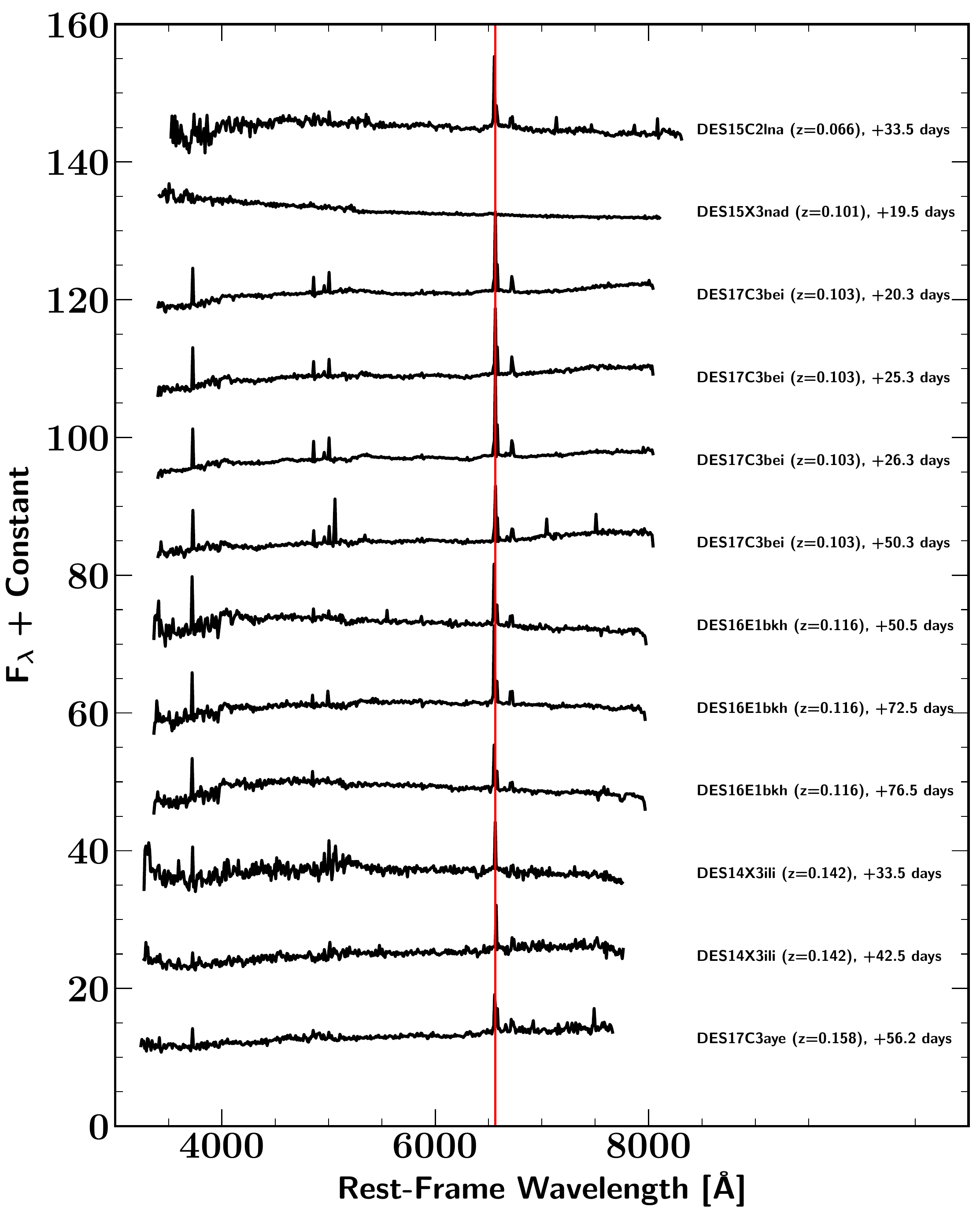}
\caption{Spectra of the 12 SNe~II from the DES-SN sample classified as ``SN~II'' or ``SN~II?'' but not included in the SCM sample because the P-Cygni profile is not clearly seen. The spectra are shown in the rest frame, and the date listed for each SN is the number of days since explosion. The redshift of each SN is labelled. The spectra were binned (10\,\AA). The red vertical line corresponds to H$\alpha$ ($\lambda$6563).}
\end{figure*}

\begin{figure*}
	\addtocounter{figure}{-1}
	\includegraphics[scale=0.8]{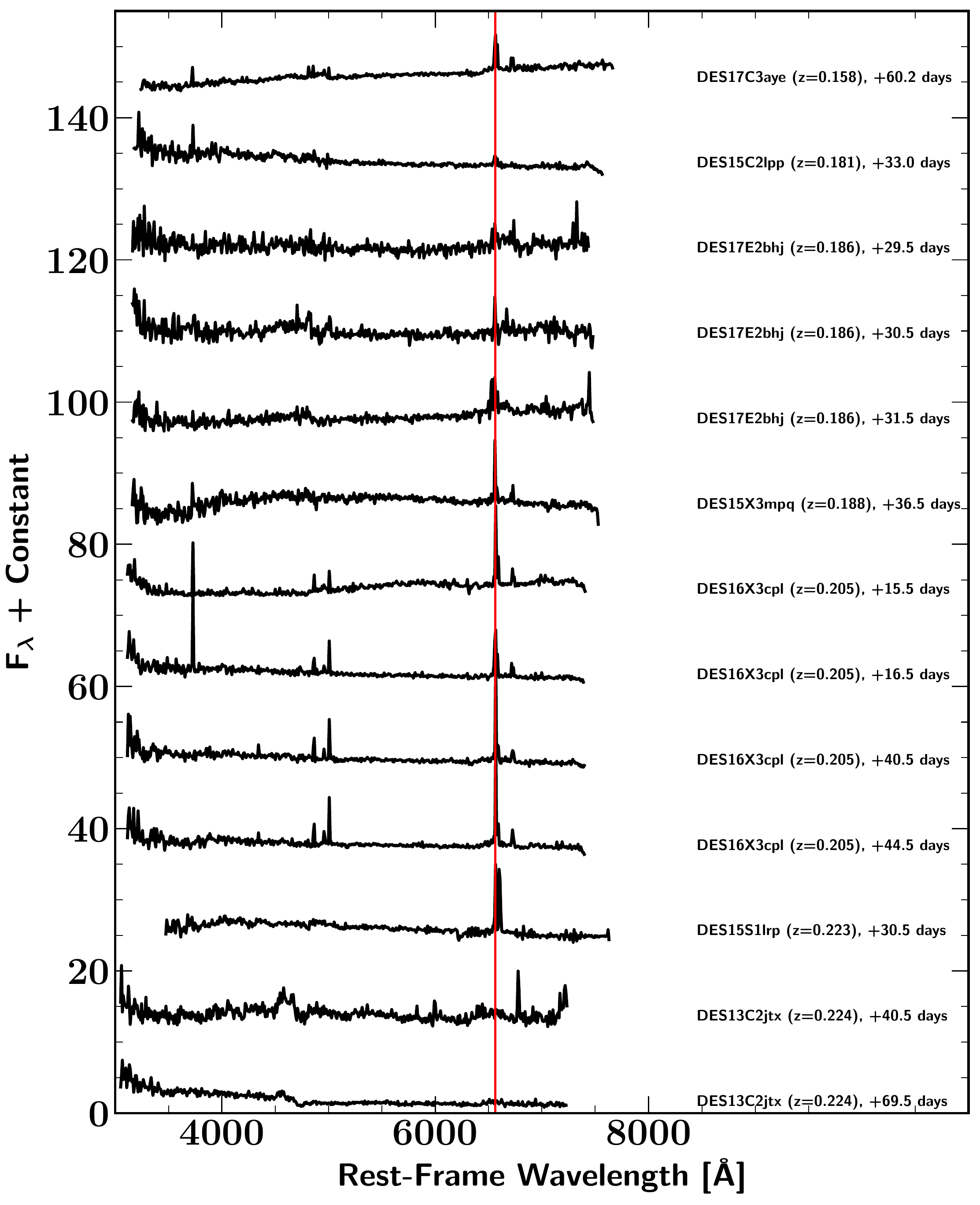}
\caption{\textit{(Cont.)} Spectra of the 12 SNe~II from the DES-SN sample classified as ``SN~II'' or ``SN~II?'' but not included in the SCM sample because the P-Cygni profile is not clearly seen. The spectra are shown in the rest frame, and the date listed for each SN is the number of days since explosion. The redshift of each SN is labelled. The spectra were binned (10\,\AA). The red vertical line corresponds to H$\alpha$ ($\lambda$6563).}
\end{figure*}

\normalsize
\onecolumn
\section{}\label{AppendixC}
All DES-SN photometric and spectroscopic SN~II data are publicly available at \url{https://github.com/tdejaeger}. The spectra are also available at the Weizmann Interactive Supernova Data Repository (\url{WISeREP;https:/wiserep.weizmann.ac.il}).

\begin{table*}
\normalsize
\caption{DES-SN sample photometry.}
\begin{threeparttable}
\begin{tabular}{cccccc}
\hline
SN name & MJD & $g$ & $r$ &$i$ &$z$ \\
 & & mag & mag &mag &mag \\
\hline
\hline
DES13C2jtx &56536.2 &(25.221) $\pm$ (0.692) &(28.770) $\pm$ (3.814) &$\cdots$ &$\cdots$ \\
DES13C2jtx &56543.2 &$\cdots$ &(26.941) $\pm$ (1.420) &(26.738) $\pm$ (1.595) &$\cdots$ \\
DES13C2jtx &56547.3 &$\cdots$ &(25.638) $\pm$ (0.806) &(27.438) $\pm$ (2.187) &$\cdots$ \\
DES13C2jtx &56551.2 &(26.517) $\pm$ (2.441) &$\cdots$ &$\cdots$ &$\cdots$ \\
DES13C2jtx &56558.2 &$\cdots$ &$\cdots$ &(24.188) $\pm$ (0.362) &$\cdots$ \\
DES13C2jtx &56563.2 &21.775 $\pm$ 0.023 &22.155 $\pm$ 0.043 &22.445 $\pm$ 0.066 &22.653 $\pm$ 0.111 \\
DES13C2jtx &56567.2 &21.406 $\pm$ 0.016 &21.554 $\pm$ 0.023 &21.747 $\pm$ 0.034 &21.786 $\pm$ 0.041 \\
DES13C2jtx &56575.2 &21.484 $\pm$ 0.015 &21.488 $\pm$ 0.018 &21.517 $\pm$ 0.023 &21.674 $\pm$ 0.032 \\
DES13C2jtx &56579.2 &21.683 $\pm$ 0.039 &21.641 $\pm$ 0.037 &21.505 $\pm$ 0.034 &21.774 $\pm$ 0.048 \\
DES13C2jtx &56590.3 &22.194 $\pm$ 0.077 &21.733 $\pm$ 0.070 &21.505 $\pm$ 0.061 &21.729 $\pm$ 0.075 \\
DES13C2jtx &56594.1 &22.406 $\pm$ 0.066 &21.834 $\pm$ 0.057 &21.544 $\pm$ 0.057 &21.781 $\pm$ 0.092 \\
DES13C2jtx &56602.1 &22.829 $\pm$ 0.042 &21.968 $\pm$ 0.027 &21.679 $\pm$ 0.028 &21.944 $\pm$ 0.045 \\
DES13C2jtx &56606.1 &(23.407) $\pm$ (0.360) &22.262 $\pm$ 0.146 &22.115 $\pm$ 0.155 &22.136 $\pm$ 0.192 \\
DES13C2jtx &56609.1 &23.173 $\pm$ 0.184 &22.004 $\pm$ 0.057 &21.699 $\pm$ 0.042 &21.878 $\pm$ 0.057 \\
DES13C2jtx &56615.0 &(23.332) $\pm$ (0.478) &22.436 $\pm$ 0.124 &22.072 $\pm$ 0.116 &22.228 $\pm$ 0.257 \\
DES13C2jtx &56625.2 &$\cdots$ &22.442 $\pm$ 0.038 &$\cdots$ &$\cdots$ \\
DES13C2jtx &56628.1 &23.980 $\pm$ 0.151 &22.437 $\pm$ 0.051 &21.994 $\pm$ 0.045 &22.117 $\pm$ 0.064 \\
DES13C2jtx &56635.1 &24.069 $\pm$ 0.193 &22.654 $\pm$ 0.066 &22.103 $\pm$ 0.055 &22.219 $\pm$ 0.089 \\
DES13C2jtx &56645.1 &(24.638) $\pm$ (0.472) &22.648 $\pm$ 0.090 &22.198 $\pm$ 0.063 &22.368 $\pm$ 0.076 \\
DES13C2jtx &56649.1 &(25.097) $\pm$ (0.479) &22.832 $\pm$ 0.082 &$\cdots$ &$\cdots$ \\
DES13C2jtx &56649.2 &$\cdots$ &$\cdots$ &22.377 $\pm$ 0.072 &22.448 $\pm$ 0.105 \\
	   &        &                        &$\cdots$                        &                   &\\
	   &        &                        &$\cdots$                        &                   &\\
\hline
\end{tabular}
\textit{Notes:} DES-SN SN~II photometry. The values in parentheses are real points with flux/err $< 3$. The table is only a fraction of a much larger table which covers each epoch of photometry for in SN. The full table is available in the online version of this article. 
\label{tab:sample_photo}
\end{threeparttable}
\end{table*}

\begin{table*}
\normalsize
\caption{Journal of spectroscopic observations of SN~II DES-SN sample.}
\begin{threeparttable}
\begin{tabular}{ccccccc}
\hline
SN name & Date & MJD & Epoch &Telescope &Instrument &range \\
 & & UT &Days & & &\AA \\
\hline
\hline
DES13C2jtx &2013-11-01 &56597.0 &40.5 &AAT &AAOmega/2dF &3733--8855 \\
DES13C2jtx &2013-11-30 &56626.0 &69.5 &AAT &AAOmega/2dF &3733--8846 \\
DES13X3fca &2013-10-30 &56595.0 &53.5 &AAT &AAOmega/2dF &3733--8855 \\
DES13X3fca &2013-11-03 &56599.0 &57.5 &AAT &AAOmega/2dF &3733--8855 \\
DES14C3aol &2014-10-29 &56959.0 &65.3 &AAT &AAOmega/2dF &3734--8854 \\
DES14C3rhw &2015-1-28 &57050.0 &47.5 &VLT &X-Shooter &3400--10000 \\
DES14X3ili &2014-11-18 &56979.0 &33.5 &AAT &AAOmega/2dF &3734--8856 \\
DES14X3ili &2014-11-27 &56988.0 &42.5 &AAT &AAOmega/2dF &3734--8868 \\
DES15C2eaz &2015-11-13 &57339.0 &68.5 &AAT &AAOmega/2dF &3754--8922 \\
DES15C2lna &2015-11-13 &57339.0 &33.5 &AAT &AAOmega/2dF &3754--8859 \\
DES15C2lpp &2015-11-13 &57339.0 &33.0 &AAT &AAOmega/2dF &3754--8936 \\
DES15C2npz &2016-01-11 &57398.0 &38.5 &Magellan &LDSS3 &4251--8669 \\
DES15E1iuh &2015-10-13 &57308.0 &27.4 &AAT &AAOmega/2dF &3754--8945 \\
DES15E1iuh &2015-11-14 &57340.0 &59.4 &AAT &AAOmega/2dF &3754--8945 \\
DES15S1lrp &2015-11-12 &57338.0 &30.5 &Magellan &LDSS3 &4250--9330 \\
DES15S2eaq &2015-12-03 &57359.0 &89.5 &AAT &AAOmega/2dF &3757--8920 \\
DES15X2mku &2015-12-14 &57370.0 &41.5 &AAT &AAOmega/2dF &3753--8944 \\
DES15X3mpq &2015-12-12 &57368.0 &36.5 &AAT &AAOmega/2dF &3753--8944 \\
DES15X3nad &2015-12-12 &57368.0 &19.5 &AAT &AAOmega/2dF &3757--8920 \\
DES16C2cbv &2016-11-03 &57695.0 &38.5 &AAT &AAOmega/2dF &3905--8945 \\
DES16C2cbv &2016-11-28 &57720.0 &63.5 &AAT &AAOmega/2dF &3753--8944 \\
DES16C2cbv &2016-11-29 &57721.0 &64.5 &AAT &AAOmega/2dF &3753--8944 \\
DES16E1bkh &2016-11-03 &57695.0 &50.5 &AAT &AAOmega/2dF &3753--8906 \\
DES16E1bkh &2016-11-25 &57717.0 &72.5 &AAT &AAOmega/2dF &3753--8898 \\
DES16E1bkh &2016-11-29 &57721.0 &76.5 &AAT &AAOmega/2dF &3753--8898 \\
DES16S1gn &2016-10-05 &57666.0 &46.5 &AAT &AAOmega/2dF &3753--8945 \\
DES16X2bkr &2016-11-03 &57695.0 &48.5 &AAT &AAOmega/2dF &3753--8945 \\
DES16X3cpl &2016-10-31 &57692.0 &15.5 &AAT &AAOmega/2dF &3753--8931 \\
DES16X3cpl &2016-11-01 &57693.0 &16.5 &AAT &AAOmega/2dF &3753--8912 \\
DES16X3cpl &2016-11-25 &57717.0 &40.5 &AAT &AAOmega/2dF &3753--8922 \\
DES16X3cpl &2016-11-29 &57721.0 &44.5 &AAT &AAOmega/2dF &3753--8923 \\
DES17C3aye &2017-11-16 &58073.0 &56.2 &AAT &AAOmega/2dF &3753--8877 \\
DES17C3aye &2017-11-20 &58077.0 &60.2 &AAT &AAOmega/2dF &3753--8881 \\
DES17C3bei &2017-10-17 &58043.0 &20.3 &AAT &AAOmega/2dF &3752--8875 \\
DES17C3bei &2017-10-22 &58048.0 &25.3 &AAT &AAOmega/2dF &3752--8875 \\
DES17C3bei &2017-10-23 &58049.0 &26.3 &AAT &AAOmega/2dF &3752--8875 \\
DES17C3bei &2017-11-16 &58073.0 &50.3 &AAT &AAOmega/2dF &3753--8874 \\
DES17E2bhj &2017-10-21 &58047.0 &29.5 &AAT &AAOmega/2dF &3752--8825 \\
DES17E2bhj &2017-10-22 &58048.0 &30.5 &AAT &AAOmega/2dF &3752--8875 \\
DES17E2bhj &2017-10-23 &58049.0 &31.5 &AAT &AAOmega/2dF &3752--8875 \\
DES17S1bxt &2017-11-16 &58073.0 &35.5 &Keck-II &Deimos &4600--9300 \\
DES17X1aow &2017-11-19 &58076.0 &71.5 &AAT &AAOmega/2dF &3753--8874 \\
DES17X1axb &2017-10-22 &58048.0 &31.5 &AAT &AAOmega/2dF &3752--8875 \\
DES17X1axb &2017-11-19 &58076.0 &59.5 &AAT &AAOmega/2dF &3753--8874 \\
DES17X3dub &2017-11-16 &58073.0 &14.5 &AAT &AAOmega/2dF &3753--8874 \\
\hline
\end{tabular}
\textit{Note:} Column 1: SN Name. Column 2: UT observation date. Column 3: epoch after explosion in days. Column 4 and 5: the telescope and instrument used to obtain the spectrum. Columns 6: wavelength range (\AA). AAT: Anglo-Australian 3.9\,m telescope at the Siding Spring Observatory in Australia, VLT: 8.2\,m Unit Telescope 2 of the Very Large Telescope at the Paranal Observatory in Chile, Keck-II: 10\,m Keck-II telescope on the Mauna Kea Observatory in Hawaii, and Magellan: 6.5\,m Magellan Telescopes at Las Campanas Observatory in Chile.
\label{tab:sample_spec}
\end{threeparttable}
\end{table*}

\newpage
\onecolumn
\section{}\label{AppendixD}

In Table \ref{SN_sample}, the relevant information for all SNe~II used in the Hubble diagram is displayed. The first column gives the SN name, followed (in Column 2) by its reddening owing to dust in our Galaxy \citep{schlafly11}. In Column 3, we list the host-galaxy velocity in the CMB frame using the CMB dipole model presented by \citet{fixsen96}. The explosion epoch is given in Column 4. In Column 5, the magnitude in the $i$ band at epoch 43\,d post-explosion is listed, followed by the $r-i$ colour at the same epoch in Column 6. Column 7 gives the plateau slope $s_{2}$ while Column 8 the H$\beta$ velocity at epoch 43\,d. In Columns 9 and 10 we respectively present the distance modulus measured using SCM and the PCM. Finally, in Column 11 we give the survey from which the SN~II originates.

\scriptsize
\setlength\LTleft{-1.0cm}
\begin{longtable}{ccccccccccc}
\caption{The supernova sample.}\\
\hline
\hline 
\multicolumn{1}{c}{SN} & \multicolumn{1}{c}{$A_{V,G}$} & \multicolumn{1}{c}{$z_{\rm CMB}$} & \multicolumn{1}{c}{Explosion date} & \multicolumn{1}{c}{$m_{i}$} & \multicolumn{1}{c}{$r-i$} & \multicolumn{1}{c}{$s_{\rm 2}$} & \multicolumn{1}{c}{$v_{{\rm H}\beta}$} & \multicolumn{1}{c}{$\mu_{\rm SCM}$} & \multicolumn{1}{c}{$\mu_{\rm PCM}$} & \multicolumn{1}{c}{Campaign}\\
\multicolumn{1}{c}{} & \multicolumn{1}{c}{mag} & \multicolumn{1}{c}{} & \multicolumn{1}{c}{MJD} & \multicolumn{1}{c}{mag} & \multicolumn{1}{c}{mag} & \multicolumn{1}{c}{mag (100\,d)$^{-1}$} & \multicolumn{1}{c}{km\,s$^{-1}$} & \multicolumn{1}{c}{mag} & \multicolumn{1}{c}{mag} & \multicolumn{1}{c}{}\\
\hline
\endfirsthead

\hline
\hline 
\multicolumn{1}{c}{SN} & \multicolumn{1}{c}{$A_{V,G}$} & \multicolumn{1}{c}{$z_{\rm CMB}$} & \multicolumn{1}{c}{Explosion date} & \multicolumn{1}{c}{$m_{i}$} & \multicolumn{1}{c}{$r-i$} & \multicolumn{1}{c}{$s_{\rm 2}$} & \multicolumn{1}{c}{$v_{{\rm H}\beta}$} & \multicolumn{1}{c}{$\mu_{\rm SCM}$} & \multicolumn{1}{c}{$\mu_{\rm PCM}$} & \multicolumn{1}{c}{Campaign}\\
\multicolumn{1}{c}{} & \multicolumn{1}{c}{mag} & \multicolumn{1}{c}{} & \multicolumn{1}{c}{MJD} & \multicolumn{1}{c}{mag} & \multicolumn{1}{c}{mag} & \multicolumn{1}{c}{mag (100\,d)$^{-1}$} & \multicolumn{1}{c}{km\,s$^{-1}$} & \multicolumn{1}{c}{mag} & \multicolumn{1}{c}{mag} & \multicolumn{1}{c}{}\\
\hline
\endhead

\hline 
\endlastfoot
SN2004er &0.070 &0.014 (0.0005) &53271.8 (2.0) &16.72 (0.01) &0.191 (0.011) &0.41 (0.01) &7567 (545) &33.85 (0.15) &33.42 (0.09) &CSP-I \\
SN2004fb &0.173 &0.021 (0.0005) &53258.6 (7.0) &18.08 (0.03) &0.024 (0.030) &0.48 (0.04) &6065 (769) &34.97 (0.22) &34.88 (0.08) &CSP-I \\
SN2005J &0.075 &0.015 (0.0005) &53379.8 (7.0) &16.97 (0.01) &$-$0.057 (0.009) &0.57 (0.01) &6324 (391) &33.99 (0.13) &33.84 (0.09) & CSP-I \\
SN2005K &0.108 &0.028 (0.0005) &53369.8 (8.0) &18.79 (0.01) &$-$0.112 (0.019) &1.16 (0.05) &5551 (706) &35.63 (0.22) &35.83 (0.06) & CSP-I \\
SN2005Z &0.076 &0.019 (0.0005) &53396.7 (6.0) &17.47 (0.01) &0.022 (0.009) &1.26 (0.02) &7123 (401) &34.63 (0.12) &34.46 (0.07) & CSP-I \\
SN2005an &0.262 &0.012 (0.0005) &53431.8 (6.0) &16.79 (0.01) &$-$0.017 (0.007) &1.70 (0.02) &5858 (419) &33.66 (0.16) &33.91 (0.11) & CSP-I \\
SN2005dk &0.134 &0.016 (0.0005) &53601.5 (6.0) &16.81 (0.01) &$-$0.083 (0.018) &0.77 (0.04) &6420 (530) &33.88 (0.16) &33.74 (0.09) & CSP-I \\
SN2005dt &0.079 &0.025 (0.0005) &53605.6 (9.0) &18.56 (0.01) &0.045 (0.014) &$-$0.20 (0.04) &4898 (463) &35.09 (0.17) &35.19 (0.07) & CSP-I \\
SN2005dw &0.062 &0.017 (0.0005) &53603.6 (9.0) &17.64 (0.01) &0.034 (0.013) &0.69 (0.02) &5559 (562) &34.38 (0.18) &34.49 (0.08) & CSP-I \\
SN2005dx &0.066 &0.026 (0.0005) &53611.8 (7.0) &19.25 (0.03) &0.080 (0.038) &0.28 (0.09) &4728 (398) &35.70 (0.16) &35.98 (0.08) & CSP-I \\
SN2005dz &0.223 &0.019 (0.0005) &53619.5 (4.0) &17.94 (0.01) &$-$0.043 (0.017) &0.37 (0.02) &5735 (498) &34.79 (0.16) &34.75 (0.08) & CSP-I \\
SN2005es &0.228 &0.036 (0.0005) &53638.7 (5.0) &18.94 (0.02) &0.030 (0.025) &0.09 (0.06) &$\cdots$ &$\cdots$ &35.65 (0.06) &CSP-I \\
SN2005gk &0.154 &0.029 (0.0005) &53650.2 (5.0) &18.58 (0.03) &0.151 (0.047) &0.65 (0.04) &$\cdots$ &$\cdots$ &35.36 (0.08) &CSP-I \\
SN2006Y &0.354 &0.033 (0.0005) &53766.5 (4.0) &18.57 (0.02) &$-$0.052 (0.032) &1.13 (0.08) &6912 (444) &35.73 (0.13) &35.57 (0.07) & CSP-I \\
SN2006ai &0.347 &0.015 (0.0005) &53781.6 (5.0) &16.80 (0.01) &$-$0.097 (0.015) &1.11 (0.04) &6296 (438) &33.84 (0.14) &33.82 (0.09) & CSP-I \\
SN2006ee &0.167 &0.014 (0.0005) &53961.9 (4.0) &17.47 (0.01) &0.016 (0.017) &$-$0.58 (0.03) &3484 (340) &33.47 (0.18) &34.03 (0.09) & CSP-I \\
SN2006iw &0.137 &0.030 (0.0005) &54010.7 (1.0) &18.74 (0.01) &0.000 (0.018) &0.36 (0.03) &5934 (557) &35.62 (0.17) &35.53 (0.06) & CSP-I \\
SN2006ms &0.095 &0.014 (0.0005) &54028.5 (6.0) &17.79 (0.01) &$-$0.021 (0.018) &$-$0.57 (0.06) &4543 (817) &34.25 (0.31) &34.37 (0.09) & CSP-I \\
SN2006qr &0.126 &0.016 (0.0005) &54062.8 (7.0) &18.13 (0.01) &0.067 (0.014) &0.63 (0.03) &4606 (536) &34.55 (0.21) &34.95 (0.08) & CSP-I \\
SN2007P &0.111 &0.042 (0.0005) &54118.7 (5.0) &18.96 (0.02) &$-$0.097 (0.020) &0.58 (0.06) &6206 (630) &35.98 (0.18) &35.85 (0.06) & CSP-I \\
SN2007U &0.145 &0.025 (0.0005) &54133.6 (6.0) &17.70 (0.01) &$-$0.078 (0.014) &1.39 (0.04) &6954 (407) &34.89 (0.12) &34.78 (0.06) & CSP-I \\
SN2007W &0.141 &0.010 (0.0005) &54130.8 (7.0) &17.36 (0.01) &$-$0.032 (0.013) &$-$0.70 (0.05) &3862 (387) &33.56 (0.20) &33.92 (0.12) & CSP-I \\
SN2007hm &0.172 &0.024 (0.0005) &54336.6 (6.0) &18.78 (0.01) &$-$0.088 (0.016) &1.34 (0.04) &6161 (332) &35.78 (0.12) &35.85 (0.07) & CSP-I \\
SN2007il &0.129 &0.022 (0.0005) &54349.8 (4.0) &17.79 (0.01) &$-$0.005 (0.015) &$-$0.43 (0.02) &6224 (416) &34.74 (0.13) &34.39 (0.07) & CSP-I \\
SN2007ld &0.255 &0.025 (0.0005) &54376.5 (8.0) &18.28 (0.01) &$-$0.152 (0.016) &1.38 (0.02) &5535 (706) &35.15 (0.22) &35.39 (0.07) & CSP-I \\
SN2007sq &0.567 &0.017 (0.0005) &54422.8 (6.0) &17.83 (0.01) &0.346 (0.010) &0.79 (0.02) &7183 (599) &34.76 (0.16) &34.54 (0.08) & CSP-I \\
SN2008F &0.135 &0.018 (0.0005) &54469.6 (6.0) &18.36 (0.02) &0.092 (0.023) &$-$0.68 (0.06) &$\cdots$ &$\cdots$ &34.85 (0.08) &CSP-I \\
SN2008W &0.267 &0.021 (0.0005) &54483.8 (8.0) &17.91 (0.01) &0.042 (0.023) &0.33 (0.03) &5814 (391) &34.72 (0.14) &34.67 (0.08) & CSP-I \\
SN2008ag &0.229 &0.015 (0.0005) &54477.9 (8.0) &16.83 (0.01) &$-$0.031 (0.015) &$-$0.23 (0.01) &5079 (402) &33.48 (0.16) &33.50 (0.09) & CSP-I \\
SN2008bh &0.060 &0.016 (0.0005) &54543.5 (5.0) &17.87 (0.01) &0.194 (0.011) &0.68 (0.03) &6267 (571) &34.69 (0.17) &34.63 (0.08) & CSP-I \\
SN2008br &0.255 &0.012 (0.0005) &54555.7 (9.0) &17.82 (0.01) &0.101 (0.016) &$-$0.66 (0.04) &2773 (579) &33.39 (0.36) &34.32 (0.11) & CSP-I \\
SN2008bu &1.149 &0.022 (0.0005) &54566.8 (7.0) &18.45 (0.03) &$-$0.268 (0.046) &1.44 (0.17) &5562 (680) &35.41 (0.22) &35.63 (0.09) & CSP-I \\
SN2008ga &1.865 &0.015 (0.0005) &54711.5 (7.0) &17.14 (0.01) &$-$0.054 (0.02) &0.84 (0.04) &5762 (435) &34.01 (0.15) &34.07 (0.09) & CSP-I \\
SN2008gi &0.181 &0.024 (0.0005) &54742.7 (9.0) &17.78 (0.01) &0.086 (0.010) &1.26 (0.02) &6021 (589) &34.62 (0.17) &34.74 (0.07) & CSP-I \\
SN2008gr &0.039 &0.022 (0.0005) &54769.6 (6.0) &17.45 (0.01) &$-$0.111 (0.010) &0.98 (0.03) &7124 (487) &34.70 (0.13) &34.45 (0.07) & CSP-I \\
SN2008hg &0.050 &0.019 (0.0005) &54779.8 (5.0) &18.50 (0.02) &0.059 (0.021) &$-$1.83 (0.08) &4437 (774) &34.86 (0.30) &34.73 (0.08) & CSP-I \\
SN2008if &0.090 &0.013 (0.0005) &54807.8 (5.0) &16.45 (0.01) &$-$0.144 (0.014) &1.22 (0.02) &6864 (328) &33.67 (0.13) &33.52 (0.10) & CSP-I \\
SN2009ao &0.106 &0.012 (0.0005) &54890.7 (4.0) &16.86 (0.01) &0.341 (0.012) &$-$0.45 (0.06) &5481 (372) &33.36 (0.15) &33.27 (0.10) & CSP-I \\
SN2009bu &0.070 &0.012 (0.0005) &54901.9 (8.0) &16.96 (0.01) &0.105 (0.007) &$-$0.37 (0.03) &6048 (428) &33.79 (0.16) &33.52 (0.11) & CSP-I \\
SN2009bz &0.110 &0.011 (0.0005) &54915.8 (4.0) &16.85 (0.01) &$-$0.053 (0.006) &0.03 (0.02) &5710 (433) &33.71 (0.16) &33.59 (0.11) & CSP-I \\
8321 &0.080 &0.107 (0.0007) &54353.6 (5.0) &21.18 (0.06) &$-$0.392 (0.084) &0.47 (0.64) &6900 (417) &38.58 (0.17) &38.20 (0.18) & SDSS-II \\
SN06gq &0.096 &0.069 (0.0007) &53992.4 (3.0) &20.40 (0.04) &$-$0.167 (0.061) &$-$0.05 (0.11) &4768 (877) &37.04 (0.32) &37.18 (0.09) & SDSS-II \\
SN06kn &0.194 &0.119 (0.0007) &54007.0 (1.5) &21.20 (0.12) &$-$0.120 (0.176) &1.20 (0.83) &6282 (508) &38.26 (0.26) &38.26 (0.27) & SDSS-II \\
SN06kv &0.080 &0.062 (0.0050) &54016.5 (4.0) &20.20 (0.08) &$-$0.009 (0.123) &1.18 (0.15) &5259 (451) &36.88 (0.28) &37.19 (0.23) & SDSS-II \\
SN07kw &0.074 &0.067 (0.0007) &54361.6 (2.5) &19.95 (0.03) &$-$0.043 (0.044) &0.83 (0.11) &5909 (506) &36.85 (0.17) &36.88 (0.07) & SDSS-II \\
SN07ky &0.105 &0.073 (0.0007) &54363.5 (3.0) &20.68 (0.05) &$-$0.064 (0.075) &0.62 (0.20) &5109 (420) &37.36 (0.18) &37.56 (0.10) & SDSS-II \\
SN07kz &0.320 &0.127 (0.0007) &54362.6 (3.5) &21.49 (0.10) &$-$0.185 (0.147) &0.74 (0.92) &6050 (419) &38.53 (0.22) &38.47 (0.27) & SDSS-II \\
SN07lb &0.496 &0.039 (0.0007) &54368.8 (7.0) &18.58 (0.01) &0.019 (0.024) &0.18 (0.08) &7593 (473) &35.85 (0.13) &35.32 (0.07) & SDSS-II \\
SN07lj &0.118 &0.049 (0.0050) &54370.2 (3.5) &19.69 (0.03) &$-$0.05 (0.047) &0.88 (0.08) &5836 (469) &36.57 (0.28) &36.63 (0.24) & SDSS-II \\
SN07lx &0.120 &0.056 (0.0007) &54374.5 (8.0) &20.15 (0.04) &0.010 (0.068) &0.06 (0.14) &5320 (487) &36.85 (0.18) &36.87 (0.09) & SDSS-II \\
SN07nr &0.079 &0.139 (0.0007) &54353.5 (5.0) &21.98 (0.13) &$-$0.123 (0.191) &1.06 (0.96) &5263 (385) &38.75 (0.29) &39.01 (0.32) & SDSS-II \\
SN07nw &0.204 &0.056 (0.0007) &54372.2 (7.0) &20.43 (0.06) &0.019 (0.088) &0.39 (0.26) &6469 (871) &37.43 (0.25) &37.22 (0.12) & SDSS-II \\
SN07ny &0.080 &0.142 (0.0007) &54367.8 (7.0) &21.78 (0.13) &$-$0.289 (0.196) &1.34 (1.47) &6424 (445) &39.01 (0.27) &38.96 (0.41) & SDSS-II \\
03D4bl &0.072 &0.317 (0.0011) &52822.0 (3.0) &24.36 (0.13) &$-$0.303 (0.156) &0.71 (1.43) &$\cdots$ &$\cdots$ &41.39 (0.38) &SNLS \\
04D1ha &0.073 &0.483 (0.0011) &53233.0 (3.0) &25.14 (0.21) &$-$0.2 (0.242) &0.05 (0.27) &$\cdots$ &$\cdots$ &41.96 (0.28) &SNLS \\
04D1ln &0.071 &0.206 (0.0011) &53274.0 (5.0) &23.29 (0.07) &$-$0.161 (0.071) &0.49 (0.23) &$\cdots$ &$\cdots$ &40.19 (0.11) &SNLS \\
04D1nz &0.072 &0.262 (0.0011) &53264.0 (4.0) &24.55 (0.15) &0.054 (0.195) &0.65 (0.27) &$\cdots$ &$\cdots$ &41.38 (0.22) &SNLS \\
04D1pj &0.076 &0.155 (0.0011) &53304.0 (8.0) &22.39 (0.04) &$-$0.055 (0.048) &0.20 (0.23) &7033 (392) &39.58 (0.15) &39.17 (0.09) & SNLS \\
04D1qa &0.072 &0.171 (0.0011) &53300.0 (3.0) &23.2 (0.10) &0.004 (0.116)&1.07 (0.30) &$\cdots$ &$\cdots$ &40.16 (0.15) &SNLS \\
04D4fu &0.072 &0.132 (0.0011) &53213.0 (6.0) &22.37 (0.04) &$-$0.095 (0.040) &0.83 (0.18) &6218 (389) &39.39 (0.15) &39.33 (0.08) & SNLS \\
05D1je &0.071 &0.308 (0.0011) &53647.0 (5.0) &24.79 (0.16) &$-$0.076 (0.183) &$-$1.82 (0.51) &$\cdots$ &$\cdots$ &41.09 (0.24) &SNLS \\
05D2ed &0.053 &0.197 (0.0011) &53417.0 (5.0) &22.72 (0.1) &0.104 (0.112) &$-$0.51 (0.59) &$\cdots$ &$\cdots$ &39.24 (0.19) &SNLS \\
05D4cb &0.073 &0.199 (0.0011) &53563.0 (3.0) &23.01 (0.06) &0.012 (0.072) &0.30 (0.09) &$\cdots$ &$\cdots$ &39.78 (0.09) &SNLS \\
05D4dn &0.073 &0.190 (0.0011) &53605.0 (7.0) &23.40 (0.08) &$-$0.078 (0.089) &0.63 (0.28) &5722 (1018) &40.28 (0.33) &40.17 (0.15) & SNLS \\
05D4du &0.072 &0.309 (0.0011) &53585.0 (5.0) &24.46 (0.14) &$-$0.389 (0.158) &$-$0.08 (0.2) &$\cdots$ &$\cdots$ &41.35 (0.18) &SNLS \\
06D1jx &0.079 &0.134 (0.0011) &54068.0 (6.0) &22.23 (0.02) &$-$0.094 (0.025) &$-$0.44 (0.25) &5923 (462) &39.18 (0.16) &38.86 (0.08) & SNLS \\
06D2ci &0.053 &0.221 (0.0011) &53768.0 (4.0) &23.42 (0.18) &0.043 (0.199) &0.91 (0.21) &$\cdots$ &$\cdots$ &40.32 (0.23) &SNLS \\
DES13X3fca &0.073 &0.095 (0.0011) &56542.0 (5.0) &21.53 (0.04) &$-$0.011 (0.048) &$-$0.59 (0.02) &5940 (545) &38.41 (0.18) &38.09 (0.08) & DES-SN \\
DES14C3aol &0.030 &0.076 (0.0011) &56894.2 (9.0) &21.58 (0.03) &$-$0.034 (0.035) &$-$0.60 (0.04) &3121 (470) &37.44 (0.26) &38.16 (0.12) & DES-SN \\
DES14C3rhw &0.033 &0.341 (0.0007) &57003.0 (2.0) &23.98 (0.12) &$-$0.383 (0.189) &$-$1.19 (0.37) &7362 (520) &41.48 (0.29) &40.60 (0.21)) & DES-SN \\
DES14X3ili &0.068 &0.141 (0.0011) &56946.0 (5.0) &22.46 (0.03) &0.013 (0.054) &$-$0.26 (0.10) &$\cdots$ &$\cdots$ &39.10 (0.08) &DES-SN \\
DES15C2eaz &0.034 &0.061 (0.0011) &57271.0 (5.0) &20.38 (0.02) &$-$0.109 (0.037) &1.55 (0.04) &5060 (523) &37.08 (0.19) &37.51 (0.07) & DES-SN \\
DES15C2lna &0.038 &0.065 (0.0011) &57306.0 (5.0) &21.06 (0.03) &0.093 (0.049) &0.57 (0.03) &$\cdots$ &$\cdots$ &37.86 (0.08) &DES-SN \\
DES15C2lpp &0.032 &0.181 (0.0011) &57306.5 (5.0) &22.81 (0.09) &0.234 (0.118) &0.19 (0.12) &$\cdots$ &$\cdots$ &39.44 (0.14) &DES-SN \\
DES15C2npz &0.026 &0.122 (0.0011) &57360.0 (7.0) &21.92 (0.07) &0.109 (0.103) &0.89 (0.09) &6509 (503) &38.87 (0.20) &38.78 (0.12) & DES-SN \\
DES15E1iuh &0.017 &0.104 (0.0011) &57281.1 (4.0) &21.50 (0.03) &$-$0.019 (0.045) &0.62 (0.04) &6768 (443) &38.60 (0.15) &38.36 (0.07) & DES-SN \\
DES15S1lrp &0.164 &0.222 (0.005) &57308.0 (4.0) &22.56 (0.1) &$-$0.042 (0.138) &2.05 (0.13) &$\cdots$ &$\cdots$ &39.77 (0.17) &DES-SN \\
DES15S2eaq &0.093 &0.067 (0.0011) &57270.0 (5.0) &21.10 (0.03) &0.088 (0.052) &$-$0.11 (0.03) &3724 (729) &37.16 (0.33) &37.73 (0.08) & DES-SN \\
DES15X2mku &0.068 &0.081 (0.0050) &57329.0 (3.0) &21.53 (0.05) &$-$0.064 (0.067) &$-$0.27 (0.09) &4885 (551) &38.14 (0.25) &38.21 (0.16) & DES-SN \\
DES16C2cbv &0.023 &0.109 (0.0011) &57657.0 (4.0) &21.36 (0.03) &$-$0.017 (0.043) &0.14 (0.06) &7189 (524) &38.56 (0.16) &38.10 (0.07) & DES-SN \\
DES16E1bkh &0.021 &0.115 (0.005) &57645.0 (8.0) &22.05 (0.1) &0.402 (0.136) &1.18 (0.16) &$\cdots$ &$\cdots$ &38.82 (0.19) &DES-SN \\
DES16S1gn &0.137 &0.190 (0.0011) &57620.0 (7.0) &22.48 (0.07) &0.202 (0.098) &$-$0.24 (0.10) &7606 (306) &39.61 (0.18) &39.02 (0.12) & DES-SN \\
DES16X2bkr &0.065 &0.158 (0.0011) &57647.0 (6.0) &22.22 (0.08) &$-$0.113 (0.098) &1.23 (0.15) &6165 (481) &39.24 (0.21) &39.28 (0.13) & DES-SN \\
DES16X3cpl &0.077 &0.204 (0.0011) &57677.0 (6.0) &22.99 (0.05) &$-$0.075 (0.059) &$-$0.07 (0.12) &$\cdots$ &$\cdots$ &39.72 (0.09) &DES-SN \\
DES17E2bhj &0.02 &0.186 (0.0011) &58018.0 (3.0) &23.09 (0.11) &0.038 (0.142) &0.69 (0.17)&$\cdots$ &$\cdots$ &39.94 (0.18) &DES-SN \\
DES17S1bxt &0.174 &0.355 (0.0011) &58038.0 (5.0) &24.68 (0.18) &$-$0.370 (0.277) &0.81 (0.39) &7219 (501) &42.15 (0.36) &41.77 (0.29) & DES-SN \\
DES17X1aow &0.055 &0.138 (0.0011) &58005.0 (9.0) &21.64 (0.04) &$-$0.007 (0.058) &0.05 (0.05) &5298 (795) &38.34 (0.27) &38.36 (0.08) & DES-SN \\
DES17X1axb &0.053 &0.138 (0.0011) &58017.0 (5.0) &22.44 (0.06) &$-$0.034 (0.084) &0.12 (0.10) &5513 (497) &39.22 (0.21) &39.19 (0.11) & DES-SN \\
DES17X3dub &0.072 &0.121 (0.0011) &58059.0 (4.0) &22.87 (0.04) &$-$0.090 (0.056) &0.03 (0.15) &4850 (550) &39.48 (0.22) &39.63 (0.08) & DES-SN \\
SN2016jhj &0.0515 &0.341 (0.0011) &57719.6 (2.0) &23.27 (0.05) &$-$0.142 (0.052) &3.24 (0.17) &9103 (534) &40.95 (0.19) &40.83 (0.09) & HSC \\

\hline
\label{SN_sample}
\end{longtable}
\setlength\LTleft{+0.8cm}
\vspace{1cm}

\bsp	
\label{lastpage}
\end{document}